\newcommand{\labl}[1]{}

\newcommand{\sEction}[1]
{\section{\fontsize{14}{16pt}\fontseries{bx}\fontshape{n}\selectfont #1}}
\newcommand{\subsEction}[1]
{\subsection{\fontsize{12}{14pt}\selectfont #1}}

\newcommand{\okrug}
{\unitlength=1mm
\makebox[10mm][l]{
\raisebox{-3mm}[4mm][3mm]{
\put(4,4){\circle{6}}
\put(1,4){\circle*{1.50}}
}}}

\newcommand{\lintwopt}[2]
{\unitlength=0.7mm
\makebox[23mm][l]{
\raisebox{-0mm}[5mm][0mm]{
\put(2,1){\line(1,0){27}}
\put(11,1){\circle*{1.5}}
\put(20,1){\circle*{1.5}}
\put(10,4){\makebox(0,0)[cc]{$#1$}}
\put(21,4){\makebox(0,0)[cc]{$#2$}}
}}}

\newcommand{\lintwoptright}[2]
{\unitlength=0.7mm
\makebox[23mm][l]{
\raisebox{-0mm}[5mm][0mm]{
\put(2,1){\line(1,0){27}}
\put(7,1){\vector(1,0){0}}
\put(16,1){\vector(1,0){0}}
\put(25,1){\vector(1,0){0}}
\put(11,1){\circle*{1.5}}
\put(20,1){\circle*{1.5}}
\put(10,4){\makebox(0,0)[cc]{$#1$}}
\put(21,4){\makebox(0,0)[cc]{$#2$}}
}}}

\newcommand{\anntwo}[2]
{\unitlength=1mm
\makebox[34mm][l]{
\raisebox{-15mm}[20mm][16mm]{
\put(16,16){\circle{10}}
\put(16,16){\oval(30,30)[]}
\put(7,30){\line(1,0){7}}
\put(18,30){\line(1,0){7}}
\put(10,34){\makebox(0,0)[cc]{$#1$}}
\put(22,34){\makebox(0,0)[cc]{$#2$}}
}}}

\newcommand{\anntwoline}[2]
{\unitlength=1mm
\makebox[34mm][l]{
\raisebox{-15mm}[20mm][16mm]{
\put(16,16){\circle{10}}
\put(16,16){\oval(30,30)[]}
\put(7,30){\line(1,0){7}}
\put(18,30){\line(1,0){7}}
\put(10,34){\makebox(0,0)[cc]{$#1$}}
\put(22,34){\makebox(0,0)[cc]{$#2$}}
\put(1,24){\line(1,0){30}}
}}}

\newcommand{\anntwogamdel}[2]
{\unitlength=1mm
\makebox[34mm][l]{
\raisebox{-15mm}[20mm][16mm]{
\put(16,16){\circle{10}}
\put(16,16){\oval(30,30)[]}
\put(7,30){\line(1,0){7}}
\put(18,30){\line(1,0){7}}
\put(10,34){\makebox(0,0)[cc]{$#1$}}
\put(22,34){\makebox(0,0)[cc]{$#2$}}
\put(16,11){\line(0,-1){10}}
\put(19,6){\makebox(0,0)[cc]{$\delta$}}
\put(16,21){\line(0,1){10}}
\put(19,24){\makebox(0,0)[cc]{$\gamma$}}
}}}

\newcommand{\diskfoursix}[6]
{\unitlength=1mm
\makebox[28mm][l]{
\raisebox{-14mm}[15mm][13mm]{
\put(1,10){\line(1,-1){7}}
\put(8,3){\line(1,0){10}}
\put(18,3){\line(1,1){7}}
\put(25,10){\line(0,1){10}}
\put(25,20){\line(-1,1){7}}
\put(18,27){\line(-1,0){10}}
\put(8,27){\line(-1,-1){7}}
\put(1,20){\line(0,-1){10}}
\put(1,11){\line(1,-1){8}}
\put(9,27){\line(-1,-1){8}}
\put(17,3){\line(1,1){8}}
\put(2,26){\makebox(0,0)[cc]{$#1$}}
\put(2,4){\makebox(0,0)[cc]{$#2$}}
\put(24,4){\makebox(0,0)[cc]{$#3$}}
\put(17,27){\line(1,-1){8}}
\put(24,26){\makebox(0,0)[cc]{$#4$}}
\put(13,3){\line(0,1){24}}
\put(1,15){\line(1,0){24}}
\put(11,23){\makebox(0,0)[cc]{$#5$}}
\put(21,17){\makebox(0,0)[cc]{$#6$}}
}}}

\newcommand{\annone}[1]
{\unitlength=1mm
\makebox[34mm][l]{
\raisebox{-16mm}[20mm][16mm]{
\put(16,16){\circle{10}}
\put(16,16){\oval(30,30)[]}
\put(7,30){\line(1,0){18}}
\put(16,34){\makebox(0,0)[cc]{$#1$}}
}}}

\newcommand{\annonetwo}[2]
{\unitlength=1mm
\makebox[34mm][l]{
\raisebox{-16mm}[20mm][16mm]{
\put(16,16){\circle{10}}
\put(16,16){\oval(30,30)[]}
\put(7,30){\line(1,0){18}}
\put(16,34){\makebox(0,0)[cc]{$#1$}}
\put(16,1){\line(0,1){10}}
\put(19,6){\makebox(0,0)[cc]{$#2$}}
}}}

\newcommand{\disktre}[3]
{\unitlength=0.75mm
\makebox[27mm][l]{
\raisebox{-11mm}[13mm][11mm]{
\put(1,15){\line(2,-3){8}}
\put(9,3){\line(1,0){14}}
\put(23,3){\line(2,3){8}}
\put(31,15){\line(-2,3){8}}
\put(23,27){\line(-1,0){14}}
\put(9,27){\line(-2,-3){8}}
\put(1,5){\makebox(0,0)[cc]{$#2$}}
\put(16,30){\makebox(0,0)[cc]{$#1$}}
\put(30,5){\makebox(0,0)[cc]{$#3$}}
\put(2,16){\line(2,-3){8.67}}
\put(30,16){\line(-2,-3){8.67}}
\put(8,26){\line(1,0){16}}
}}}

\newcommand{\disktwo}[2]
{\unitlength=0.5mm
\makebox[14mm][l]{
\raisebox{-6mm}[9mm][6mm]{
\put(11,26){\makebox(0,0)[cc]{$#1$}}
\put(11,3){\makebox(0,0)[cc]{$#2$}}
\put(1,7){\framebox(20,15)[cc]{}}
\put(1,8){\line(1,0){20}}
\put(21,21){\line(-1,0){20}}
}}}

\newcommand{\diskone}[1]
{\unitlength=0.66mm
\linethickness{0.4pt}
\makebox[12mm][l]{
\raisebox{-4mm}[6mm][4mm]{ 
\put(7.50,8){\oval(13,14)[b]}
\put(1,8){\line(1,0){13}}
\put(14,9){\line(-1,0){13}}
\put(8,12){\makebox(0,0)[cc]{$#1$}}
}}}

\newcommand{\torchokur}[2]
{\unitlength=0.75pt
\makebox[36 pt][l]{
\raisebox{-4 pt}[15 pt][7.5 pt]{
\put(0,0){\line(1,0){40}}
\put(12,0){\vector(1,0){0}}
\put(32,0){\vector(1,0){0}}
\put(20,0){\line(0,1){20}}
\put(20,12){\vector(0,1){0}}
\put(20,20){\circle*{4}}
\put(20,0){\circle*{4}}
\put(3,4){$#1$}
\put(28,4){$#2$}
}}}

\newcommand{\torchokdr}[2]
{\unitlength=0.75pt
\makebox[36 pt][l]{
\raisebox{-4 pt}[15 pt][7.5 pt]{
\put(0,0){\line(1,0){40}}
\put(12,0){\vector(1,0){0}}
\put(32,0){\vector(1,0){0}}
\put(20,0){\line(0,1){20}}
\put(20,8){\vector(0,-1){0}}
\put(20,20){\circle*{4}}
\put(20,0){\circle*{4}}
\put(3,4){$#1$}
\put(28,4){$#2$}
}}}

\newcommand{\torchokul}[2]
{\unitlength=0.75pt
\makebox[36 pt][l]{
\raisebox{-4 pt}[15 pt][7.5 pt]{
\put(0,0){\line(1,0){40}}
\put(8,0){\vector(-1,0){0}}
\put(28,0){\vector(-1,0){0}}
\put(20,0){\line(0,1){20}}
\put(20,12){\vector(0,1){0}}
\put(20,20){\circle*{4}}
\put(20,0){\circle*{4}}
\put(3,4){$#1$}
\put(28,4){$#2$}
}}}

\newcommand{\lezhak}[2]
{\unitlength=0.75pt
\makebox[36 pt][l]{
\raisebox{3 pt}[16 pt][2 pt]{
\put(0,0){\line(1,0){40}}
\put(20,0){\circle*{4}}
\put(3,4){$#1$}
\put(28,4){$#2$}
}}}

\newcommand{\lezhakright}[2]
{\unitlength=0.75pt
\makebox[36 pt][l]{
\raisebox{3 pt}[16 pt][2 pt]{
\put(0,0){\line(1,0){40}}
\put(12,0){\vector(1,0){0}}
\put(32,0){\vector(1,0){0}}
\put(20,0){\circle*{4}}
\put(3,4){$#1$}
\put(28,4){$#2$}
}}}

\newcommand{\lezhakleft}[2]
{\unitlength=0.75pt
\makebox[36 pt][l]{
\raisebox{3 pt}[16 pt][2 pt]{
\put(0,0){\line(1,0){40}}
\put(8,0){\vector(-1,0){0}}
\put(28,0){\vector(-1,0){0}}
\put(20,0){\circle*{4}}
\put(3,4){$#1$}
\put(28,4){$#2$}
}}}

\newcommand{\lezhakpt}[2]
{\unitlength=0.75pt
\makebox[36 pt][l]{
\raisebox{3 pt}[16 pt][2 pt]{
\put(0,0){\line(1,0){40}}
\put(20,0){\circle*{4}}
\put(40,0){\circle*{4}}
\put(3,4){$#1$}
\put(28,4){$#2$}
}}}

\newcommand{\lezhakptright}[2]
{\unitlength=0.75pt
\makebox[36 pt][l]{
\raisebox{3 pt}[16 pt][2 pt]{
\put(0,0){\line(1,0){40}}
\put(12,0){\vector(1,0){0}}
\put(32,0){\vector(1,0){0}}
\put(20,0){\circle*{4}}
\put(40,0){\circle*{4}}
\put(3,4){$#1$}
\put(28,4){$#2$}
}}}

\newcommand{\ptlezhak}[2]
{\unitlength=0.75pt
\makebox[36 pt][l]{
\raisebox{3 pt}[16 pt][2 pt]{
\put(0,0){\line(1,0){40}}
\put(0,0){\circle*{4}}
\put(20,0){\circle*{4}}
\put(3,4){$#1$}
\put(28,4){$#2$}
}}}

\newcommand{\ptlezhakright}[2]
{\unitlength=0.75pt
\makebox[36 pt][l]{
\raisebox{3 pt}[16 pt][2 pt]{
\put(0,0){\line(1,0){40}}
\put(12,0){\vector(1,0){0}}
\put(32,0){\vector(1,0){0}}
\put(0,0){\circle*{4}}
\put(20,0){\circle*{4}}
\put(3,4){$#1$}
\put(28,4){$#2$}
}}}

\newcommand{\pointedge}[1]
{\unitlength=0.75pt
\makebox[22 pt][l]{
\raisebox{3 pt}[16 pt][2 pt]{
\put(2,0){\line(1,0){20}}
\put(2,0){\circle*{4}}
\put(10,4){$#1$}
}}}

\newcommand{\pointedger}[1]
{\unitlength=0.75pt
\makebox[22 pt][l]{
\raisebox{3 pt}[16 pt][2 pt]{
\put(2,0){\line(1,0){20}}
\put(14,0){\vector(1,0){0}}
\put(2,0){\circle*{4}}
\put(10,4){$#1$}
}}}

\newcommand{\pointedgel}[1]
{\unitlength=0.75pt
\makebox[22 pt][l]{
\raisebox{3 pt}[16 pt][2 pt]{
\put(2,0){\line(1,0){20}}
\put(10,0){\vector(-1,0){0}}
\put(2,0){\circle*{4}}
\put(10,4){$#1$}
}}}

\newcommand{\edgepoint}[1]
{\unitlength=0.75pt
\makebox[22 pt][l]{
\raisebox{3 pt}[16 pt][2 pt]{
\put(2,0){\line(1,0){20}}
\put(22,0){\circle*{4}}
\put(10,4){$#1$}
}}}

\newcommand{\edgepointright}[1]
{\unitlength=0.75pt
\makebox[22 pt][l]{
\raisebox{3 pt}[16 pt][2 pt]{
\put(2,0){\line(1,0){20}}
\put(14,0){\vector(1,0){0}}
\put(22,0){\circle*{4}}
\put(10,4){$#1$}
}}}

\newcommand{\twopointedge}
{\unitlength=0.75pt
\makebox[24 pt][l]{
\raisebox{3 pt}[16 pt][2 pt]{
\put(2,0){\line(1,0){20}}
\put(2,0){\circle*{4}}
\put(22,0){\circle*{4}}
}}}

\newcommand{\twopointedger}
{\unitlength=0.75pt
\makebox[24 pt][l]{
\raisebox{3 pt}[16 pt][2 pt]{
\put(2,0){\line(1,0){20}}
\put(14,0){\vector(1,0){0}}
\put(2,0){\circle*{4}}
\put(22,0){\circle*{4}}
}}}

\newcommand{\twopointedgel}
{\unitlength=0.75pt
\makebox[24 pt][l]{
\raisebox{3 pt}[16 pt][2 pt]{
\put(2,0){\line(1,0){20}}
\put(10,0){\vector(-1,0){0}}
\put(2,0){\circle*{4}}
\put(22,0){\circle*{4}}
}}}

\newcommand{\stoyak}[2]
{\unitlength=0.75pt
\makebox[23 pt][l]{
\raisebox{3 pt}[22 pt][16 pt]{
\put(20,-20){\line(0,1){40}}
\put(20,0){\circle*{4}}
\put(5,7){$#1$}
\put(5,-17){$#2$}
}}}

\newcommand{\stoyakd}[2]
{\unitlength=0.75pt
\makebox[23 pt][l]{
\raisebox{3 pt}[22 pt][16 pt]{
\put(20,-20){\line(0,1){40}}
\put(20,8){\vector(0,-1){0}}
\put(20,-12){\vector(0,-1){0}}
\put(20,0){\circle*{4}}
\put(5,7){$#1$}
\put(5,-17){$#2$}
}}}

\newcommand{\trileftl}[3]
{\unitlength=0.75pt
\makebox[39 pt][l]{
\raisebox{3 pt}[19 pt][13 pt]{
\put(22,0){\line(-1,0){23}}
\put(8,0){\vector(-1,0){0}}
\put(22,0){\line(3,-5){12}}
\put(22,0){\line(3,5){12}}
\put(22,0){\circle*{4}}
\put(2,4){$#1$}
\put(35,-17){$#2$}
\put(35,10){$#3$}
}}}

\newcommand{\trileftr}[3]
{\unitlength=0.75pt
\makebox[39 pt][l]{
\raisebox{3 pt}[19 pt][13 pt]{
\put(22,0){\line(-1,0){23}}
\put(13,0){\vector(1,0){0}}
\put(22,0){\line(3,-5){12}}
\put(22,0){\line(3,5){12}}
\put(22,0){\circle*{4}}
\put(2,4){$#1$}
\put(35,-17){$#2$}
\put(35,10){$#3$}
}}}

\newcommand{\triup}[3]
{\unitlength=0.75pt
\makebox[33 pt]{
\raisebox{3 pt}[22.5 pt][16.5 pt]{
\put(-2,0){\line(0,1){23}}
\put(-2,0){\line(-5,-3){20}}
\put(-2,0){\line(5,-3){20}}
\put(-2,0){\circle*{4}}
\put(-17,16){$#1$}
\put(-21,-23){$#2$}
\put(7,-23){$#3$}
}}}

\newcommand{\triuprpt}[3]
{\unitlength=0.75pt
\makebox[33 pt]{
\raisebox{3 pt}[22.5 pt][16.5 pt]{
\put(-2,0){\line(0,1){23}}
\put(-2,0){\line(-5,-3){20}}
\put(-2,0){\line(5,-3){20}}
\put(-2,0){\circle*{4}}
\put(18,-12){\circle*{4}}
\put(-17,16){$#1$}
\put(-21,-23){$#2$}
\put(7,-23){$#3$}
}}}

\newcommand{\triupioo}[3]
{\unitlength=0.75pt
\makebox[33 pt]{
\raisebox{3 pt}[22.5 pt][16.5 pt]{
\put(-2,0){\line(0,1){23}}
\put(-2,12.5){\vector(0,-1){0}}
\put(-2,0){\line(-5,-3){20}}
\put(-12,-6){\vector(-3,-2){0}}
\put(-2,0){\line(5,-3){20}}
\put(8,-6){\vector(3,-2){0}}
\put(-2,0){\circle*{4}}
\put(-17,16){$#1$}
\put(-21,-23){$#2$}
\put(7,-23){$#3$}
}}}

\newcommand{\triupioolpt}[3]
{\unitlength=0.75pt
\makebox[33 pt]{
\raisebox{3 pt}[22.5 pt][16.5 pt]{
\put(-2,0){\line(0,1){23}}
\put(-2,12.5){\vector(0,-1){0}}
\put(-2,0){\line(-5,-3){20}}
\put(-22,-12){\circle*{4}}
\put(-12,-6){\vector(-3,-2){0}}
\put(-2,0){\line(5,-3){20}}
\put(8,-6){\vector(3,-2){0}}
\put(-2,0){\circle*{4}}
\put(-17,16){$#1$}
\put(-21,-23){$#2$}
\put(7,-23){$#3$}
}}}

\newcommand{\triupioorpt}[3]
{\unitlength=0.75pt
\makebox[33 pt]{
\raisebox{3 pt}[22.5 pt][16.5 pt]{
\put(-2,0){\line(0,1){23}}
\put(-2,12.5){\vector(0,-1){0}}
\put(-2,0){\line(-5,-3){20}}
\put(-12,-6){\vector(-3,-2){0}}
\put(-2,0){\line(5,-3){20}}
\put(18,-12){\circle*{4}}
\put(8,-6){\vector(3,-2){0}}
\put(-2,0){\circle*{4}}
\put(-17,16){$#1$}
\put(-21,-23){$#2$}
\put(7,-23){$#3$}
}}}

\newcommand{\triupiio}[3]
{\unitlength=0.75pt
\makebox[33 pt]{
\raisebox{3 pt}[22.5 pt][16.5 pt]{
\put(-2,0){\line(0,1){23}}
\put(-2,12.5){\vector(0,-1){0}}
\put(-2,0){\line(-5,-3){20}}
\put(-12,-6){\vector(-3,-2){0}}
\put(-2,0){\line(5,-3){20}}
\put(8,-6){\vector(-3,2){0}}
\put(-2,0){\circle*{4}}
\put(-17,16){$#1$}
\put(-21,-23){$#2$}
\put(7,-23){$#3$}
}}}

\newcommand{\triupoio}[3]
{\unitlength=0.75pt
\makebox[33 pt]{
\raisebox{3 pt}[22.5 pt][16.5 pt]{
\put(-2,0){\line(0,1){23}}
\put(-2,12.5){\vector(0,1){0}}
\put(-2,0){\line(-5,-3){20}}
\put(-12,-6){\vector(-3,-2){0}}
\put(-2,0){\line(5,-3){20}}
\put(8,-6){\vector(-3,2){0}}
\put(-2,0){\circle*{4}}
\put(-17,16){$#1$}
\put(-21,-23){$#2$}
\put(7,-23){$#3$}
}}}

\newcommand{\triupoii}[3]
{\unitlength=0.75pt
\makebox[33 pt]{
\raisebox{3 pt}[22.5 pt][16.5 pt]{
\put(-2,0){\line(0,1){23}}
\put(-2,12.5){\vector(0,1){0}}
\put(-2,0){\line(-5,-3){20}}
\put(-12,-6){\vector(3,2){0}}
\put(-2,0){\line(5,-3){20}}
\put(8,-6){\vector(-3,2){0}}
\put(-2,0){\circle*{4}}
\put(-17,16){$#1$}
\put(-21,-23){$#2$}
\put(7,-23){$#3$}
}}}

\newcommand{\triupooi}[3]
{\unitlength=0.75pt
\makebox[33 pt]{
\raisebox{3 pt}[22.5 pt][16.5 pt]{
\put(-2,0){\line(0,1){23}}
\put(-2,12.5){\vector(0,1){0}}
\put(-2,0){\line(-5,-3){20}}
\put(-12,-6){\vector(3,2){0}}
\put(-2,0){\line(5,-3){20}}
\put(8,-6){\vector(3,-2){0}}
\put(-2,0){\circle*{4}}
\put(-17,16){$#1$}
\put(-21,-23){$#2$}
\put(7,-23){$#3$}
}}}

\newcommand{\triupioi}[3]
{\unitlength=0.75pt
\makebox[33 pt]{
\raisebox{3 pt}[22.5 pt][16.5 pt]{
\put(-2,0){\line(0,1){23}}
\put(-2,12.5){\vector(0,-1){0}}
\put(-2,0){\line(-5,-3){20}}
\put(-12,-6){\vector(3,2){0}}
\put(-2,0){\line(5,-3){20}}
\put(8,-6){\vector(3,-2){0}}
\put(-2,0){\circle*{4}}
\put(-17,16){$#1$}
\put(-21,-23){$#2$}
\put(7,-23){$#3$}
}}}

\newcommand{\trirightl}[3]
{\unitlength=0.75pt
\makebox[39 pt][l]{
\raisebox{3 pt}[19 pt][13 pt]{
\put(22,0){\line(1,0){23}}
\put(31,0){\vector(-1,0){0}}
\put(22,0){\line(-3,5){12}}
\put(22,0){\line(-3,-5){12}}
\put(22,0){\circle*{4}}
\put(35,4){$#1$}
\put(-3,10){$#2$}
\put(-3,-17){$#3$}
}}}

\newcommand{\trirightr}[3]
{\unitlength=0.75pt
\makebox[39 pt][l]{
\raisebox{3 pt}[19 pt][13 pt]{
\put(22,0){\line(1,0){23}}
\put(36,0){\vector(1,0){0}}
\put(22,0){\line(-3,5){12}}
\put(22,0){\line(-3,-5){12}}
\put(22,0){\circle*{4}}
\put(35,4){$#1$}
\put(-3,10){$#2$}
\put(-3,-17){$#3$}
}}}

\newcommand{\tridowniio}[3]
{\unitlength=0.75pt
\makebox[33 pt]{
\raisebox{3 pt}[22.5 pt][16.5 pt]{
\put(-2,0){\line(0,-1){23}}
\put(-2,-13){\vector(0,-1){0}}
\put(-2,0){\line(5,3){20}}
\put(8,6){\vector(-3,-2){0}}
\put(-2,0){\line(-5,3){20}}
\put(-12,6){\vector(3,-2){0}}
\put(-2,0){\circle*{4}}
\put(-14,-23){$#1$}
\put(8,13){$#2$}
\put(-20,13){$#3$}
}}}

\newcommand{\tridownioo}[3]
{\unitlength=0.75pt
\makebox[33 pt]{
\raisebox{3 pt}[22.5 pt][16.5 pt]{
\put(-2,0){\line(0,-1){23}}
\put(-2,-13){\vector(0,-1){0}}
\put(-2,0){\line(5,3){20}}
\put(8,6){\vector(3,2){0}}
\put(-2,0){\line(-5,3){20}}
\put(-12,6){\vector(3,-2){0}}
\put(-2,0){\circle*{4}}
\put(-14,-23){$#1$}
\put(8,13){$#2$}
\put(-20,13){$#3$}
}}}

\newcommand{\tridownoio}[3]
{\unitlength=0.75pt
\makebox[33 pt]{
\raisebox{3 pt}[22.5 pt][16.5 pt]{
\put(-2,0){\line(0,-1){23}}
\put(-2,-13){\vector(0,-1){0}}
\put(-2,0){\line(5,3){20}}
\put(8,6){\vector(-3,-2){0}}
\put(-2,0){\line(-5,3){20}}
\put(-12,6){\vector(-3,2){0}}
\put(-2,0){\circle*{4}}
\put(-14,-23){$#1$}
\put(8,13){$#2$}
\put(-20,13){$#3$}
}}}

\newcommand{\tarahor}[5]
{\unitlength=0.75pt
\makebox[45 pt]{
\raisebox{3 pt}[27 pt][21 pt]{
\put(-19,0){\line(-3,5){12}}
\put(-19,0){\line(-3,-5){12}}
\put(15,0){\line(3,-5){12}}
\put(15,0){\line(3,5){12}}
\put(-19,0){\line(1,0){34}}
\put(-19,0){\circle*{4}}
\put(15,0){\circle*{4}}
\put(-25,20){$#1$}
\put(-25,-28){$#2$}
\put(11,-28){$#3$}
\put(11,20){$#4$}
\put(-7,5){$#5$}
}}}

\newcommand{\tarahorl}[5]
{\unitlength=0.75pt
\makebox[45 pt]{
\raisebox{3 pt}[27 pt][21 pt]{
\put(-19,0){\line(-3,5){12}}
\put(-19,0){\line(-3,-5){12}}
\put(15,0){\line(3,-5){12}}
\put(15,0){\line(3,5){12}}
\put(-19,0){\line(1,0){34}}
\put(-2,0){\vector(-1,0){0}}
\put(-19,0){\circle*{4}}
\put(15,0){\circle*{4}}
\put(-25,20){$#1$}
\put(-25,-28){$#2$}
\put(11,-28){$#3$}
\put(11,20){$#4$}
\put(-7,5){$#5$}
}}}

\newcommand{\tarahorr}[5]
{\unitlength=0.75pt
\makebox[45 pt]{
\raisebox{3 pt}[27 pt][21 pt]{
\put(-19,0){\line(-3,5){12}}
\put(-19,0){\line(-3,-5){12}}
\put(15,0){\line(3,-5){12}}
\put(15,0){\line(3,5){12}}
\put(-19,0){\line(1,0){34}}
\put(-2,0){\vector(1,0){0}}
\put(-19,0){\circle*{4}}
\put(15,0){\circle*{4}}
\put(-25,20){$#1$}
\put(-25,-28){$#2$}
\put(11,-28){$#3$}
\put(11,20){$#4$}
\put(-7,5){$#5$}
}}}

\newcommand{\tarahorioiol}[5]
{\unitlength=0.75pt
\makebox[45 pt]{
\raisebox{3 pt}[27 pt][21 pt]{
\put(-19,0){\line(-3,5){12}}
\put(-25,10){\vector(2,-3){0}}
\put(-19,0){\line(-3,-5){12}}
\put(-25,-10){\vector(-2,-3){0}}
\put(15,0){\line(3,-5){12}}
\put(21,-10){\vector(-2,3){0}}
\put(15,0){\line(3,5){12}}
\put(21,10){\vector(2,3){0}}
\put(-19,0){\line(1,0){34}}
\put(-2,0){\vector(-1,0){0}}
\put(-19,0){\circle*{4}}
\put(15,0){\circle*{4}}
\put(-25,20){$#1$}
\put(-25,-28){$#2$}
\put(11,-28){$#3$}
\put(11,20){$#4$}
\put(-7,5){$#5$}
}}}

\newcommand{\tarahoriooil}[5]
{\unitlength=0.75pt
\makebox[45 pt]{
\raisebox{3 pt}[27 pt][21 pt]{
\put(-19,0){\line(-3,5){12}}
\put(-25,10){\vector(2,-3){0}}
\put(-19,0){\line(-3,-5){12}}
\put(-25,-10){\vector(-2,-3){0}}
\put(15,0){\line(3,-5){12}}
\put(21,-10){\vector(2,-3){0}}
\put(15,0){\line(3,5){12}}
\put(21,10){\vector(-2,-3){0}}
\put(-19,0){\line(1,0){34}}
\put(-2,0){\vector(-1,0){0}}
\put(-19,0){\circle*{4}}
\put(15,0){\circle*{4}}
\put(-25,20){$#1$}
\put(-25,-28){$#2$}
\put(11,-28){$#3$}
\put(11,20){$#4$}
\put(-7,5){$#5$}
}}}

\newcommand{\tarahoriooir}[5]
{\unitlength=0.75pt
\makebox[45 pt]{
\raisebox{3 pt}[27 pt][21 pt]{
\put(-19,0){\line(-3,5){12}}
\put(-25,10){\vector(2,-3){0}}
\put(-19,0){\line(-3,-5){12}}
\put(-25,-10){\vector(-2,-3){0}}
\put(15,0){\line(3,-5){12}}
\put(21,-10){\vector(2,-3){0}}
\put(15,0){\line(3,5){12}}
\put(21,10){\vector(-2,-3){0}}
\put(-19,0){\line(1,0){34}}
\put(-2,0){\vector(1,0){0}}
\put(-19,0){\circle*{4}}
\put(15,0){\circle*{4}}
\put(-25,20){$#1$}
\put(-25,-28){$#2$}
\put(11,-28){$#3$}
\put(11,20){$#4$}
\put(-7,5){$#5$}
}}}

\newcommand{\tarahoriooor}[5]
{\unitlength=0.75pt
\makebox[45 pt]{
\raisebox{3 pt}[27 pt][21 pt]{
\put(-19,0){\line(-3,5){12}}
\put(-25,10){\vector(2,-3){0}}
\put(-19,0){\line(-3,-5){12}}
\put(-25,-10){\vector(-2,-3){0}}
\put(15,0){\line(3,-5){12}}
\put(21,-10){\vector(2,-3){0}}
\put(15,0){\line(3,5){12}}
\put(21,10){\vector(2,3){0}}
\put(-19,0){\line(1,0){34}}
\put(-2,0){\vector(1,0){0}}
\put(-19,0){\circle*{4}}
\put(15,0){\circle*{4}}
\put(-25,20){$#1$}
\put(-25,-28){$#2$}
\put(11,-28){$#3$}
\put(11,20){$#4$}
\put(-7,5){$#5$}
}}}

\newcommand{\tarahoriobor}[5]
{\unitlength=0.75pt
\makebox[45 pt]{
\raisebox{3 pt}[27 pt][21 pt]{
\put(-19,0){\line(-3,5){12}}
\put(-25,10){\vector(2,-3){0}}
\put(-19,0){\line(-3,-5){12}}
\put(-25,-10){\vector(-2,-3){0}}
\put(15,0){\line(3,5){12}}
\put(21,10){\vector(2,3){0}}
\put(-19,0){\line(1,0){34}}
\put(-2,0){\vector(1,0){0}}
\put(-19,0){\circle*{4}}
\put(15,0){\circle*{4}}
\put(-25,20){$#1$}
\put(-25,-28){$#2$}
\put(11,-28){$#3$}
\put(11,20){$#4$}
\put(-7,5){$#5$}
}}}

\newcommand{\tarahoriooorptdr}[5]
{\unitlength=0.75pt
\makebox[45 pt]{
\raisebox{3 pt}[27 pt][21 pt]{
\put(-19,0){\line(-3,5){12}}
\put(-25,10){\vector(2,-3){0}}
\put(-19,0){\line(-3,-5){12}}
\put(-25,-10){\vector(-2,-3){0}}
\put(15,0){\line(3,-5){12}}
\put(27,-20){\circle*{4}}
\put(21,-10){\vector(2,-3){0}}
\put(15,0){\line(3,5){12}}
\put(21,10){\vector(2,3){0}}
\put(-19,0){\line(1,0){34}}
\put(-2,0){\vector(1,0){0}}
\put(-19,0){\circle*{4}}
\put(15,0){\circle*{4}}
\put(-25,20){$#1$}
\put(-25,-28){$#2$}
\put(11,-28){$#3$}
\put(11,20){$#4$}
\put(-7,5){$#5$}
}}}

\newcommand{\tarahorioior}[5]
{\unitlength=0.75pt
\makebox[45 pt]{
\raisebox{3 pt}[27 pt][21 pt]{
\put(-19,0){\line(-3,5){12}}
\put(-25,10){\vector(2,-3){0}}
\put(-19,0){\line(-3,-5){12}}
\put(-25,-10){\vector(-2,-3){0}}
\put(15,0){\line(3,-5){12}}
\put(21,-10){\vector(-2,3){0}}
\put(15,0){\line(3,5){12}}
\put(21,10){\vector(2,3){0}}
\put(-19,0){\line(1,0){34}}
\put(-2,0){\vector(1,0){0}}
\put(-19,0){\circle*{4}}
\put(15,0){\circle*{4}}
\put(-25,20){$#1$}
\put(-25,-28){$#2$}
\put(11,-28){$#3$}
\put(11,20){$#4$}
\put(-7,5){$#5$}
}}}

\newcommand{\tarahoroooil}[5]
{\unitlength=0.75pt
\makebox[45 pt]{
\raisebox{3 pt}[27 pt][21 pt]{
\put(-19,0){\line(-3,5){12}}
\put(-25,10){\vector(-2,3){0}}
\put(-19,0){\line(-3,-5){12}}
\put(-25,-10){\vector(-2,-3){0}}
\put(15,0){\line(3,-5){12}}
\put(21,-10){\vector(2,-3){0}}
\put(15,0){\line(3,5){12}}
\put(21,10){\vector(-2,-3){0}}
\put(-19,0){\line(1,0){34}}
\put(-2,0){\vector(-1,0){0}}
\put(-19,0){\circle*{4}}
\put(15,0){\circle*{4}}
\put(-25,20){$#1$}
\put(-25,-28){$#2$}
\put(11,-28){$#3$}
\put(11,20){$#4$}
\put(-7,5){$#5$}
}}}

\newcommand{\tarahorooiol}[5]
{\unitlength=0.75pt
\makebox[45 pt]{
\raisebox{3 pt}[27 pt][21 pt]{
\put(-19,0){\line(-3,5){12}}
\put(-25,10){\vector(-2,3){0}}
\put(-19,0){\line(-3,-5){12}}
\put(-25,-10){\vector(-2,-3){0}}
\put(15,0){\line(3,-5){12}}
\put(21,-10){\vector(-2,3){0}}
\put(15,0){\line(3,5){12}}
\put(21,10){\vector(2,3){0}}
\put(-19,0){\line(1,0){34}}
\put(-2,0){\vector(-1,0){0}}
\put(-19,0){\circle*{4}}
\put(15,0){\circle*{4}}
\put(-25,20){$#1$}
\put(-25,-28){$#2$}
\put(11,-28){$#3$}
\put(11,20){$#4$}
\put(-7,5){$#5$}
}}}

\newcommand{\taraver}[5]
{\unitlength=0.75pt
\makebox[45 pt]{
\raisebox{3 pt}[27 pt][21 pt]{
\put(-2,17){\line(-5,3){24}}
\put(-2,-17){\line(-5,-3){24}}
\put(-2,-17){\line(5,-3){24}}
\put(-2,17){\line(5,3){24}}
\put(-2,17){\line(0,-1){34}}
\put(-2,17){\circle*{4}}
\put(-2,-17){\circle*{4}}
\put(-32,17){$#1$}
\put(-32,-25){$#2$}
\put(18,-25){$#3$}
\put(18,17){$#4$}
\put(2,-3){$#5$}
}}}

\newcommand{\taraverooiou}[5]
{\unitlength=0.75pt
\makebox[45 pt]{
\raisebox{3 pt}[27 pt][21 pt]{
\put(-2,17){\line(-5,3){24}}
\put(-12,23){\vector(-3,2){0}}
\put(-2,-17){\line(-5,-3){24}}
\put(-12,-23){\vector(-3,-2){0}}
\put(-2,-17){\line(5,-3){24}}
\put(8,-23){\vector(-3,2){0}}
\put(-2,17){\line(5,3){24}}
\put(8,23){\vector(3,2){0}}
\put(-2,17){\line(0,-1){34}}
\put(-2,0){\vector(0,1){0}}
\put(-2,17){\circle*{4}}
\put(-2,-17){\circle*{4}}
\put(-32,17){$#1$}
\put(-32,-25){$#2$}
\put(18,-25){$#3$}
\put(18,17){$#4$}
\put(2,-3){$#5$}
}}}

\newcommand{\taraverioiou}[5]
{\unitlength=0.75pt
\makebox[45 pt]{
\raisebox{3 pt}[27 pt][21 pt]{
\put(-2,17){\line(-5,3){24}}
\put(-12,23){\vector(3,-2){0}}
\put(-2,-17){\line(-5,-3){24}}
\put(-12,-23){\vector(-3,-2){0}}
\put(-2,-17){\line(5,-3){24}}
\put(8,-23){\vector(-3,2){0}}
\put(-2,17){\line(5,3){24}}
\put(8,23){\vector(3,2){0}}
\put(-2,17){\line(0,-1){34}}
\put(-2,0){\vector(0,1){0}}
\put(-2,17){\circle*{4}}
\put(-2,-17){\circle*{4}}
\put(-32,17){$#1$}
\put(-32,-25){$#2$}
\put(18,-25){$#3$}
\put(18,17){$#4$}
\put(2,-3){$#5$}
}}}

\newcommand{\taraverioiod}[5]
{\unitlength=0.75pt
\makebox[45 pt]{
\raisebox{3 pt}[27 pt][21 pt]{
\put(-2,17){\line(-5,3){24}}
\put(-12,23){\vector(3,-2){0}}
\put(-2,-17){\line(-5,-3){24}}
\put(-12,-23){\vector(-3,-2){0}}
\put(-2,-17){\line(5,-3){24}}
\put(8,-23){\vector(-3,2){0}}
\put(-2,17){\line(5,3){24}}
\put(8,23){\vector(3,2){0}}
\put(-2,17){\line(0,-1){34}}
\put(-2,0){\vector(0,-1){0}}
\put(-2,17){\circle*{4}}
\put(-2,-17){\circle*{4}}
\put(-32,17){$#1$}
\put(-32,-25){$#2$}
\put(18,-25){$#3$}
\put(18,17){$#4$}
\put(2,-3){$#5$}
}}}

\newcommand{\taraveroooid}[5]
{\unitlength=0.75pt
\makebox[45 pt]{
\raisebox{3 pt}[27 pt][21 pt]{
\put(-2,17){\line(-5,3){24}}
\put(-12,23){\vector(-3,2){0}}
\put(-2,-17){\line(-5,-3){24}}
\put(-12,-23){\vector(-3,-2){0}}
\put(-2,-17){\line(5,-3){24}}
\put(8,-23){\vector(3,-2){0}}
\put(-2,17){\line(5,3){24}}
\put(8,23){\vector(-3,-2){0}}
\put(-2,17){\line(0,-1){34}}
\put(-2,0){\vector(0,-1){0}}
\put(-2,17){\circle*{4}}
\put(-2,-17){\circle*{4}}
\put(-32,17){$#1$}
\put(-32,-25){$#2$}
\put(18,-25){$#3$}
\put(18,17){$#4$}
\put(2,-3){$#5$}
}}}

\newcommand{\taraveriooid}[5]
{\unitlength=0.75pt
\makebox[45 pt]{
\raisebox{3 pt}[27 pt][21 pt]{
\put(-2,17){\line(-5,3){24}}
\put(-12,23){\vector(3,-2){0}}
\put(-2,-17){\line(-5,-3){24}}
\put(-12,-23){\vector(-3,-2){0}}
\put(-2,-17){\line(5,-3){24}}
\put(8,-23){\vector(3,-2){0}}
\put(-2,17){\line(5,3){24}}
\put(8,23){\vector(-3,-2){0}}
\put(-2,17){\line(0,-1){34}}
\put(-2,0){\vector(0,-1){0}}
\put(-2,17){\circle*{4}}
\put(-2,-17){\circle*{4}}
\put(-32,17){$#1$}
\put(-32,-25){$#2$}
\put(18,-25){$#3$}
\put(18,17){$#4$}
\put(2,-3){$#5$}
}}}

\newcommand{\taraverioood}[5]
{\unitlength=0.75pt
\makebox[45 pt]{
\raisebox{3 pt}[27 pt][21 pt]{
\put(-2,17){\line(-5,3){24}}
\put(-12,23){\vector(3,-2){0}}
\put(-2,-17){\line(-5,-3){24}}
\put(-12,-23){\vector(-3,-2){0}}
\put(-2,-17){\line(5,-3){24}}
\put(8,-23){\vector(3,-2){0}}
\put(-2,17){\line(5,3){24}}
\put(8,23){\vector(3,2){0}}
\put(-2,17){\line(0,-1){34}}
\put(-2,0){\vector(0,-1){0}}
\put(-2,17){\circle*{4}}
\put(-2,-17){\circle*{4}}
\put(-32,17){$#1$}
\put(-32,-25){$#2$}
\put(18,-25){$#3$}
\put(18,17){$#4$}
\put(2,-3){$#5$}
}}}

\newcommand{\taraveriobod}[5]
{\unitlength=0.75pt
\makebox[45 pt]{
\raisebox{3 pt}[27 pt][21 pt]{
\put(-2,17){\line(-5,3){24}}
\put(-12,23){\vector(3,-2){0}}
\put(-2,-17){\line(-5,-3){24}}
\put(-12,-23){\vector(-3,-2){0}}
\put(-2,17){\line(5,3){24}}
\put(8,23){\vector(3,2){0}}
\put(-2,17){\line(0,-1){34}}
\put(-2,0){\vector(0,-1){0}}
\put(-2,17){\circle*{4}}
\put(-2,-17){\circle*{4}}
\put(-32,17){$#1$}
\put(-32,-25){$#2$}
\put(18,-25){$#3$}
\put(18,17){$#4$}
\put(2,-3){$#5$}
}}}

\newcommand{\taraveriooodptdr}[5]
{\unitlength=0.75pt
\makebox[45 pt]{
\raisebox{3 pt}[27 pt][21 pt]{
\put(-2,17){\line(-5,3){24}}
\put(-12,23){\vector(3,-2){0}}
\put(-2,-17){\line(-5,-3){24}}
\put(-12,-23){\vector(-3,-2){0}}
\put(-2,-17){\line(5,-3){24}}
\put(22,-31.6){\circle*{4}}
\put(8,-23){\vector(3,-2){0}}
\put(-2,17){\line(5,3){24}}
\put(8,23){\vector(3,2){0}}
\put(-2,17){\line(0,-1){34}}
\put(-2,0){\vector(0,-1){0}}
\put(-2,17){\circle*{4}}
\put(-2,-17){\circle*{4}}
\put(-32,17){$#1$}
\put(-32,-25){$#2$}
\put(18,-25){$#3$}
\put(18,17){$#4$}
\put(2,-3){$#5$}
}}}

\newcommand{\celoup}[4]
{\unitlength=0.75pt
\makebox[42 pt][l]{
\raisebox{-39 pt}[45 pt][39 pt]{
\put(24,76){\oval(40,40)[]}
\put(24,28){\line(0,1){28}}
\put(24,28){\line(-5,-3){24}}
\put(24,28){\line(5,-3){24}}
\put(24,28){\circle*{4}}
\put(24,56){\circle*{4}}
\put(5,5){$#1$}
\put(34,5){$#2$}
\put(29,38){$#3$}
\put(32,96){$#4$}
}}}

\newcommand{\celodown}[4]
{\unitlength=0.75pt
\makebox[42 pt][l]{
\raisebox{2 pt}[45 pt][39 pt]{
\put(24,-20){\oval(40,40)[]}
\put(24,28){\line(0,-1){28}}
\put(24,28){\line(5,3){24}}
\put(24,28){\line(-5,3){24}}
\put(24,28){\circle*{4}}
\put(24,0){\circle*{4}}
\put(6,42){$#1$}
\put(34,42){$#2$}
\put(29,11){$#3$}
\put(30,-49){$#4$}
}}}

\newcommand{\celodownor}[4]
{\unitlength=0.75pt
\makebox[42 pt][l]{
\raisebox{2 pt}[45 pt][39 pt]{
\put(24,-20){\oval(40,40)[]}
\put(26,-40){\vector(1,0){0}}
\put(24,28){\line(0,-1){28}}
\put(24,12){\vector(0,-1){0}}
\put(24,28){\line(5,3){24}}
\put(34,34){\vector(-3,-2){0}}
\put(24,28){\line(-5,3){24}}
\put(14,34){\vector(3,-2){0}}
\put(24,28){\circle*{4}}
\put(24,0){\circle*{4}}
\put(6,42){$#1$}
\put(34,42){$#2$}
\put(29,11){$#3$}
\put(30,-49){$#4$}
}}}

\newcommand{\tennis}[2]
{\unitlength=0.75pt
\makebox[67 pt][l]{
\raisebox{3 pt}[22.5 pt][16.5 pt]{
\put(0,0){\line(1,0){28}}
\put(48,0){\oval(40,40)[]}
\put(28,0){\circle*{4}}
\put(3,4){$#1$}
\put(70,4){$#2$}
}}}

\newcommand{\tennisu}[2]
{\unitlength=0.75pt
\makebox[67 pt][l]{
\raisebox{3 pt}[22.5 pt][16.5 pt]{
\put(0,0){\line(1,0){28}}
\put(16,0){\vector(1,0){0}}
\put(48,0){\oval(40,40)[]}
\put(68,0){\vector(0,1){0}}
\put(28,0){\circle*{4}}
\put(3,4){$#1$}
\put(70,4){$#2$}
}}}

\newcommand{\tennisd}[2]
{\unitlength=0.75pt
\makebox[67 pt][l]{
\raisebox{3 pt}[22.5 pt][16.5 pt]{
\put(0,0){\line(1,0){28}}
\put(16,0){\vector(1,0){0}}
\put(48,0){\oval(40,40)[]}
\put(68,0){\vector(0,-1){0}}
\put(28,0){\circle*{4}}
\put(3,4){$#1$}
\put(70,4){$#2$}
}}}

\newcommand{\sun}[4]
{\unitlength=0.75pt
\makebox[78 pt][l]{
\raisebox{3 pt}[30 pt][24 pt]{
\put(0,0){\line(1,0){28}}
\put(48,0){\oval(40,40)[]}
\put(28,0){\circle*{4}}
\put(68,0){\circle*{4}}
\put(68,0){\line(1,0){28}}
\put(3,4){$#1$}
\put(84,4){$#2$}
\put(55,21){$#3$}
\put(55,-30){$#4$}
}}}

\newcommand{\sunor}[4]
{\unitlength=0.75pt
\makebox[78 pt][l]{
\raisebox{3 pt}[30 pt][24 pt]{
\put(0,0){\line(1,0){28}}
\put(16,0){\vector(1,0){0}}
\put(48,0){\oval(40,40)[]}
\put(46,20){\vector(-1,0){0}}
\put(50,-20){\vector(1,0){0}}
\put(28,0){\circle*{4}}
\put(68,0){\circle*{4}}
\put(68,0){\line(1,0){28}}
\put(80,0){\vector(-1,0){0}}
\put(3,4){$#1$}
\put(84,4){$#2$}
\put(55,21){$#3$}
\put(55,-30){$#4$}
}}}

\newcommand{\cylindre}{
\unitlength=1mm
\makebox[31mm][l]{\raisebox{-15mm}[18mm][18mm]{ 
\put(15,15){\circle{10}}
\put(15,15){\oval(30,30)[]}
}}}

\newcommand{\spherepantstriline}[3]{
\unitlength=0.5mm
\makebox[30mm][l]{\raisebox{-13mm}[13.5mm][13mm]{
\put(30,45){\circle{10}}
\put(15,15){\circle{10}}
\put(45,15){\circle{10}}
\put(15,15){\makebox(0,0)[cc]{$#1$}}
\put(45,15){\makebox(0,0)[cc]{$#2$}}
\put(30,45){\makebox(0,0)[cc]{$#3$}}
\put(30,40){\line(0,-1){14}}
\put(30,26){\line(-4,-3){11}}
\put(30,26){\line(4,-3){11}}
}}}

\newcommand{\spherepantsrightlines}[3]{
\unitlength=0.5mm
\makebox[38mm][l]{\raisebox{-13mm}[13.5mm][13mm]{
\put(30,45){\circle{10}}
\put(15,15){\circle{10}}
\put(45,15){\circle{10}}
\put(15,15){\makebox(0,0)[cc]{$#1$}}
\put(45,15){\makebox(0,0)[cc]{$#2$}}
\put(30,45){\makebox(0,0)[cc]{$#3$}}
\put(42,19){\line(2,3){6.67}}
\put(48.67,29){\line(-5,3){18.67}}
\bezier{472}(49,29)(74,-17)(19,18)
}}}

\newcommand{\fourier}{
\unitlength=0.7mm
\linethickness{0.4pt}
\begin{picture}(50,43)
\put(1,22){\makebox(0,0)[cc]{$S\ =$}}
\put(26,23){\oval(20,18)[b]}
\put(40,19.50){\oval(20,19)[t]}
\put(50,20){\line(0,-1){18}}
\put(30,2){\line(0,1){8}}
\put(36,33){\line(0,1){9}}
\put(16,23){\line(0,1){19}}
\put(26,43){\makebox(0,0)[cc]{$\f$}}
\put(40,1){\makebox(0,0)[cc]{$\f$}}
\put(40,29){\line(0,1){8}}
\put(45,35){\makebox(0,0)[cc]{$\mu$}}
\end{picture}
}

\newcommand{\invfourier}{
\unitlength=0.7mm
\linethickness{0.4pt}
\begin{picture}(51,44)
\put(41,25){\oval(20,18)[b]}
\put(25.50,20){\oval(21,20)[t]}
\put(31,34){\line(0,1){10}}
\put(51,25){\line(0,1){19}}
\put(36,12){\line(0,-1){10}}
\put(15,20){\line(0,-1){18}}
\put(25,30){\line(0,1){8}}
\put(41,44){\makebox(0,0)[cc]{$\f$}}
\put(26,1){\makebox(0,0)[cc]{$\f$}}
\put(20,36){\makebox(0,0)[cc]{$\mu$}}
\put(1,22){\makebox(0,0)[cc]{$S^{-1} \ =$}}
\end{picture}
}

\newcommand{\CA}{{\cal A}} \newcommand{\CC}{{\cal C}}
\newcommand{\CF}{{\cal F}} \newcommand{\CS}{{\cal S}}
\newcommand{\CM}{{\cal M}} \newcommand{\CZ}{{\cal Z}}
 \newcommand{\CN}{{\cal N}}
 
 \newcommand{\CG}{{\cal G}}
\newcommand{\cd}{{\cal D}} \newcommand{\CR}{{\cal R}}
 
\newcommand{\TN}{{\cal T\cal N}} \newcommand{\RG}{{\cal R\cal G}}
 \newcommand{\ON}{{\cal O\cal N}}
\newcommand{\f}{{\bold f}}

\newcommand{\R}{{\Bbb R}}
\newcommand{\C}{{\Bbb C}}

\newcommand{\Z}{{\Bbb Z}}

\newcommand{\be}{\begin{equation}}

\newcommand{\vect}{\text{-vect}}
\newcommand{\Vect}{\text{\fontshape{n}\selectfont-Vect}}
\newcommand{\tens}{\otimes}

\newcommand{\Aut}{\operatorname{Aut}}
\newcommand{\End}{\operatorname{End}}
\newcommand{\Hom}{\operatorname{Hom}}
\newcommand{\Mor}{\operatorname{Mor}}
\newcommand{\ev}{\operatorname{ev}}
\newcommand{\coev}{\operatorname{coev}}

\newcommand{\Ker}{\operatorname{Ker}}
\newcommand{\Ob}{\operatorname{Ob}}

\newcommand{\id}{\operatorname{id}}

\newcommand{\Int}{\operatorname{Int}}
\newcommand{\Card}{\operatorname{Card}}

\newcommand{\Surf}{Sur\!f}

\newcommand{\qqquad}{\qquad\quad}
\newcommand{\nquad}{\!\!\!\!\!\!}
\newcommand{\nqquad}{\nquad\nquad}
\newcommand{\pti}{\haj{{\ }}}
\newcommand{\haj}[1]{{\mathaccent20 #1}}
\newcommand{\und}[1]{{\underline {#1}}}

\newcommand{\aow}{\hbox to 20pt {\rightarrowfill}}
\newcommand{\arow}{\hbox to 30pt {\rightarrowfill}}
\newcommand{\arrow}{\hbox to 45pt {\rightarrowfill}}
\newcommand{\arrrow}{\hbox to 60pt {\rightarrowfill}}
\newcommand{\arrrrow}{\hbox to 100pt {\rightarrowfill}}
\newcommand{\lfaow}{\hbox to 20pt {\leftarrowfill}}
\newcommand{\lfarow}{\hbox to 30pt {\leftarrowfill}}
\newcommand{\lfarrow}{\hbox to 45pt {\leftarrowfill}}
\newcommand{\lfarrrow}{\hbox to 60pt {\leftarrowfill}}
\newcommand{\lfarrrrow}{\hbox to 100pt {\leftarrowfill}}

\documentstyle[amscd,amssymb,12pt,righttag,ctagsplt,bezier]{amsart}

\newtheorem{thm}{Theorem}[section]
\newtheorem{cor}[thm]{Corollary}
\newtheorem{lem}[thm]{Lemma}
\newtheorem{prop}[thm]{Proposition}

\theoremstyle{definition}
\newtheorem{defn}{Definition}[section]

\newtheorem{acknowledgement}{Acknowledgements}

\theoremstyle{remark}
\newtheorem{rem}{Remark}[section]
\newtheorem{notation}{Notation}
\newtheorem{conjecture}{Conjecture}

\numberwithin{equation}{section}

\newcommand{\thmref}[1]{Theorem~\ref{#1}}
\newcommand{\secref}[1]{Section~\ref{#1}}

\newcommand{\propref}[1]{Proposition~\ref{#1}}

\renewcommand{\le}{\leqslant}
\renewcommand{\ge}{\geqslant}

\begin{document}

\title[\fontsize{9}{11pt}\selectfont Modular properties of
ribbon abelian categories]
{\fontsize{17}{25pt}\bf\selectfont Modular properties of ribbon
abelian categories}

\author[\fontsize{9}{11pt}\selectfont V.~Lyubashenko]
{\fontsize{14}{16pt}\selectfont Volodimir Lyubashenko}

\thanks
{The research was supported in part by the SERC research grant GR/G 42976.}
\thanks{The detailed version of this paper has been submitted
for publication elsewhere.}

\date {\ \\ \ }

\maketitle

\noindent {\fontsize{11}{13pt}\fontseries{bx}\selectfont Abstract.}
{\fontsize{11}{13pt}\selectfont
A category $N$ of labeled (oriented) trivalent graphs (nets) or ribbon
graphs is extended by new generators called fusing, braiding, twist and
switch with relations which can be called Moore--Seiberg relations. A
functor to $N$ is constructed from the category $\Surf$ of oriented surfaces
with labeled boundary and their homeomorphisms. Given an (eventually
non-semisimple) $k$-linear abelian ribbon braided category $\CC$ with some
finiteness conditions we construct a functor from a central extension of $N$
with the set of labels Ob$\CC$ to $k$-vector spaces. Composing the functors
we get a modular functor from a central extension of $\Surf$ to $k$-vector
spaces.}

\vskip 12pt

\noindent{\fontsize{11}{13pt}\selectfont 1991 Mathematics Subject
Classification: 18B30, 18D10, 57N05}

\vskip 12pt plus 1pt

\fontsize{12}{14.4pt}\selectfont
\noindent Moore and Seiberg's study \cite{MooSei} on conformal field theory
was continued and developed by Walker \cite{Wal} from the topological point
of view. The first systematic study in this direction of the example of
$\widehat{\frak s\frak l}(2)$ was made by Kohno \cite{Ko:inv,Ko:3man}. A
different topological approach was proposed by Reshetikhin and Turaev~
\cite{ResTur:3} (see also \cite{Tur:q3}). The aim of this article is
to present a categorical point of view on the subject. We use freely
notations and results from the previous papers \cite{Lyu:tan,Lyu:mod}.

We consider a category of surfaces $\CS$ labeled by a set $\CC$, which
Grothendieck calls the Teichm\"uller's tower. Its subcategory $\Surf$
consists of labeled surfaces and isotopy classes of their homeomorphisms.
Its central extension is denoted $E\Surf$. We give also a definition of
a modular functor.

We show that any ribbon abelian category $\CC$,  satisfying the axioms of
modularity \cite{Lyu:mod} yields a modular functor
$Z_{\CC}:E\Surf\to k$-vect. Thus, such category deserves to be called
{\sl modular}. Precisely, modularity means the following: $\CC$ is a
noetherian abelian $k$-linear ribbon tensor category with finite
dimensional $k$-vector spaces $\Hom_{\CC}(A,B)$. In a cocompletion of
$\CC$ there exists a Hopf algebra $F=\int^{X\in \CC} X\tens X\pti$, an
automorphism $T:F\to F$ and a Hopf pairing $\omega:F\tens F\to I$ (see
\cite{Lyu:mod}) with the kernel $\Ker\omega\subset F$. We require that

(M1) $\f \stackrel{\text{def}}{=} F/\Ker\omega$ is an object of $\CC$
    (``$\f$ is finite dimensional'');

(M2) $T(\Ker\omega)\subset\Ker\omega$.

Vice versa, with some assumptions of representability any modular functor
$Z:E\Surf \to k$-vect induce on $\CC$ the structure of a ribbon
category and the functor $Z$ is isomorphic to $Z_{\CC}$. So, up to some
extent, our conditions are not only sufficient, but also necessary.

In the case when $\CC$ is a semisimple abelian category with finite number of
simple objects and $\Ker\omega=0$, obtained results generalize those of Moore
and Seiberg~\cite{MooSei}. Practically, in this case the obtained functor is
the same as constructed by Walker~\cite{Wal} (in dimension two) and probably
coincides with the one of Reshetikhin and Turaev~\cite{ResTur:3}. A new
feature is a possibility to work with non-semisimple categories.

\begin{acknowledgement} I wish to thank  V.G.~Drinfeld, A.~Joyal,
T.~Kohno, S.~Majid, Y.~Soibelman, R.~Street for multiple helpful discussions.
I am grateful to L.~Breen for his kind hospitality and opportunity to work
in Universit\'e Paris-Nord, where some results of this paper were obtained.
More results were obtained, when the author visited C.~De Concini at Scuola
Normale Superiore and I express my deep gratitude to him.
\end{acknowledgement}

\sEction{Introduction} 
In \secref{surfaces} a category of labeled surfaces $\CS$ and isotopy classes
of their glueing maps is considered. It encompasses the usual category of
surfaces $\Surf$ with isotopy classes of homeomorphisms as morphisms. This
is a topological counterpart of Teichm\"uller's tower, considered by
Grothendieck \cite{Gro:esq}. He conjectured that the
Teichm\"uller's tower can be described by few essential relations in small
genera. These relations were written by Moore and Seiberg \cite{MooSei}.

They are essential relations of the category of ribbon graphs $RG$ defined
in \secref{Ribgra}. Ribbon graphs are surfaces with distinguished intervals
at the boundary. The category has infinite number of generators combined into
finite number of classes: topological morphisms (homeomorphisms and glueings
of ribbon graphs), braiding, twists and switches. Also it has infinite number
of relations, which are relatively trivial and can be considered as
commutation relations or as identification of generators.

There is a functor $dupl:RG \to \CS$. We construct in \secref{compared to} a
functor $\Surf\to RG$ using the presentations of the mapping class group
$M_{g,0}$ by Wajnryb~\cite{Waj} and of the braid group of a closed surface by
Scott~\cite{Scott}. The composition $\Surf\to RG\to\CS$ is the natural
inclusion. The conjecture is that the functor $dupl$ is an equivalence
between $\CS$ and $RG$.

We describe another category which objects are 1-complexes (graphs) called
nets and morphisms besides maps of graphs include morphisms of insertion or
deletion of a vertex, fusing, braiding, twists and switches. This category
comes in two versions---unoriented $N$ and oriented $ON$ (Sections
\ref{trivalent}--\ref{oriented}). Both are equivalent to the category of
ribbon graphs $RG$.

In \secref{Extensions} we study central extensions of the last category (in
any form) and define a universal extension $EN$.

Given an abelian ribbon tensor category $\CC$ with some finiteness properties
we define a functor on a category $EN$ labeled by objects of $\CC$ in
\secref{to a functor}. Composing the functors we get a representation of the
central extension of the category $\Surf$. This is a direct generalization
of Moore and Seiberg's results \cite{MooSei}.

Examples of such categories $\CC$ can be found in \cite{LyuMaj}. Besides
familiar semisimple abelian categories with finite number of simple objects,
there are other, non-semisimple examples, such as the category of
representations of a quantum group at a root of unity.

All the relevant categories and functors make part of the following diagram
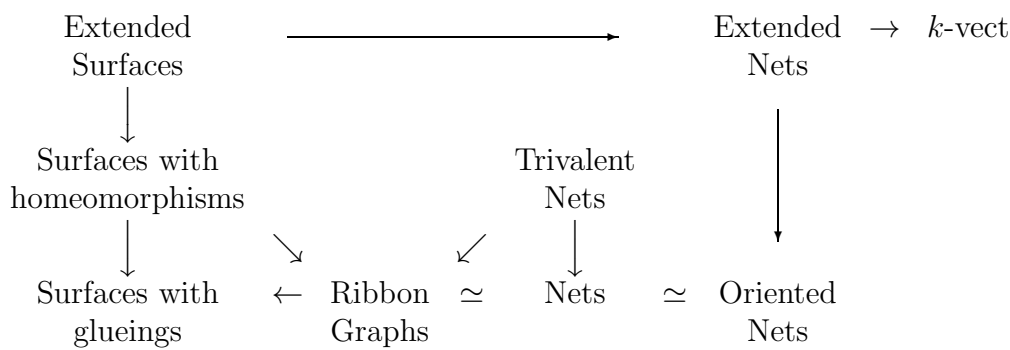
\[
\begin{array}{ccccccccc}
\text{Extended} & \makebox[0mm][l]{\put(0,0){\vector(1,0){125}}} &&&&&
\text{Extended} & \to & k\vect \\
\text{Surfaces}  &&&&&&  \text{Nets} \\
\Big\downarrow&&&&&& \raisebox{0mm}[1mm][1mm]{\put(0,5){\vector(0,-1){50}}}\\
\text{Surfaces with} &&&& \text{Trivalent} \\
\text{homeomorphisms} &&&& \text{Nets} \\
\Big\downarrow & \searrow && \swarrow & \Big\downarrow \\
\text{Surfaces with} & \leftarrow & \text{Ribbon}
& \simeq & \text{Nets} & \simeq & \text{Oriented} \\
\text{glueings} && \text{Graphs} &&&& \text{Nets}
\end{array}
\]

Proofs of the obtained results will be published
elsewhere~\cite{Lyu:rib=mod}.

\sEction{Surfaces}\label{surfaces}
\subsEction{Labeled surfaces} 
Let $\CC=\{A,B,C,...\}$ be a set of labels with an involution
$\cdot\pti: \CC\to\CC$. By a {\sl labeled surface} we shall understand the
following: a compact oriented surface $\Sigma$ with a boundary, with a
labelling of boundary circles
$L: \pi_0(\partial\Sigma)\to \CC$, $i\mapsto A_i$ and
with a chosen point $x_i$ on $i^{\text{th}}$ boundary circle, i.e. a section
$x:\pi_0(\partial\Sigma)\to\partial\Sigma$ of the projection
$\partial\Sigma\to \pi_0(\partial\Sigma)$ is fixed. So we write
$\partial\Sigma=\coprod_i \und{C}_i$, where a circle is described as
$\und{C}_i=(C_i,A_i,x_i)$.

\begin{defn}
Let $\Sigma,\tilde\Sigma$ be two labeled surfaces. A continuous surjective
mapping $f:\Sigma\to\tilde\Sigma$ is called {\sl glueing}, if its
restriction to interior part is an orientation preserving homeomorphism
$\Int\Sigma\to f(\Int\Sigma)$ and $\tilde\Sigma$ is a quotient space of
$\Sigma$ with respect to the following equivalence relation $\sim$ in
$\partial\Sigma$. We choose in $\pi_0(\partial\Sigma)$ some mutually
disjoint pairs $(C_j',A_j,x_j')$ and $(C_j'',A_j\pti,x_j'')$ and
homeomorphisms $\phi_j:C_j'\to C_j''$ such that $\phi_j(x_j')=x_j''$. If
$\phi_j(x')=x''$, we write $x'\sim x''$. For any boundary circle
$(C_i,A_i,x_i)$ in $\tilde\Sigma$ its preimage must have the same labels
$(f^{-1}(C_i),A_i,f^{-1}(x_i))$.
\end{defn}

Note that a particular case of a glueing is an orientation, labelling and
distinguished points preserving homeomorphism. Since a composition of
glueings is again a glueing, they form a category.

\begin{defn}
Consider a category $\CS= \CS_{\CC}$ which objects are labeled surfaces and
morphisms are isotopy classes of glueings (an isotopy here is a continuous
family of glueings). The category $\CS$ is a symmetric monoidal category
with disjoint union as a monoidal product.
\end{defn}

\begin{prop}
The equality $gf_1=gf_2$ of morphisms in the category $\CS$ implies $f_1=f_2$
(we shall say that the category $\CS$ has {\sl left cancellations} property).
\end{prop}

\begin{pf} Glueings are surjective maps.
\end{pf}

\subsEction{Ribbon graphs} \label{Ribgra} 
\begin{defn}
A ribbon graph is an oriented surface $X$ such that its each component has
a non-empty boundary, equipped with a subset $B\subset\partial X$
homeomorphic to finite disjoint sum of closed intervals, with a labelling
$L:\pi_0(B)\to \CC$, and with a chosen point in the interior of each
component of $B$, i.e. with a section $x:\pi_0(B)\to \Int B$ of the
projection $B\to \pi_0(B)$.
\end{defn}

\begin{defn}
Let $\CR\CG$ be a category which objects are ribbon graphs and morphisms
are isotopy classes of glueings of ribbon graphs. A {\em glueing}
$f:(X_1,B_1,L_1)\to (X_2,B_2,L_2)$ is a continuous surjective orientation
preserving mapping $f:X_1\to X_2$, such that the preimage of each point
consists of one or two points and the induced relation $\sim$ in $X_1$
($x,y\in X_1$ are equivalent if $f(x)=f(y)$) reduces to  pairwise
identification of some components of $B_1$ with other components having dual
labels. Distinguished points must be pairwise identified or mapped to
distinguished points, preimage $f^{-1}(B_2)$ must be a union of components
of $B_1$, and homeomorphism $f:f^{-1}(B_2)\to B_2$ must preserve labeling.
The category $\RG$ is a symmetric monoidal category with respect to disjoint
union.
\end{defn}

The category $\CR\CG$ has left cancellation property.

There is a functor $dupl:\CR\CG\to \CS$ called duplication. Having a ribbon
graph $X$, construct a surface $\Sigma=X\cup\bar X$, where $\bar X$ is a
second copy of $X$ with reversed orientation and $\partial X-\Int B$ is
identified with $\partial\bar X-\Int \bar B$ via ``identity map''. Boundary
of $\Sigma$ is the suspension of $B$ and it obtains labeling from the
labeling of $B$. The chosen points in $B$ become chosen points in
$\partial\Sigma$. To each glueing $f:X\to Y$ of ribbon graphs corresponds
the glueing $f\cup\bar f:X\cup\bar X\to Y\cup\bar Y$ of surfaces.

The duplication functor is injective on morphisms and essentially surjective
on objects. Duplication of the following ribbon graphs (double lines mark the
subset $B\subset\partial X$)
\be\label{D0123}
D_0= \put(20,3){\circle{32}} \qquad \qquad, \ D_1=\diskone A ,
\ D_2=\disktwo AE ,\ D_3=\disktre CAE
\end{equation}
\be\label{A0A1}
A_0=\cylindre ,  \qquad A_1= \annone A
\end{equation}
gives sphere, disk, annulus, pants, torus and torus with one hole
correspondingly.

The category $\CS$ has more morphisms than $\CR\CG$. For instance, the
following are automorphisms of annulus, pants (sphere $S^2=\bar\C$ with 3
holes) and torus with one hole
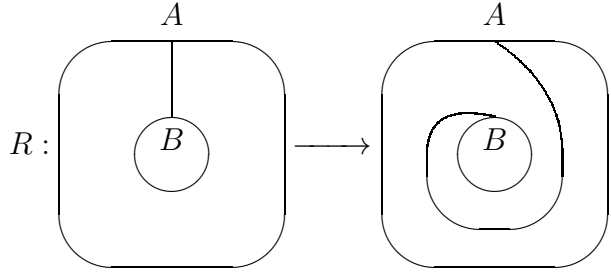
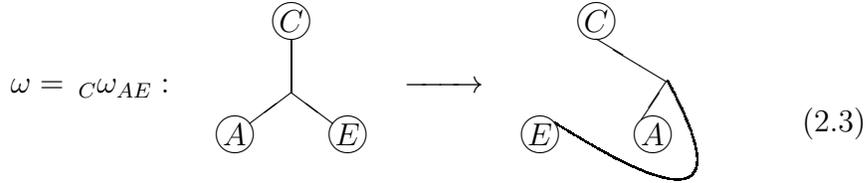
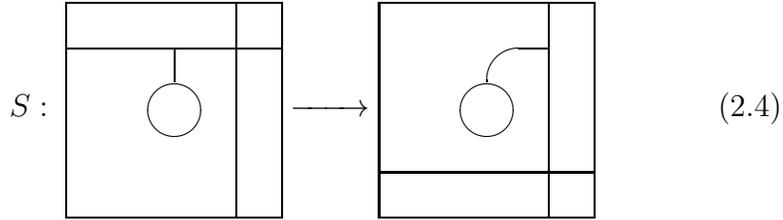
\[ R:
\unitlength=1mm
\makebox[31mm][l]{\raisebox{-15mm}[21mm][16mm]{
\put(15,15){\circle{10}}
\put(15,15){\oval(30,30)[]}
\put(15,35){\makebox(0,0)[ct]{$A$}}
\put(15,17){\makebox(0,0)[cc]{$B$}}
\put(15,20){\line(0,1){10}}
}}
\arow\
\unitlength=1mm
\makebox[31mm][l]{\raisebox{-15mm}[21mm][16mm]{
\put(15,15){\circle{10}}
\put(15,15){\oval(30,30)[]}
\put(15,35){\makebox(0,0)[ct]{$A$}}
\put(15,17){\makebox(0,0)[cc]{$B$}}
\bezier{80}(15,30)(24,24)(24,15)
\put(15,15){\oval(18,20)[b]}
\bezier{64}(6,15)(6,22)(15,20)
}}
\]
\be\label{omega}
\omega=\, _C\omega_{AE} :\spherepantstriline AEC \arow
\spherepantsrightlines EAC
\end{equation}
\be\label{Shomeo}
S:
\unitlength=0.75mm
\makebox[31mm][l]{ \raisebox{-13.5mm}{
\put(0,0){\framebox(38,38)[cc]{}}
\put(19,19){\circle{10}}
\put(0,30){\line(1,0){38}}
\put(30,38){\line(0,-1){38}}
\put(19,24){\line(0,1){6}}
}}
\arow
\unitlength=0.75mm
\makebox[31mm][l]{ \raisebox{-13.5mm}{
\put(0,0){\framebox(38,38)[cc]{}}
\put(19,19){\circle{10}}
\put(30,38){\line(0,-1){38}}
\put(38,8){\line(-1,0){38}}
\put(30,24){\oval(22,12)[lt]}
}}
\end{equation}
but not of the ribbon graph they came from. Here additional lines on annulus,
pants and torus start from the chosen points on the boundary. The additional
lines on the left hand side go to the additional lines in the right hand side
under the homeomorphism $R,\omega$ or $S$, which completely determines its
isotopy class. Indeed, when the surface (annulus, sphere $S^2$ with 3 holes,
or torus with 1 hole represented by a square with identified edges) is cut
along additional lines it becomes a disk, and one knows that a homeomorphism
of the boundary of the disk can be extended to a homeomorphism of the disk
unique up to isotopy.

$R$ is called inverse Dehn  twist, or shortly twist, $\omega$ is called
braiding, $S$ is called switch. Our first goal will be to give a presentation
of the category $\CS$ over the category $\CR\CG$, i.e. to add new generators
of type $R,\omega,S$ to $\CR\CG$ and to find relations, which give
a category, equivalent to $\CS$.

When we present a symmetric monoidal category with left cancellations $\cd$
by generators and relations we mean the following: take a free category $\CF$
on given generators; consider a minimal multiplicative equivalence relation
$\sim$ in $\Mor\CF$ such that in $\CF/\sim$ given relations are satisfied;
enlarge the multiplicative equivalence relation $\sim$ to $\sim_1$ so that
$\CF/\sim_1$ were symmetric monoidal category for a priori given tensor
product; enlarge the multiplicative equivalence relation $\sim_1$ to
$\sim_2$ so that in $\cd=\CF/\sim_2$ any equation $gf_1=gf_2$ with
$g,f_1,f_2\in \Mor\cd$ would imply $f_1=f_2$.

\begin{defn}\label{defrib}
Let $RG$ be a symmetric monoidal category with left cancellations having
ribbon graphs as objects and with morphisms generated over $\CR\CG$ by
new morphisms described as follows.

Let $\gamma:[0,1]\hookrightarrow X$ be a curve in a ribbon graph $(X,B,L)$
such that $\gamma(]0,1[)\subset \Int X$, $\gamma (0),\gamma (1)\in
\partial X-B$. Such curve will be called a cut. We associate with
$\gamma ([0,1])$ a new morphism $Tw_{\gamma}:X\to X$ called twist.

Let $h:D_3\hookrightarrow X$ be a continuous orientation preserving map such
that $h(\partial D_3-B_3)\subset \partial X-B$, where hexagon ribbon graph
$(D_3,B_3,L_3)$ is from \eqref{D0123} with distinguished interval $C$. We
associate with $h$ a new morphism $Br_h:X\to X'$ called braiding. Here $X'$
is obtained from $X$ by a surgery. Denote by $A,E$ subintervals of $B_3$
labeled by $A,E$. Cut $X$ along $h(A)$ and $h(E)$ and reglue the edges
cross-wise preserving the orientation.

Let $j:A_1\hookrightarrow X$ be a continuous orientation preserving map such
that $j(\partial A_1-B_1)\subset\partial X-B$, where annulus ribbon graph
$(A_1,B_1,L_1)$ is from \eqref{A0A1}. We associate with it a new morphism
$S_j:X\to X$ called switch.

These generators are subject to the following relations. If the curve
$\gamma :[0,1]\hookrightarrow X$ is shrinkable in the class of such curves
that $\gamma(0),\gamma(1)\in\partial X-B$, we set
\[Tw_\gamma=\id_X.\]
If the map $h:D_3\hookrightarrow X$ can be deformed inside the class of maps
with condition $h(\partial D_3-B_3)\subset\partial X-B$ so that $h(A)$ or
$h(E)$ shrinks to one point we set
\[ Br_h=\phi:X\to X' \]
if $h(C)$ shrinks to one point we set
\[Br_h=Tw_\gamma^{-1}\cdot\phi:X\to X'\]
where $\phi:X\to X'\in \CR\CG$ is the unique up to isotopy homeomorphism
between $X$ and $X'$ which is identity outside of the region $h(D_3)\cup F$
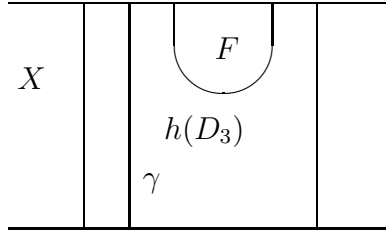
\[ 
\unitlength=1mm
\begin{picture}(50,30)
\put(0,0){\line(1,0){50}}
\put(50,30){\line(-1,0){50}}
\put(10,30){\line(0,-1){30}}
\put(41,0){\line(0,1){30}}
\put(16,30){\line(0,-1){30}}
\put(3,20){\makebox(0,0)[cc]{$X$}}
\put(19,6){\makebox(0,0)[cc]{$\gamma$}}
\put(28.50,30){\oval(13,24)[b]}
\put(29,24){\makebox(0,0)[cc]{$F$}}
\put(26,13){\makebox(0,0)[cc]{$h(D_3)$}}
\end{picture}
\]
and $\gamma:[0,1]\hookrightarrow h(D_3)$ is a curve isotopic to one of
non-shrinkable $h(A),h(E),h(C)$.

For each glueing $g:X\to Y$  and a map
$\gamma:[0,1]\hookrightarrow X, h:D_3\hookrightarrow X$ or $j:A_1
\hookrightarrow X$ the following diagrams commute:
\begin{equation}\label{gTw=Twg}
\begin{CD}
X @>Tw_\gamma>> X \\
@VgVV   @VVgV \\
Y @>Tw_{\gamma g}>> Y
\end{CD}
\end{equation}
\begin{equation}\labl{gBr=Brg}
\begin{CD}
X @>Br_h>> X' \\
@VgVV   @VVg'V \\
Y @>Br_{hg}>> Y'
\end{CD}
\end{equation}
\begin{equation}\label{gS=Sg}
\begin{CD}
X @>S_j>> X \\
@VgVV   @VVgV \\
Y @>S_{jg}>> Y
\end{CD}
\end{equation}
where the glueing $g'$ is defined as composition of surgery in $h(A),h(B)$,
glueing $g$ and surgery in $g\circ h(A)$, $g\circ h(B)$.
In these diagrams $g,g'$ denote isotopy classes of glueings $g,g'$.
Particular case when $g$ is isotopic to identity shows that $Tw_\gamma,
Br_h,S_j$ depend only on isotopy classes of $\gamma,h,j$.

In the case $X=D_3$ there are the following relations
\begin{equation}\label{Br=TwTwTw}
Br^2_{BC} = Tw_A\ Tw_B^{-1} \ Tw_C^{-1} : \disktre ABC \aow \disktre ABC
\end{equation}
\begin{multline}\labl{TwA=BrBr}
Tw_A^{-1} = \left( \disktre CAB @>Br_{AB}>> \disktre CBA @>>> \right. \\
\left. @>Br_{AC}>> \disktre ABC @>rot>> \disktre CAB \right)
\end{multline}
\begin{multline}\labl{TwB=BrBr}
Tw_B^{-1}= \left( \disktre CAB @>Br_{AB}>> \disktre CBA @>>> \right. \\
\left. @>Br_{CB}>> \disktre BCA @>rot^{-1}>> \disktre CAB \right)
\end{multline}
Here $Br_{AB}$ stands for $Br_{\id}$, similarly $Br_{AC},Br_{CD}$ are defined
and $rot$ denotes a morphism from $\CR\CG$, rotation by $2\pi/3$.

There are two relations for $X=D_4=$ a disk with 4 intervals marked on the
boundary:
\be\labl{Br=BrBr}
\begin{array}{rcl}
\diskfoursix DABCZY & @>_DBr_{YC}^{\pm1}>> & \diskfoursix DCABYU \\
_ZBr_{BC}^{\pm1} \searrow && \nearrow\, _UBr_{AC}^{\pm1} \\
& \diskfoursix DABCZU &
\end{array}
\end{equation}
Here three inclusions $D_3\hookrightarrow D_4$ are determined by 3 indices
appended to braidings. Image of $D_3$ covers a half of $D_4$.

Relations for $X=A_1$ are
\begin{equation}\labl{ST3=S2}
(S T)^3 =S^2 : \annonetwo M\gamma \aow \annonetwo M\gamma ,
\end{equation}
where $T=Tw_\gamma$, and \eqref{S2=Br-1Tw-1} (see Figure~\ref{A_1}).
\begin{figure}[htb]
\be\label{S2=Br-1Tw-1}
\begin{CD}
\annonetwo X\gamma @>S^2>> \annone X \\
@V_XBr_{\gamma\gamma}^{-1}VV  @| \\
\annonetwo X\gamma @>Tw_\gamma^{-1}>> \annonetwo X\gamma
\end{CD}
\end{equation}
\caption{A relation for annulus ribbon graph $A_1$\label{A_1}}
\end{figure}

Finally, there is a relation for annulus ribbon graph $A_2$ with two
intervals marked on the boundary (see diagram~\ref{A2mainS},
Figure~\ref{A_2}).
\begin{figure}[htbp]
\be\label{A2mainS}
\begin{CD}
\anntwoline YX @>Br_{XY}>> \anntwo XY \\
@VS^{-1}VV  @VVrot_\pi V \\
\anntwogamdel YX @.
\unitlength=1mm
\makebox[31mm][l]{\raisebox{-15mm}[21mm][15mm]{
\put(15,20){\circle{10}}
\put(15,20){\oval(30,30)[]}
\put(6,6){\line(1,0){7}}
\put(17,6){\line(1,0){7}}
\put(9,0){\makebox(0,0)[cb]{$Y$}}
\put(21,0){\makebox(0,0)[cb]{$X$}}
}}
\\
@VTw^{-1}_\gamma\ Tw_\delta VV  @VV\phi V \\
\anntwoline YX @>S>> \anntwo YX
\end{CD}
\end{equation}
\caption{A relation for annulus ribbon graph $A_2$\label{A_2}}
\end{figure}
Here $rot_\pi\in \CR\CG$ is the rotation by $\pi$, and $\phi\in\CR\CG$ is a
homeomorphism, which is identity in a neighbourhood of the hole and slides
intervals $X,Y$ along the boundary.
\end{defn}

\begin{rem}\labl{locprince}
The requirement that $RG$ is a symmetric monoidal category with left
cancellations produces more relations than just in a category with relations
\eqref{gTw=Twg}--\eqref{A2mainS}. For instance, any two generators $f_1,f_2$
with essentially non-intersecting supports $(\gamma([0,1]),h(D_3),j(A_1))$
commute (this can be called a ``locality principle''). Essentially
non-intersecting means that images of some $\gamma',h',j'$ isotopic to
$\gamma,h,j$ do not intersect. Indeed, cut the ribbon graph $X$ into pieces
so that one of the pieces contained support of $f_1$ and another contained
support of $f_2$. Denote the resulting ribbon graph $\tilde X$. The liftings
$\tilde{f_1},\tilde{f_2}: \tilde X\to \tilde X$ commute because they are
tensor products (disjoint unions) of a generator with identity map. The
axioms \eqref{gTw=Twg}--\eqref{gS=Sg} imply that
$gf_1f_2 = gf_2f_1 : \tilde X\to X$. The left cancellation property implies
that $f_1,f_2:X\to X$ commute.

Also \eqref{Br=TwTwTw}--\eqref{A2mainS} imply relations of the type
\eqref{Br=TwTwTw}--\eqref{A2mainS} in $X$
associated with embeddings of $D_3$, $D_4$, $A_1$, $A_2$ into $X$.
\end{rem}

\begin{prop}
The duplication functor $dupl:\CR\CG\to \CS$ extends to a functor
$dupl:RG\to \CS$. It sends $Tw_\gamma:X\to X$ to the inverse Dehn twist $R$
in $X\cup\bar X$ performed in neighbourhood of the cycle
$\gamma\cup\bar\gamma$. The braiding $Br_h:X\to X'$  goes to a braiding
homeomorphism $\omega:X\cup\bar X\to X'\cup\bar X'$, supported in
$h(D_3)\cup \bar{h(D_3)}$, looking as \eqref{omega}
\[ \omega:
\unitlength=0.5mm
\makebox[30mm][l]{\raisebox{-13mm}[13.5mm][13mm]{
\put(30,45){\circle{10}}
\put(15,15){\circle{10}}
\put(45,15){\circle{10}}
\put(15,15){\makebox(0,0)[cc]{$A$}}
\put(45,15){\makebox(0,0)[cc]{$B$}}
\put(30,45){\makebox(0,0)[cc]{$C$}}
\put(30,40){\line(0,-1){14}}
\put(30,26){\line(-4,-3){11}}
\put(30,26){\line(4,-3){11}}
\put(45,20){\line(-1,2){11}}
\put(26,42){\line(-1,-2){11}}
\put(20,15){\line(1,0){20}}
}}
\arow
\unitlength=0.5mm
\makebox[38mm][l]{\raisebox{-13mm}[13.5mm][13mm]{
\put(30,45){\circle{10}}
\put(15,15){\circle{10}}
\put(45,15){\circle{10}}
\put(15,15){\makebox(0,0)[cc]{$B$}}
\put(45,15){\makebox(0,0)[cc]{$A$}}
\put(30,45){\makebox(0,0)[cc]{$C$}}
\put(42,19){\line(2,3){6.67}}
\put(48.67,29){\line(-5,3){18.67}}
\bezier{472}(49,29)(74,-17)(19,18)
\put(45,20){\line(-1,2){11}}
\put(26,42){\line(-1,-2){11}}
\put(20,15){\line(1,0){20}}
}}
\]
(exterior of the first figure is $\bar{h(D_3)}$).
The switch $S_j:X\to X$ goes to a switching homeomorphism
$S:X\cup\bar X\to X\cup\bar X$ supported in $j(A_1)\cup\bar {j(A_1)}$ looking
as \eqref{Shomeo}
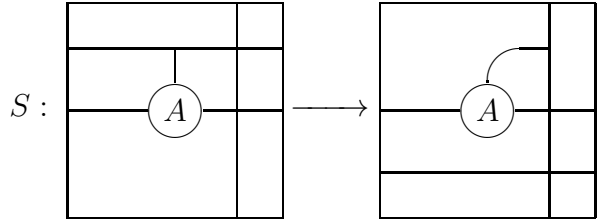
\[ S:
\unitlength=0.75mm
\makebox[31mm][l]{ \raisebox{-13.5mm}{
\put(0,0){\framebox(38,38)[cc]{}}
\put(19,19){\circle{10}}
\put(0,30){\line(1,0){38}}
\put(30,38){\line(0,-1){38}}
\put(19,24){\line(0,1){6}}
\put(19,19){\makebox(0,0)[cc]{$A$}}
\put(14,19){\line(-1,0){14}}
\put(24,19){\line(1,0){14}}
}}
\arow
\unitlength=0.75mm
\makebox[31mm][l]{ \raisebox{-13.5mm}{
\put(0,0){\framebox(38,38)[cc]{}}
\put(19,19){\circle{10}}
\put(30,38){\line(0,-1){38}}
\put(38,8){\line(-1,0){38}}
\put(30,24){\oval(22,12)[lt]}
\put(19,19){\makebox(0,0)[cc]{$A$}}
\put(24,19){\line(1,0){14}}
\put(14,19){\line(-1,0){14}}
}}
\]
(the upper half of the square with identified edges is $j(A_1)$, the lower
half is $\bar{j(A_1)}$).
\end{prop}

\sEction{Ribbon graphs compared to surfaces} \label{compared to} 
We say that a ribbon graph $(X,B)$ has $g$ loops with $n$ entries if
$\dim H_1(X,\R)=g, \Card\pi_0(B)=n$. Duplicated, it gives a surface
$X\cup\bar X$ of genus $g$ with $n$ holes.

\begin{thm} \label{StoRGthm}
There exists a functor $\Phi: \Surf\to RG$ such that the composition
$dupl\circ \Phi:\Surf\to\CS$ is isomorphic to the inclusion functor
$\Surf \hookrightarrow\CS$.
\end{thm}

\begin{pf}
The theorem reduces to constructing a splitting of the homomorphism
\[ \Aut_{RG}(Y_{g,n}) \to \Aut_{\CS}(dupl(Y_{g,n})) \simeq
\Aut_{\Surf}(dupl(Y_{g,n})) .\]
Here $Y_{g,n}$ means a graph with $g$ loops (of genus $g$) and with $n$
components of $B$ (entries). In general, $Y_{g,n}$ is labeled by $n_1$
copies of $L_1\in\CC$, $n_2$ copies of $L_2\in\CC$, etc. Each morphism
induces a permutation of the set
$\pi_0(B)\simeq\pi_0(\partial(dupl(Y_{g,n})))$, so there are group
homomorphisms
\[ \Aut_{RG}(Y_{g,n})\to \Aut_\CS(dupl(Y_{g,n}))\to \frak S_n.\]
In particular case when all labels coincide, the automorphism groups
$G_{RG},G_\CS$ will be the biggest, and the general case groups are just
preimages under projections $G_{RG}\to \frak S_n$, $G_\CS\to\frak S_n$ of the
subgroup of permutations leaving invariant $n_k$-element subset of $\pi_0(B)$
with label $L_k, k\ge 1, n=n_1+n_2+\dots$. Thus it suffices to consider the
case when all labels $L\in\CC$ coincide. We have to construct a splitting of
the homomorphism
\[ \Aut_{RG}(Y_{g,n})\to\Aut_{\CS}(dupl(Y_{g,n}))=M_{g,n}, \]
where the last group is the mapping class group of a surface
of genus $g$ with $n$ holes.

Using braiding isomorphisms any ribbon graph $Y_{g,n}$ can be reduced to a
standard form
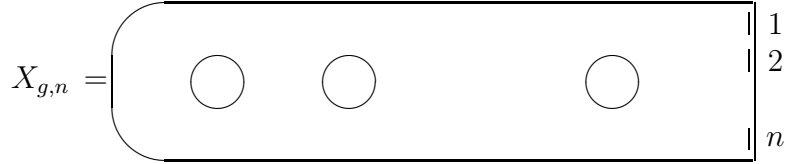
\[
\unitlength=0.7mm
\begin{picture}(143,30)
\put(0,15){\makebox(0,0)[cc]{$X_{g,n}\,=$}}
\put(131.50,15){\oval(243,30)[l]}
\put(30,15){\circle{10}}
\put(55,15){\circle{10}}
\put(105,15){\circle{10}}
\put(132,30){\line(0,-1){30}}
\put(131,28){\line(0,-1){4}}
\put(131,21){\line(0,-1){4}}
\put(131,2){\line(0,1){4}}
\put(136,26){\makebox(0,0)[cc]{1}}
\put(136,19){\makebox(0,0)[cc]{2}}
\put(136,4){\makebox(0,0)[cc]{$n$}}
\end{picture}
\]
so we have to consider only standard case $Y_{g,n}=X_{g,n}$. Using results of
Dehn \cite{Dehn} and Lickorish \cite{Lic:3,Lic:gen}, Birman \cite{Bir:mcg}
gave a set of generators of $M_{g,n}$. These consist of braidings
permuting the holes, some Dehn twists coming from $Tw_\gamma\in RG$, and
some Dehn twists along cycles in $dupl(X_{g,n})$ which can be obtained from
cycles of the form $\gamma\cup\bar\gamma$ by action of homeomorphisms $S$.
This implies that the map
\[ \Aut_{RG}(X_{g,n})\to M_{g,n} \]
is an epimorphism. More details will be given below.

When $g\le 1$ or $n\le 1$ not only generators of $M_{g,n}$ but also relations
are known (Magnus \cite{Mag} for $g=0$, Birman~\cite{Bir:mcg} for $g=1$ and
Wajnryb \cite{Waj} for $n=0,1$). For generic values of $g,n$ defining
relations were not written, but it is known (Birman~\cite{Bir:mcg}, Scott
\cite{Scott}) that there is an exact sequence
\[ 1 \to \bar B_n(\Sigma_g) \to M_{g,n} \to M_{g,0} \to 1 ,\]
where $\bar B_n(\Sigma_g)$ is the extension of the braid group of a closed
surface of genus $g$ by $\Z^n$. Its defining relations are found by Scott
\cite{Scott}. This information is sufficient to show case by case that
generators of $M_{g,n}$ admit liftings to $\Aut_{RG}(X_{g,n})$ and the
relations in $M_{g,n}$ lift to relations in $\Aut_{RG}(X_{g,n})$. This
implies the theorem.
\end{pf}

The topological version of the Grothendieck conjecture
about Teichm\"uller's tower is
\begin{conjecture}
The duplication functor $dupl:RG\to \CS$ is an equivalence of categories.
\end{conjecture}

\sEction{A category of trivalent nets}\label{trivalent}
\subsEction{Trivalent nets} 
\begin{defn}
Let a trivalent net $\Gamma$ be a 1-complex with set of vertices
$V\sqcup B$ and set of edges $E$. Elements of $V$, called 3-vertices, occur
thrice as endpoints of edges, and elements of $B$, called ends, occur once as
an endpoint of an edge. Each edge has at least one endpoint in $V$ and
$\Gamma$ is equipped with labeling of ends $B\to \CC$. For each vertex
$v\in V$ belonging to $3$ edges $a,b,c$ denote $a',b',c'$ 3 different isotopy
classes of embeddings $\gamma: [0,1] \hookrightarrow a,b,c$ such that
$\gamma(0)=a$. A cyclic order in the set $\{a',b',c'\}$
is chosen (which is one of the two cyclic permutations $(a',b',c')$ or
$(a',c',b')$).
\end{defn}

An edge connecting a 3-vertex and an end is called external leg, others are
internal edges. Trivalent nets will be drawn in the plane in such a way
that cyclic orientation of edges around a 3-vertex is coherent with the
orientation of the plane of drawing.

Let $\Gamma_1,\Gamma_2$ be trivalent nets.

\begin{defn}
A glueing $g:\Gamma_1\to\Gamma_2$ is a surjective continuous mapping of
1-complexes (sending vertices to vertices and edges into edges), bijective on
3-vertices and preserving cyclic order for edges incident to each 3-vertex,
and such that preimage of an end is an end with the same
label from $\CC$, preimage of an internal edge is an internal edge or a pair
of external legs with dual labels.
\end{defn}

Denote $\TN_0$ the category of trivalent nets with glueings as morphisms. It
is a symmetric monoidal category with disjoint union as a monoidal product.

\begin{defn}\label{deftrivalnet}
Let $\TN$ be a symmetric monoidal category with left cancellations having
trivalent nets as objects
and morphisms, generated over $\Mor\TN_0$ by fusing moves: for each internal
edge $a\in\Gamma\in\Ob \TN_0$ with different endpoints the fusing move
$fus_a:\Gamma\to\Gamma'$ is a new morphism, where $\Gamma'$ differs from
$\Gamma$ in a neighbourhood of $a$ as shown in the figure
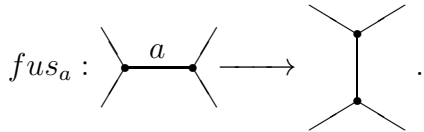
\[ fus_a: \tarahor{}{}{}{}a \arow \taraver{}{}{}{}{} .\]

New morphisms are subject to the following relations (commutative diagrams).
For any glueing $g$ and any internal edge $a\in \Gamma_1$
\begin{equation}\label{gfus=fusg}
\begin{CD}
\Gamma_1 @>fus_a>> \Gamma_1' \\
@VgVV              @VVg'V \\
\Gamma_2 @>fus_{g(a)}>> \Gamma_2'
\end{CD}
\end{equation}
The following relations are written literally, without any complementary part
\be\labl{fusfus}
\left( \tarahor{}{}{}{}a @>fus_a>> \taraver{}{}{}{}b @>fus_b>>
\tarahor{}{}{}{}{} \right) =\id
\end{equation}
and \eqref{pentagon}, Figure~\ref{5horses}.
\begin{figure}[htb]
\be\label{pentagon}
\unitlength=0.75mm
\begin{picture}(159,131)
\put(80,131){\line(2,-3){8}}
\put(88,119){\line(5,-2){12}}
\put(88,119){\circle*{2}}
\put(88,119){\line(-2,-1){14}}
\put(74,112){\line(1,-2){5.67}}
\put(79.67,101){\line(5,-3){11.33}}
\put(74,112){\circle*{2}}
\put(80,100){\circle*{2}}
\put(80,100){\line(-2,-1){11}}
\put(74,112){\line(-5,1){14}}
\put(20,91){\line(3,-5){7.67}}
\put(27.67,78){\line(5,-2){12.33}}
\put(28,78){\circle*{2}}
\put(28,78){\line(-1,-3){4.33}}
\put(23.67,65){\line(-4,1){12.67}}
\put(11,68){\line(-2,1){11}}
\put(11,68){\circle*{2}}
\put(23,65){\circle*{2}}
\put(23,65){\line(2,-3){8}}
\put(11,68){\line(-1,-3){5}}
\put(140,91){\line(-3,-5){7.67}}
\put(132.33,78){\line(3,-2){13.67}}
\put(146,69){\line(-4,-5){7.33}}
\put(138.67,60){\line(5,-3){12.33}}
\put(133,78){\circle*{2}}
\put(133,78){\line(-5,-2){13}}
\put(146,69){\circle*{2}}
\put(146,69){\line(3,1){13}}
\put(139,60){\circle*{2}}
\put(139,60){\line(-3,-2){10}}
\put(115,36){\line(-3,-5){7.33}}
\put(107.67,24){\line(-3,-1){12.67}}
\put(108,24){\circle*{2}}
\put(108,24){\line(1,-3){4}}
\put(112,12){\line(-5,-6){9}}
\put(112,12){\circle*{2}}
\put(112,12){\line(5,1){14}}
\put(126,14.67){\line(5,3){9}}
\put(126,15){\circle*{2}}
\put(126,15){\line(1,-5){2.67}}
\put(45,36){\line(0,-1){13}}
\put(45,23){\line(-6,-5){10}}
\put(45,23){\circle*{2}}
\put(45,23){\line(6,-5){9}}
\put(54,15.67){\line(1,-5){3}}
\put(54,15){\circle*{2}}
\put(54,15){\line(2,1){11}}
\put(35,15){\circle*{2}}
\put(35,15){\line(-2,1){10}}
\put(35,15){\line(-1,-5){2.67}}
\put(70,14){\vector(1,0){20}}
\put(27,50){\vector(2,-3){11.33}}
\put(40,87){\vector(2,1){19}}
\put(104,97){\vector(2,-1){19}}
\put(126,33){\vector(3,4){12.67}}
\end{picture}
\end{equation}
\caption{The pentagon relation\label{5horses}}
\end{figure}
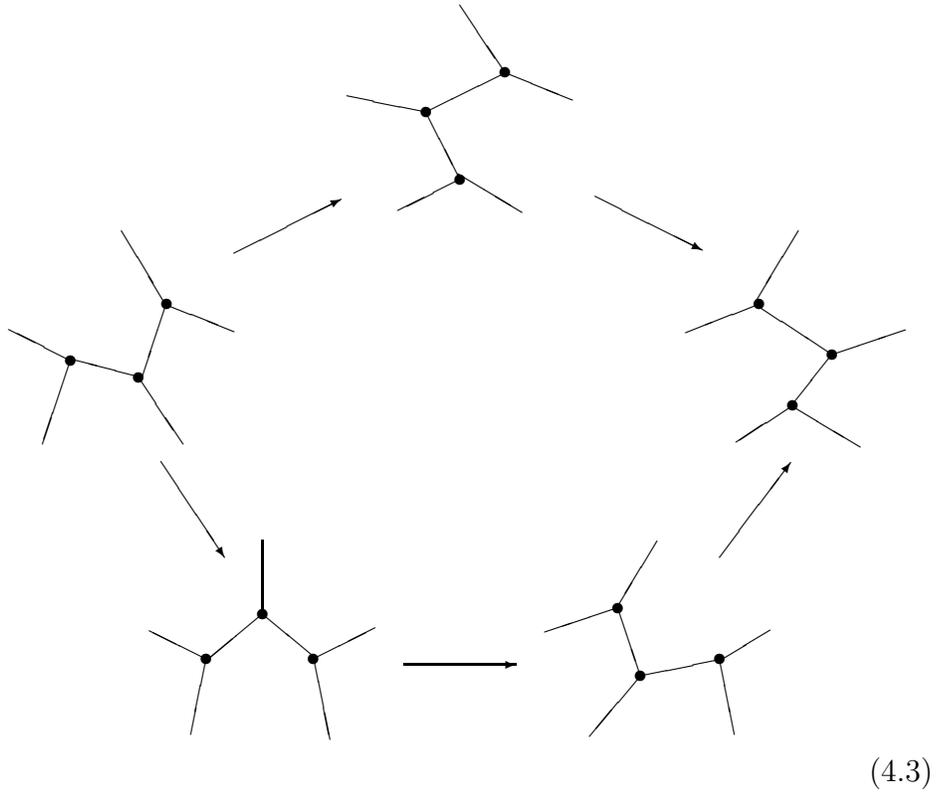
\end{defn}

There is a functor $fat:\TN\to\CR\CG$ called fattening. Given a trivalent
net $\Gamma$ with a set of 3-vertices $V$, we take $\Card V$ copies of a
ribbon graph $D_3$, glue pairwise intervals corresponding to internal edges
of $\Gamma$, and put on $fat(\Gamma)$ the same labels as on $\Gamma$. It is
important to glue respecting the orientation of $D_3$'s, and to make a
bijection between $\pi_0(B_3)$ and the set of edges incident to a 3-vertex
respecting the cyclic ordering ($\pi_0(B_3)$ has a canonical cyclic ordering
coming from the orientation of $D_3$). Fattening of a glueing is a glueing
and fattening of the fusing move $fus_a$ is the isotopy class of a
homeomorphism $fat(\Gamma)\to fat(\Gamma')$, which is identity outside of
the two $D_3$'s glued by $a$.
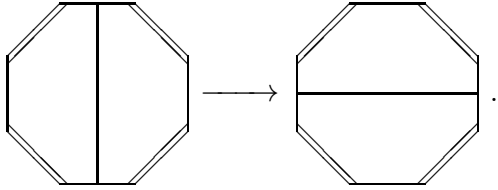
\[
\unitlength=1mm
\makebox[28mm][l]{
\raisebox{-14mm}[15mm][13mm]{
\put(1,10){\line(1,-1){7}}
\put(8,3){\line(1,0){10}}
\put(18,3){\line(1,1){7}}
\put(25,10){\line(0,1){10}}
\put(25,20){\line(-1,1){7}}
\put(18,27){\line(-1,0){10}}
\put(8,27){\line(-1,-1){7}}
\put(1,20){\line(0,-1){10}}
\put(1,11){\line(1,-1){8}}
\put(9,27){\line(-1,-1){8}}
\put(17,3){\line(1,1){8}}
\put(17,27){\line(1,-1){8}}
\put(13,3){\line(0,1){24}}
}}
\arow
\unitlength=1mm
\makebox[28mm][l]{
\raisebox{-14mm}[15mm][13mm]{
\put(1,10){\line(1,-1){7}}
\put(8,3){\line(1,0){10}}
\put(18,3){\line(1,1){7}}
\put(25,10){\line(0,1){10}}
\put(25,20){\line(-1,1){7}}
\put(18,27){\line(-1,0){10}}
\put(8,27){\line(-1,-1){7}}
\put(1,20){\line(0,-1){10}}
\put(1,11){\line(1,-1){8}}
\put(9,27){\line(-1,-1){8}}
\put(17,3){\line(1,1){8}}
\put(17,27){\line(1,-1){8}}
\put(1,15){\line(1,0){24}}
}} .\]

The ribbon graphs $fat(\Gamma)$ are canonically marked. A {\em marking} $M$
of a ribbon graph $(X,B)$ is by definition an isotopy class of
$\bigcup_k\gamma_k([0,1])$ for non-intersecting cuts $\gamma_k:[0,1]
\hookrightarrow X$ (such that $\gamma_k(]0,1[)\subset\Int X$,
$\gamma_k(0),\ \gamma_k(1)\in\partial X-B$) and such that
$X-\bigcup_k\gamma_k([0,1])$ is a union of hexagons $D_3$.

\subsEction{Coherence theorem for markings of ribbon graphs} 
\begin{thm}\labl{markingfusing}
Any marking of a ribbon graph can be obtained from any other marking by a
sequence of fusing moves. Any such sequence can be transformed into any
other using the pentagon, quadrilateral and $fus^2=\id$ identity.
\end{thm}

\begin{cor}
For all trivalent nets $\Gamma_1,\Gamma_2$
\[ \TN(\Gamma_1,\Gamma_2)\to \CR\CG(fat(\Gamma_1),fat(\Gamma_2)) \]
is bijective.
\end{cor}

\begin{pf}
Any morphism $f$ in $\TN$ or $\CR\CG$ is uniquely decomposed as $gh$,
where $g$ is a canonical glueing determined by the set of pairs of ends glued
by $f$, and $h$ is an isomorphism. Hence, we can consider only isomorphisms.
Let $\phi:fat(\Gamma_1)\to fat(\Gamma_2)$ be a homeomorphism. Use it to
transfer the canonical marking of $fat(\Gamma_2)$ to a marking $M$ of
$fat(\Gamma_1)$. By Theorem it can be obtained from the canonical marking of
$fat(\Gamma_1)$ by a sequence of fusing moves. This sequence determines a
morphism $j:\Gamma_1\to\Gamma\in\TN$ where $\Gamma$ is the net determined
by the marking $M$. It is identified with $\Gamma_2$ by an isomorphism
$i:\Gamma\to\Gamma_2\in\TN_0$. Finally, for $h=ji$ we have $\phi=fat(h)$.

For the second statement,
 it suffices to consider the case of isomorphic $\Gamma_1$ and
$\Gamma_2$. Furthermore, we have only to prove that any
$f:\Gamma\to\Gamma\in\TN$, such that $fat(f)$ is isotopic to
$\id_{fat(\Gamma)}$, equals $\id_\Gamma$. Represent $f$ as a product
$fus_{a_1}\ fus_{a_2}\ \dots\ fus_{a_n}$. The sequence of fusing moves in
$fat(\Gamma)$ corresponds to it. This sequence produces a new marking of
$fat(\Gamma)$, which equals to the image of the canonical marking under the
homeomorphism $fat(f)^{-1}$. Hence, it is isotopic to the canonical marking.
This gives a loop in $\CM$ which by the Theorem can be shrunk by the
pentagon, quadrilateral and $fus^2=\id$ moves. Parallelly we reduce the
expression for $f$ to 1, using the pentagon, quadrilateral and $fus^2=\id$
identities in $\TN$.
\end{pf}

This corollary implies

\begin{thm}
The functor $fat: \TN \to \RG$ is an equivalence of $\TN$ with
a full subcategory $\RG_+$ of ribbon graphs, each component of which satisfy
the inequalities
\be\label{g>n>}
n\ge3\quad \text{ or }\quad g\ge1,\, n\ge1\quad \text{ or } \quad g\ge2
\end{equation}
($g,n$ stand for genus and number of marked intervals).
\end{thm}

\begin{rem}
Formally adjoining the objects $D_0, D_1, D_2, A_0$ and some morphisms to
$\RG_+$ we can reconstruct $\RG$. We shall do a similar thing later.
\end{rem}

\subsEction{Trivalent nets with twists, braiding and switching} 
Now we add to $\TN$ new morphisms just as we added them to $\RG$. The
obtained category $TN$ will be a full subcategory of $RG$.

\begin{defn}\label{defTN}
Let $TN$ be a symmetric monoidal category with left cancellations which
objects are trivalent nets and morphisms are generated over $\TN$ by the
following morphisms.

        For an edge $A$ of a trivalent net $\Gamma$ there is a twist
automorphism $Tw_A:\Gamma \to \Gamma$.

        For a 3-vertex $a\in \Gamma$ and two incident edges $B,C$ there is
a braiding morphism $_aBr_{BC}: \Gamma \to \Gamma'$, denoted also $_ABr_{BC}$
or $Br_{BC}$ if there is no ambiguity, where $\Gamma'$ is obtained from
$\Gamma$ by a surgery cutting edges $B$ and $C$ and reglueing the ends
crosswise. $\Gamma'$ is isomorphic to $\Gamma$ with the changed cyclic
orientation at $a$. Locally it looks like
\[ _aBr_{BC}: \triup ABC \arow \triup ACB \]

For each loop-edge $A$ of $\Gamma$ there is a switch automorphism
$S_A:\Gamma \to \Gamma$, e.g.
\[ S_A: \tennis {}A @>>> \tennis {}A .\]

The morphisms are subject to the following relations. For any glueing $g$
\be\label{gTw=TN}
\begin{CD}
\Gamma_1 @>Tw_A>>   \Gamma_1 \\
@VgVV           @VVgV \\
\Gamma_2 @>Tw_{g(A)}>> \Gamma_2
\end{CD}
\end{equation}
\be
\begin{CD}
\Gamma_1 @>_aBr_{BC}>> \Gamma_1' \\
@VgVV           @VVg'V \\
\Gamma_2 @>_{g(a)}Br_{g(B),g(C)}>> \Gamma_2'
\end{CD}
\end{equation}
\be\label{gS=TN}
\begin{CD}
\Gamma_1 @>S_A>> \Gamma_1 \\
@VgVV           @VVgV \\
\Gamma_2 @>S_{g(A)}>> \Gamma_2
\end{CD}
\end{equation}
commute. Also
\begin{equation}\label{Br2=3TwTN}
Br^2_{BC} = Tw_A\ Tw_B^{-1} \ Tw_C^{-1} : \triup ABC \aow \triup ABC
\end{equation}
\begin{equation}\labl{TwA=BrBrTN}
Tw_A^{-1} = \left( \triup CAB \buildrel Br_{AB}\over\arow \triup CBA
     \buildrel Br_{AC}\over\arow  \triup ABC @>rot>> \triup CAB \right)
\end{equation}
\begin{equation}\labl{TwB=BrBrTN}
Tw_B^{-1}= \left( \triup CAB \buildrel Br_{AB}\over\arow \triup CBA
     \buildrel Br_{CB}\over\arow \triup BCA @>rot^{-1}>> \triup CAB \right)
\end{equation}
(here $rot$ is an isomorphism of rotation),
\begin{equation}\labl{hexagon}
\begin{array}{ccccc}
\tarahor XABC{} & \stackrel{Br_{BC}^{\pm1}}{\arrow} & \tarahor XACB{} &
          \stackrel{fus}{\arrow} & \taraver XACB{} \\
fus \bigg\downarrow \qquad &&&& \qquad \bigg\downarrow Br_{AC}^{\pm1} \\
\taraver XABCY & \stackrel{Br_{YC}^{\pm1}}{\arrow} & \tarahor XCABY &
          \stackrel{fus}{\arrow} & \taraver XCAB{}
\end{array}
\end{equation}
\begin{equation}\labl{ST3=S2TN}
(S_M Tw_M)^3 =S^2_M : \tennis {}M \aow \tennis {}M ,
\end{equation}
\be\labl{S2=Br-1Tw-1TN}
\begin{array}{ccc}
\tennis XM  & \buildrel S_M^2\over\arrow & \tennis XM \\
 _XBr_{MM}^{-1}\bigg\downarrow\qquad & & \quad \bigg\Vert \\
\tennis XM  & \buildrel Tw_M^{-1}\over\arrow & \tennis XM
\end{array}
\end{equation}
and \eqref{mainSdiagTN} (see Figure~\ref{A-2net}) are satisfied.
\begin{figure}[htbp]
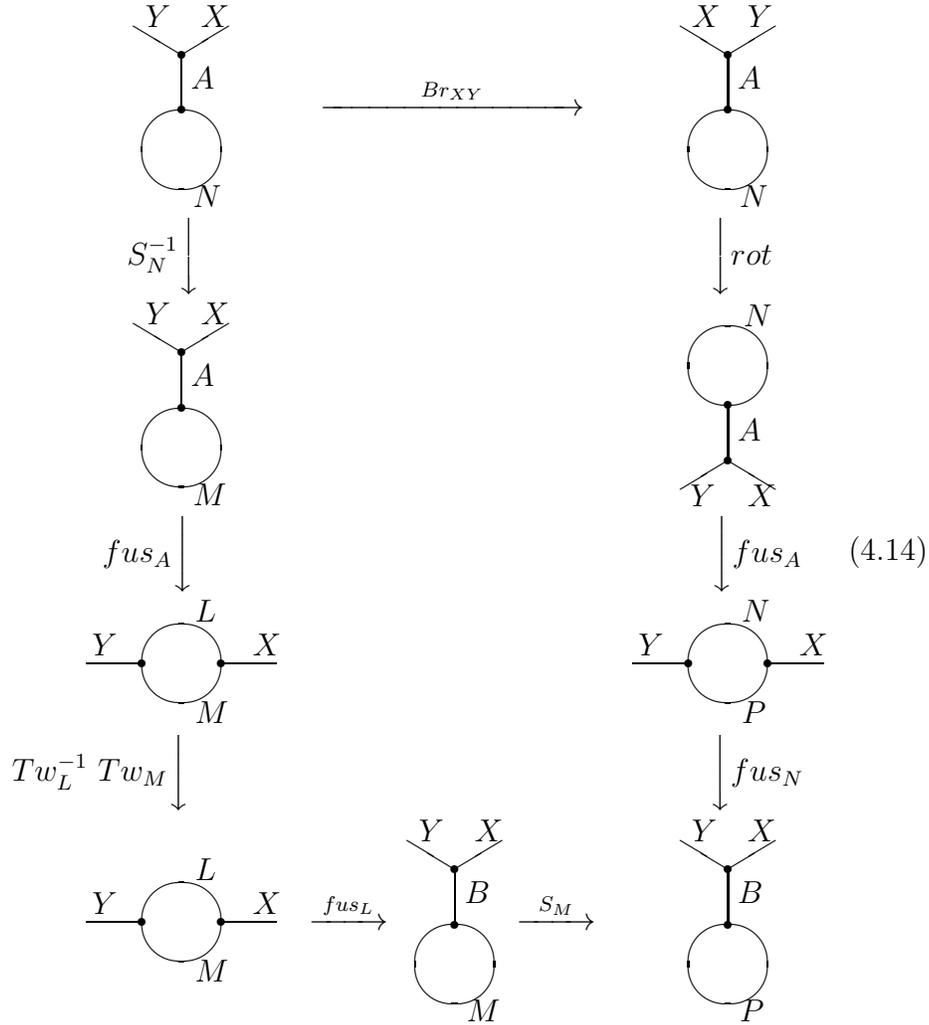

\be\label{mainSdiagTN}
\begin{array}{ccc}
\celodown YXAN & \stackrel{Br_{XY}}{\arrrrow} & \celodown XYAN \\
S_N^{-1}\bigg\downarrow\quad && \quad\bigg\downarrow rot \\
\celodown YXAM && \celoup YXAN \\
fus_A\bigg\downarrow\qquad && \qquad\bigg\downarrow fus_A \\
\sun YXLM && \sun YXNP \\
\nqquad Tw^{-1}_L\ Tw_M\bigg\downarrow\qqquad &&\qquad\bigg\downarrow fus_N\\
\sun YXLM & \stackrel{fus_L}{\arow} \celodown YXBM \stackrel{S_M}{\arow} &
\celodown YXBP
\end{array}
\end{equation}
\caption{A relation for $A_2$ net\label{A-2net}}
\end{figure}
\end{defn}

It is clear that the functor $fat: \TN \to \RG$ extends to a functor
$fat: TN \to RG$.

\begin{thm}\labl{fatTNRG}
The functor $fat:TN\to RG$ is an equivalence of $TN$ with a full subcategory
$RG_+$ of ribbon graphs satisfying condition \eqref{g>n>}.
\end{thm}

\begin{pf} A quasiinverse functor $\Phi: \RG_+ \to \TN$ is constructed like
that. For any ribbon graph $X$ choose a marking $M_X$ and set $\Phi(X)$ to be
the trivalent net $N(M_X)$ associated with $M_X$. Let $g:X\to Y$ be a
glueing, then there is a marking $g^*(M_X)$ of $Y$ consisting of $g(M_X)$
and images of those boundary intervals of $X$ which are glued by $g$. To $g$
corresponds the morphism
\[ \Phi(X) = N(M_X) @>g'>> N(g^*(M_X)) @>f>> N(M_Y) = \Phi(Y) ,\]
where $g'$ is a canonical glueing, and $f$ is a composition of fusings,
transforming $g^*(M_X)$ into $M_Y$.

The functor $\Phi$ extends to a functor $\Phi: RG_+ \to TN$. Indeed, take a
map $\gamma :[0,1] \hookrightarrow X$, or $h: (D_3,B_3) \hookrightarrow X$,
or $j:(A_1,B_1) \hookrightarrow X$ which determines a morphism $Tw_\gamma$,
$Br_h$, or $S_j$ as in Definition \ref{defrib}. There exists a marking $M$ of
$X$ containing $\gamma([0,1])$, or $h(B_3)$, or $j(B_1)$ and a composition
$f$ of fusing moves relating $M_X$ and $M$. Set
$\Phi(Tw_\gamma) = f\ Tw_A\ f^{-1}$, or $\Phi(S_j) = f\ S_A\ f^{-1}$ for $A$
from $M$ determined by $\gamma$ or $h$ and similarly for $Br_h$. The morphism
$f\in \TN$ is unique and a different choice of marking $M$, say $M'$, will
give $f'= fg$, where $g$ is a composition of fusing moves $fus_a$ with
$a\ne \gamma([0,1])$, $a\not\subset h(B_3)$, $a\ne j(B_1)$. Hence,
$\Phi(Tw_\gamma)$, $\Phi(Br_h)$, $\Phi(S_j)$ are correctly defined morphisms
of $TN$.

The relations \eqref{gTw=TN}--\eqref{gS=TN} follow from the relations
\eqref{gTw=Twg}--\eqref{gS=Sg}, and $\Phi$ applied to
\eqref{Br2=3TwTN}--\eqref{mainSdiagTN} is just
\eqref{Br=TwTwTw}--\eqref{A2mainS}. Thus the functor $\Phi:RG_+\to TN$ is
constructed and one can easily see that it is quasiinverse to
$fat:TN \to RG_+$.
\end{pf}


\sEction{A category of nets}
\subsEction{Nets} 
Now we consider nets with valency not bigger than 3.

\begin{defn}
Let a net be a 1-complex $\Gamma$ with the set of vertices
$V_3\sqcup V_2 \sqcup V_1\sqcup B$ and the set of edges $E$. The elements of
$V_i$, called $i$-vertices ($i=1,2,3$) occur $i$ times as endpoints of edges,
elements of $B$ called ends occur once as an endpoint of an edge. We assume
that each edge has at least one endpoint in $V_3\sqcup V_2 \sqcup V_1$. A net
$\Gamma$ is equipped with a labeling of ends $B\to \CC$ and for each 3-vertex
a cyclic order in the set of incident edges is chosen.
\end{defn}

Denote $\CN_0$ the symmetric monoidal category of nets with glueings as
morphisms. Glueings are defined similarly to trivalent case with requirement
that a glueing is bijective on $i$-vertices, $i=1,2,3$.

\begin{defn}
Let $\CN$ be a symmetric monoidal category with left cancellations which has
nets as objects, and morphisms are generated over $\CN_0$ by the following
morphisms:

insertion $ins_a:\Gamma\to \Gamma'$ is an isomorphism of $\Gamma$ to
$\Gamma'$ with a 1-vertex $a\in\Gamma'$ and the incident edge deleted or a
2-vertex $a\in\Gamma'$ deleted and the two incident edges combined in one;

deletion $del_a:\Gamma'\to \Gamma$ is the inverse morphism.

These morphisms are subject to the following relations. Whenever the diagram
\be\label{gins=N}
\begin{CD}
\Gamma_1 @>ins_a>>   \Gamma_1' \\
@VgVV           @VVg'V \\
\Gamma_2 @>ins_{g'(a)}>> \Gamma_2'
\end{CD}
\end{equation}
commutes as a diagram of mappings of 1-complexes, it commutes also in $\CN$.
Also the diagrams
\be
\begin{CD}
\edgepoint a @>ins_b>> \lezhakpt ab \\
@Vins_cVV            @VVins_cV  \\
\lezhakpt ac @>ins_b>>
{\unitlength=0.75pt
\makebox[39 pt][l]{
\raisebox{3 pt}[19 pt][13 pt]{
\put(22,0){\line(-1,0){23}}
\put(22,0){\line(3,-5){12}}
\put(34,-20){\circle*{4}}
\put(22,0){\line(3,5){12}}
\put(34,20){\circle*{4}}
\put(22,0){\circle*{4}}
\put(2,4){$a$}
\put(35,-17){$b$}
\put(35,10){$c$}
}}}
\end{CD}
\end{equation}
\be\labl{ins12ver}
\begin{CD}
\pointedge a @>ins_b>> \ptlezhak ba \\
@Vins_cVV \nqquad\nqquad \nearrow \phi \qqquad \\
\ptlezhak ac
\end{CD}
\end{equation}
where $\phi$ is the isomorphism of nets,
\be\label{ins2pointN}
\begin{CD}
\lezhak{}a @>ins_b>> \lintwopt ba \\
@Vins_cVV \nqquad\nqquad \nearrow \psi \qqquad\qquad \\
\lintwopt ac
\end{CD}
\end{equation}
where $\psi$ is the isomorphism of nets sending $a \mapsto b$, $c \mapsto a$,
are all commutative.

Finally, the isomorphism
\be\label{rot=1}
rot: \twopointedge \aow \twopointedge
\end{equation}
interchanging the vertices is set equal to identity morphism.
\end{defn}

Let $\CN_+$ be the full subcategory of such nets that genus $g$ and the
number of ends $n$ for each component satisfy \eqref{g>n>}. There is an
obvious inclusion functor $inc: \TN_0 \to \CN_+$. Let us construct a functor
$tri: \CN_+ \to \TN_0$. Take a net $\Gamma$. Denote $red_1(\Gamma)$ the net
$\Gamma$ with all 1-vertices and incident edges deleted. Repeat the reduction
until $\Gamma'=(red_1)^k(\Gamma)$ does not contain 1-vertices. Change each
connected string of 2-vertices in $\Gamma'$ together with incident edges to
one edge. The obtained net $tri(\Gamma)$ is trivalent. Any glueing
$g:\Gamma_1 \to \Gamma_2$ is bijective on 1-vertices, hence, induces a
glueing $red_1(\Gamma_1) \to red_1(\Gamma_2)$, therefore, a glueing
$\Gamma_1' \to \Gamma_2'$ which leads to a glueing
$tri(g): tri(\Gamma_1) \to tri(\Gamma_2)$. Any morphism $ins_a$, $del_a$ is
sent to identity by the functor $tri$.

\begin{thm}\labl{inctriCal}
The functors $inc: \TN_0 \to \CN_+$ and $tri: \CN_+ \to \TN_0$ are
quasi-inverse to each other.
\end{thm}

\subsEction{Nets with fusing, braiding, twists and switches}
Now we add more morphisms to the category of nets.

\begin{defn}
Let $N$ be a symmetric monoidal category with left cancellations having
nets as objects and having morphisms generated over $\CN$ by
$fus_a:\Gamma \to\Gamma'$, $Tw_a: \Gamma \to \Gamma$,
$_aBr_{BC}:\Gamma \to\Gamma'$, $S_a:\Gamma \to\Gamma$ as in Definitions
\ref{deftrivalnet}, \ref{defTN}. We subject them to the relations of
Definitions \ref{deftrivalnet}, \ref{defTN} and the following. There are
equations
\be\labl{TwA=TwB}
Tw_A = Tw_B : \lezhak AB \aow \lezhak AB ,
\end{equation}
\be
Tw_a = 1 : \pointedge a \aow \pointedge a .
\end{equation}
The following diagrams commute
\begin{equation}
\unitlength=0.75pt
\begin{array}{ccccc}
{\makebox[33 pt]{
\raisebox{3 pt}[22.5 pt][16.5 pt]{
\put(-2,0){\line(5,-3){20}}
\put(-2,0){\line(5,3){20}}
\put(-2,0){\line(-5,3){20}}
\put(-2,0){\circle*{4}}
}}}
& @>ins>> &
{\makebox[45 pt]{
\raisebox{3 pt}[27 pt][21 pt]{
\put(-19,0){\line(-3,5){12}}
\put(15,0){\line(3,-5){12}}
\put(15,0){\line(3,5){12}}
\put(-19,0){\line(1,0){34}}
\put(-19,0){\circle*{4}}
\put(15,0){\circle*{4}}
}}}
& @>ins>> &
{\unitlength=0.75pt
\makebox[45 pt]{
\raisebox{3 pt}[27 pt][21 pt]{
\put(-19,0){\line(-3,5){12}}
\put(-19,0){\line(-3,-5){12}}
\put(-31,-20){\circle*{4}}
\put(15,0){\line(3,-5){12}}
\put(15,0){\line(3,5){12}}
\put(-19,0){\line(1,0){34}}
\put(-19,0){\circle*{4}}
\put(15,0){\circle*{4}}
}}}
\\
& ins\searrow &&& \quad \Big\downarrow fus \\
&&
{\makebox[45 pt]{
\raisebox{3 pt}[27 pt][21 pt]{
\put(-2,17){\line(-5,3){24}}
\put(-2,-17){\line(5,-3){24}}
\put(-2,17){\line(5,3){24}}
\put(-2,17){\line(0,-1){34}}
\put(-2,17){\circle*{4}}
\put(-2,-17){\circle*{4}}
}}}
& @>ins>> &
{\unitlength=0.75pt
\makebox[45 pt]{
\raisebox{3 pt}[27 pt][21 pt]{
\put(-2,17){\line(-5,3){24}}
\put(-2,-17){\line(-5,-3){24}}
\put(-26,-31.6){\circle*{4}}
\put(-2,-17){\line(5,-3){24}}
\put(-2,17){\line(5,3){24}}
\put(-2,17){\line(0,-1){34}}
\put(-2,17){\circle*{4}}
\put(-2,-17){\circle*{4}}
}}}
\end{array}
\end{equation}
\begin{equation}\labl{braidwith1N}
\begin{array}{rcl}
{\unitlength=0.75pt
\makebox[33 pt]{
\raisebox{3 pt}[22.5 pt][16.5 pt]{
\put(-2,0){\line(0,1){23}}
\put(-2,0){\line(-5,-3){20}}
\put(-22,-12){\circle*{4}}
\put(-2,0){\line(5,-3){20}}
\put(-2,0){\circle*{4}}
\put(-17,16){$C$}
\put(-21,-23){$A$}
\put(7,-23){$B$}
}}}
& @>Br_{AB}^{\pm1}>> & \triuprpt CBA \\
del_A\searrow\nquad  && \nquad \swarrow del_A \\
& \stoyak CB  &
\end{array}
\end{equation}
\end{defn}

Denote $N_+$ a full subcategory of nets whose components satisfy
inequalities \eqref{g>n>}.

\begin{thm}\labl{triNTN}
The functor $tri$ uniquely extends to a functor $tri:N_+ \to TN$
quasi-inverse to inclusion functor $inc: TN\to N_+$.
\end{thm}

\subsEction{Nets compared with ribbon graphs} 
Extend the functor $N_+ @>tri>> TN @>fat>> RG$ to a functor $fat:N\to RG$
setting
\begin{align*}
fat(\twopointedge) &= \put(16,2){\circle{26}} \labl{2pt=disk} \\
fat(\pointedge A ) &= \diskone A , \\
fat(\lezhak AB  ) &= \disktwo AB , \\
fat(\okrug) &= \cylindre , \labl{okrugcyl}
\end{align*}
The fattening of an insertion or deletion morphism is set to the unique
``identity'' homeomorphism of the ribbon graphs. Fattenings of other
generators are already defined.

\begin{thm} \label{fatNtoRG}
The functor $fat:N\to RG$ is an equivalence of categories.
\end{thm}

\sEction{A category of oriented nets}\label{oriented}
\subsEction{Oriented nets} 
By an {\em oriented net} we mean a net $\Gamma$, equipped with such
orientation of edges that each 2-vertex or 3-vertex is a source for some edge
and a target for some edge. Given an oriented net $\Gamma$ with labeling
$L:B\to\CC$, we define its desorientation $deso(\Gamma)$ as a net $\Gamma$
with labeling $L':B\to\CC$ constructed as follows. $L'(b)=L(b)$ if the end
$b$ is a target and $L'(b)=L(b)\pti$ if the end $b$ is a source of the
incident external leg. A glueing of oriented nets is an orientation
preserving map, which is a glueing of underlying desoriented nets. In
particular, only those external legs can be glued which carry the same label
from $\CC$ and one of them has the incident end as a target, and another as
a source.

\begin{defn}
Let $\ON$ be a symmetric monoidal category with left cancellations having
oriented nets as objects and morphisms generated over glueings by the
following new morphisms:

reversal $rev_a:\Gamma\to\Gamma'$, where $a$ is an internal edge of $\Gamma$
and $\Gamma'$ differs from $\Gamma$ by a reversed orientation of $a$;

dualising morphism $du_a:\Gamma\to\Gamma'$, where $a$ is an external leg of
$\Gamma$, $\Gamma'$ differs from $\Gamma$ by a reversed orientation of $a$,
and the labels $A,A'\in\CC$ attached to $a$ are dual to each other;

$ins_a:\Gamma\to\Gamma',\ del_a:\Gamma\to\Gamma'$, insertion and deletion
of a 2-vertex.

These morphisms are subject to relations
\[ins\cdot del=1, \quad del\cdot ins=1, \quad rev_a^2=1, \quad du_a^2=1. \]
For each glueing $g:\Gamma_1\to\Gamma_2$ the diagram \eqref{gins=N} and
\begin{equation}\labl{grev=ON}
\begin{CD}
\Gamma_1 @>rev_a>> \Gamma_1' \\
@VgVV              @VVg'V \\
\Gamma_2 @>rev_{g(a)}>> \Gamma_2'
\end{CD}
\end{equation}
\be\labl{gdu=ON}
\begin{CD}
\Gamma_1 @>du_a>> \Gamma_1' \\
@VgVV             @VVg'V \\
\Gamma_2 @>du_{g(a)}>> \Gamma_2'
\end{CD}
\end{equation}
(if the external leg $a$ is not glued by $g$) commute. The diagrams
\be
\begin{CD}
\lezhakright{}a @>ins_b>> \lintwoptright ba \\
@Vins_cVV \nqquad\nqquad \nearrow \psi \qqquad\qquad \\
\lintwoptright ac
\end{CD}
\end{equation}
\begin{equation}\label{gluerev}
\begin{CD}
\trirightr XAB \trileftr XCD @>du_X\ du_X>>
\trirightl {X\pti}AB \trileftl {X\pti}CD \\
@Vg'VV @VVg''V \\
\tarahorr ABCD{} @>rev>> \tarahorl ABCD{}
\end{CD}
\end{equation}
\begin{equation}\label{1ptdudurev}
\begin{CD}
\pointedger X \ \trileftr XAB @>du_X\ du_X>>
\pointedgel {X\pti} \ \trileftl {X\pti}AB \\
@Vg'VV @VVg''V \\
{\unitlength=0.75pt
\makebox[39 pt][l]{
\raisebox{3 pt}[19 pt][13 pt]{
\put(22,0){\line(-1,0){23}}
\put(12,0){\vector(1,0){0}}
\put(-1,0){\circle*{4}}
\put(22,0){\line(3,-5){12}}
\put(22,0){\line(3,5){12}}
\put(22,0){\circle*{4}}
\put(35,-17){$A$}
\put(35,10){$B$}
}}}
@>rev>>
{\unitlength=0.75pt
\makebox[39 pt][l]{
\raisebox{3 pt}[19 pt][13 pt]{
\put(22,0){\line(-1,0){23}}
\put(10,0){\vector(-1,0){0}}
\put(-1,0){\circle*{4}}
\put(22,0){\line(3,-5){12}}
\put(22,0){\line(3,5){12}}
\put(22,0){\circle*{4}}
\put(35,-17){$A$}
\put(35,10){$B$}
}}}
\end{CD}
\end{equation}
\be\label{2ptdudurev}
\begin{CD}
\pointedger X \edgepointright X @>du_X\ du_X>>
\pointedgel {X\pti}
{\unitlength=0.75pt
\makebox[22 pt][l]{
\raisebox{3 pt}[16 pt][2 pt]{
\put(2,0){\line(1,0){20}}
\put(10,0){\vector(-1,0){0}}
\put(22,0){\circle*{4}}
\put(10,4){$X\pti$}
}}}
\\
@Vg'VV @VVg''V \\
\twopointedger @>rev>> \twopointedgel
\end{CD}
\end{equation}
\begin{equation}\label{dddddd}
\begin{array}{ccccc}
&& \triupioo CAB && \\
& du_A \nearrow && \searrow du_B & \\
\triupioi C{A\pti}B &&&& \triupiio CA{B\pti} \\
du_C\Big\uparrow\quad &&&& \quad\Big\downarrow du_C \\
\triupooi {C\pti}{A\pti}B &&&& \triupoio  {C\pti}A{B\pti} \\
& du_B \nwarrow && \swarrow du_A & \\
&& \triupoii {C\pti}{A\pti}{B\pti}
\end{array}
\end{equation}
also commute.

Finally, the morphism
\[ rev: a \twopointedger b\ \arow\ a \twopointedgel b \]
equals to the isomorphism of graphs, which sends $a$ to $b$, $b$ to $a$.
\end{defn}

\begin{notation}
Both morphisms $rev_a$ and $du_a$ will be sometimes denoted $Du_a$.
\end{notation}

Denote $\CN'\subset \CN$ the category of nets $\CN_0$ extended by morphisms
$ins_a, del_a$ of insertion and deletion of a 2-vertex with relations
(\ref{gins=N}), (\ref{ins2pointN}), (\ref{rot=1}).
There is a desorienting functor $deso:\ON\to \CN'$ which sends
$rev_a,du_a:\Gamma\to\Gamma'$ to $\id_{deso(\Gamma)}$.

\begin{thm}\labl{desoONN'}
The functor $deso:\ON\to\CN'$ is an equivalence of categories.
\end{thm}

\subsEction{Oriented nets with fusing, twists, braiding and switches} 
Now we add to $\ON$ new morphisms. The obtained category $ON$ will be made
equivalent to $N$.

\begin{defn}
Let $ON$ be a symmetric monoidal category with left cancellations which
objects are oriented nets and morphisms are generated over $\ON$  by the
following isomorphisms:

$ ins_a:\Gamma\to\Gamma',\ del_a:\Gamma'\to\Gamma $
where $\Gamma'$ differs from $\Gamma$ by a 1-vertex $a$ and an incident edge;

$ fus_a:\Gamma\to\Gamma', $
where  a neighbourhood of $a$ looks as
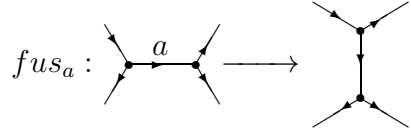
\[ fus_a: \tarahoriooor {}{}{}{}a \arow \taraverioood {}{}{}{}{} \]
(only part of $\Gamma,\Gamma'$ is drawn here, the complements of it in
$\Gamma$ and $\Gamma'$ are the same). It is important that cyclic ordering
for all 3-vertices is coherent here with the orientation of the plane of
drawing;

$Tw_a:\Gamma\to\Gamma$ for an edge $a$ of $\Gamma$;

$_aBr_{BC}:\Gamma\to\Gamma'$ for a 3-vertex $a\in\Gamma$ and two incident
edges $B,C$ with sources in $a$. The net $\Gamma'$ is obtained from $\Gamma$
as in unoriented case. Locally the morphism looks as
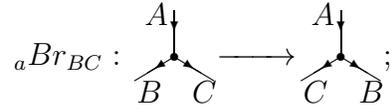
\[ _aBr_{BC}: \triupioo ABC \arow \triupioo ACB ; \]

$S_a:\Gamma\to\Gamma$ for a loop-edge $a\in\Gamma$ looking as
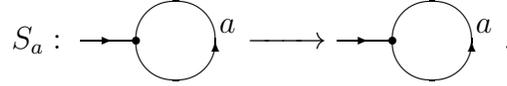
\[ S_a: \tennisu {}a \arow \tennisu {}a .\]

They are subject to the following relations. For any glueing
$g:\Gamma_1\to\Gamma_2$ whenever upper row in diagrams \eqref{gins=N},
\eqref{gfus=fusg}, \eqref{gTw=TN}--\eqref{gS=TN} is defined, the whole
diagram is defined and commute. The following diagrams commute
\begin{equation}
\begin{CD}
\lezhakright{}{} @>ins>> \torchokur{}{} \\
@| @VVrevV \\
\lezhakright{}{} @>ins>> \torchokdr{}{}
\end{CD}
\end{equation}
\begin{equation}
\begin{array}{ccccc}
\lezhakright ab & @>Ins>> \torchokur ab @>Du_b>>
  &
{\unitlength=0.75pt
\makebox[36 pt][l]{
\raisebox{-4 pt}[15 pt][7.5 pt]{
\put(0,0){\line(1,0){40}}
\put(12,0){\vector(1,0){0}}
\put(28,0){\vector(-1,0){0}}
\put(20,0){\line(0,1){20}}
\put(20,12){\vector(0,1){0}}
\put(20,20){\circle*{4}}
\put(20,0){\circle*{4}}
\put(3,4){$a$}
\put(28,4){$b$}
}}}
& @>Du_a>> & \torchokul ab
\\
Del \bigg\downarrow\quad &&&& \quad\bigg\downarrow Del \\
\begin{picture}(25,4)
\put(0,3){\line(1,0){20}}
\put(12,3){\vector(1,0){0}}
\end{picture}
& @>{d\text{ or }rev}>> &
\begin{picture}(25,4)
\put(0,3){\line(1,0){20}}
\put(8,3){\vector(-1,0){0}}
\end{picture}
& @>Ins>> &
{\unitlength=0.75pt
\makebox[36 pt][l]{
\raisebox{3 pt}[16 pt][2 pt]{
\put(0,0){\line(1,0){40}}
\put(8,0){\vector(-1,0){0}}
\put(28,0){\vector(-1,0){0}}
\put(20,0){\circle*{4}}
\put(3,4){$a$}
\put(28,4){$b$}
}}}
\end{array}
\end{equation}
\be
\begin{CD}
\edgepointright a @>ins_b>> \lezhakptright ab \\
@Vins_cVV            @VVins_cV  \\
\lezhakptright ac @>ins_b>>
{\unitlength=0.75pt
\makebox[39 pt][l]{
\raisebox{3 pt}[19 pt][13 pt]{
\put(22,0){\line(-1,0){23}}
\put(13,0){\vector(1,0){0}}
\put(22,0){\line(3,-5){12}}
\put(34,-20){\circle*{4}}
\put(28,-10){\vector(2,-3){0}}
\put(22,0){\line(3,5){12}}
\put(34,20){\circle*{4}}
\put(28,10){\vector(2,3){0}}
\put(22,0){\circle*{4}}
\put(2,4){$a$}
\put(35,-17){$b$}
\put(35,10){$c$}
}}}
\end{CD}
\end{equation}
\be
\begin{CD}
\pointedger a @>ins_b>> \ptlezhakright ba \\
@Vins_cVV \nqquad\nqquad \nearrow \phi\qqquad \\
\ptlezhakright ac
\end{CD}
\end{equation}
(where $\phi$ is the isomorphism of oriented nets)
\be\labl{TwforI}
Tw_a = 1 : \pointedger a \aow \pointedger a .
\end{equation}
and similar diagrams with opposite orientation of edges also commute.

The following diagrams commute
\begin{equation}
\unitlength=0.75pt
\begin{array}{ccccc}
{\makebox[33 pt]{
\raisebox{3 pt}[22.5 pt][16.5 pt]{
\put(-2,0){\line(5,-3){20}}
\put(8,-6){\vector(3,-2){0}}
\put(-2,0){\line(5,3){20}}
\put(8,6){\vector(3,2){0}}
\put(-2,0){\line(-5,3){20}}
\put(-12,6){\vector(-3,2){0}}
\put(-2,0){\circle*{4}}
}}}
& @>ins>> &
{\unitlength=0.75pt
\makebox[45 pt]{
\raisebox{3 pt}[27 pt][21 pt]{
\put(-19,0){\line(-3,5){12}}
\put(-25,10){\vector(2,-3){0}}
\put(15,0){\line(3,-5){12}}
\put(21,-10){\vector(2,-3){0}}
\put(15,0){\line(3,5){12}}
\put(21,10){\vector(2,3){0}}
\put(-19,0){\line(1,0){34}}
\put(-2,0){\vector(1,0){0}}
\put(-19,0){\circle*{4}}
\put(15,0){\circle*{4}}
}}}
& @>ins>> &
{\unitlength=0.75pt
\makebox[45 pt]{
\raisebox{3 pt}[27 pt][21 pt]{
\put(-19,0){\line(-3,5){12}}
\put(-25,10){\vector(2,-3){0}}
\put(-19,0){\line(-3,-5){12}}
\put(-31,-20){\circle*{4}}
\put(-25,-10){\vector(-2,-3){0}}
\put(15,0){\line(3,-5){12}}
\put(21,-10){\vector(2,-3){0}}
\put(15,0){\line(3,5){12}}
\put(21,10){\vector(2,3){0}}
\put(-19,0){\line(1,0){34}}
\put(-2,0){\vector(1,0){0}}
\put(-19,0){\circle*{4}}
\put(15,0){\circle*{4}}
}}}
\\
& ins\searrow &&& \quad \Big\downarrow fus \\
&&
{\unitlength=0.75pt
\makebox[45 pt]{
\raisebox{3 pt}[27 pt][21 pt]{
\put(-2,17){\line(-5,3){24}}
\put(-12,23){\vector(3,-2){0}}
\put(-2,-17){\line(5,-3){24}}
\put(8,-23){\vector(3,-2){0}}
\put(-2,17){\line(5,3){24}}
\put(8,23){\vector(3,2){0}}
\put(-2,17){\line(0,-1){34}}
\put(-2,0){\vector(0,-1){0}}
\put(-2,17){\circle*{4}}
\put(-2,-17){\circle*{4}}
}}}
& @>ins>> &
{\unitlength=0.75pt
\makebox[45 pt]{
\raisebox{3 pt}[27 pt][21 pt]{
\put(-2,17){\line(-5,3){24}}
\put(-12,23){\vector(3,-2){0}}
\put(-2,-17){\line(-5,-3){24}}
\put(-26,-31.6){\circle*{4}}
\put(-12,-23){\vector(-3,-2){0}}
\put(-2,-17){\line(5,-3){24}}
\put(8,-23){\vector(3,-2){0}}
\put(-2,17){\line(5,3){24}}
\put(8,23){\vector(3,2){0}}
\put(-2,17){\line(0,-1){34}}
\put(-2,0){\vector(0,-1){0}}
\put(-2,17){\circle*{4}}
\put(-2,-17){\circle*{4}}
}}}
\end{array}
\end{equation}
\begin{equation}
\unitlength=0.75pt
\begin{array}{ccccc}
{\makebox[33 pt]{
\raisebox{3 pt}[22.5 pt][16.5 pt]{
\put(-2,0){\line(-5,3){20}}
\put(-12,6){\vector(3,-2){0}}
\put(-2,0){\line(-5,-3){20}}
\put(-12,-6){\vector(-3,-2){0}}
\put(-2,0){\line(5,-3){20}}
\put(8,-6){\vector(3,-2){0}}
\put(-2,0){\circle*{4}}
}}}
& @>ins>> &
{\unitlength=0.75pt
\makebox[45 pt]{
\raisebox{3 pt}[27 pt][21 pt]{
\put(-19,0){\line(-3,5){12}}
\put(-25,10){\vector(2,-3){0}}
\put(-19,0){\line(-3,-5){12}}
\put(-25,-10){\vector(-2,-3){0}}
\put(15,0){\line(3,-5){12}}
\put(21,-10){\vector(2,-3){0}}
\put(-19,0){\line(1,0){34}}
\put(-2,0){\vector(1,0){0}}
\put(-19,0){\circle*{4}}
\put(15,0){\circle*{4}}
}}}
& @>ins>> &
{\unitlength=0.75pt
\makebox[45 pt]{
\raisebox{3 pt}[27 pt][21 pt]{
\put(-19,0){\line(-3,5){12}}
\put(-25,10){\vector(2,-3){0}}
\put(-19,0){\line(-3,-5){12}}
\put(-25,-10){\vector(-2,-3){0}}
\put(15,0){\line(3,-5){12}}
\put(21,-10){\vector(2,-3){0}}
\put(15,0){\line(3,5){12}}
\put(27,20){\circle*{4}}
\put(21,10){\vector(2,3){0}}
\put(-19,0){\line(1,0){34}}
\put(-2,0){\vector(1,0){0}}
\put(-19,0){\circle*{4}}
\put(15,0){\circle*{4}}
}}}
\\
& ins\searrow &&& \quad \Big\downarrow fus \\
&&
{\unitlength=0.75pt
\makebox[45 pt]{
\raisebox{3 pt}[27 pt][21 pt]{
\put(-2,17){\line(-5,3){24}}
\put(-12,23){\vector(3,-2){0}}
\put(-2,-17){\line(-5,-3){24}}
\put(-12,-23){\vector(-3,-2){0}}
\put(-2,-17){\line(5,-3){24}}
\put(8,-23){\vector(3,-2){0}}
\put(-2,17){\line(0,-1){34}}
\put(-2,0){\vector(0,-1){0}}
\put(-2,17){\circle*{4}}
\put(-2,-17){\circle*{4}}
}}}
& @>ins>> &
{\unitlength=0.75pt
\makebox[45 pt]{
\raisebox{3 pt}[27 pt][21 pt]{
\put(-2,17){\line(-5,3){24}}
\put(-12,23){\vector(3,-2){0}}
\put(-2,-17){\line(-5,-3){24}}
\put(-12,-23){\vector(-3,-2){0}}
\put(-2,-17){\line(5,-3){24}}
\put(8,-23){\vector(3,-2){0}}
\put(-2,17){\line(5,3){24}}
\put(22,31.6){\circle*{4}}
\put(8,23){\vector(3,2){0}}
\put(-2,17){\line(0,-1){34}}
\put(-2,0){\vector(0,-1){0}}
\put(-2,17){\circle*{4}}
\put(-2,-17){\circle*{4}}
}}}
\end{array}
\end{equation}
\begin{equation}\labl{ins4fus3}
\unitlength=0.75pt
\begin{array}{ccccc}
{\makebox[33 pt]{
\raisebox{3 pt}[22.5 pt][16.5 pt]{
\put(-2,0){\line(-5,-3){20}}
\put(-12,-6){\vector(-3,-2){0}}
\put(-2,0){\line(5,3){20}}
\put(8,6){\vector(3,2){0}}
\put(-2,0){\line(-5,3){20}}
\put(-12,6){\vector(3,-2){0}}
\put(-2,0){\circle*{4}}
}}}
& @>ins>> & \tarahoriobor{}{}{}{}{} & @>ins>> &
\tarahoriooorptdr {}{}{}{}{} \\
& ins\searrow &&& \quad \Big\downarrow fus \\
&& \taraveriobod {}{}{}{}{} & @>ins>> & \taraveriooodptdr {}{}{}{}{}
\end{array}
\end{equation}
\begin{equation}\label{fus1?}
\begin{array}{ccccccc}
\tarahoriooor ABCD{} & \buildrel d_C\over\aow & \tarahorioior AB{C\pti}D{}
    & \buildrel rev\over\aow & \tarahorioiol AB{C\pti}D{}
    & \buildrel d_A\over\aow & \tarahorooiol {A\pti}B{C\pti}D{} \\
Fus\bigg\downarrow\quad &&&&&& \quad\bigg\downarrow Fus \\
\taraverioood ABCD{} & \buildrel d_C\over\aow & \taraverioiod AB{C\pti}D{}
    & \buildrel rev\over\aow & \taraverioiou AB{C\pti}D{}
    & \buildrel d_A\over\aow & \taraverooiou {A\pti}B{C\pti}D{}
\end{array}
\end{equation}
\begin{equation}\label{fus2?}
\begin{array}{ccccc}
\nquad \tarahoroooil ABCD{}  & \buildrel d_A\over\aow
\tarahoriooil {A\pti}BCD{} & \buildrel rev\over\arrow &
\tarahoriooir {A\pti}BCD{} \buildrel d_D\over\aow
   & \tarahoriooor {A\pti}BC{D\pti}{}    \\
\nquad Fus \bigg\uparrow\quad &&&& \quad\bigg\downarrow Fus \\
\nquad \taraveroooid ABCD{} & \buildrel d_A\over\lfarrrow &
\taraveriooid {A\pti}BCD{}
    & \buildrel d_D\over\lfarrrow & \taraverioood {A\pti}BC{D\pti}{}
\end{array}
\end{equation}
\be\labl{TwA=TwB_or}
Tw_A = Tw_B : \lezhakright AB \aow \lezhakright AB ,
\end{equation}
\begin{figure}[htb]
\be\label{pentagon_or}
\unitlength=0.75mm
\begin{picture}(160,131)
\put(80,131){\line(2,-3){8}}
\put(88,119){\line(5,-2){12}}
\put(88,119){\circle*{2}}
\put(88,119){\line(-2,-1){14}}
\put(74,112){\line(1,-2){5.67}}
\put(79.67,101){\line(5,-3){11.33}}
\put(74,112){\circle*{2}}
\put(80,100){\circle*{2}}
\put(80,100){\line(-2,-1){11}}
\put(74,112){\line(-5,1){14}}
\put(20,91){\line(3,-5){7.67}}
\put(27.67,78){\line(5,-2){12.33}}
\put(28,78){\circle*{2}}
\put(28,78){\line(-1,-3){4.33}}
\put(23.67,65){\line(-4,1){12.67}}
\put(11,68){\line(-2,1){11}}
\put(11,68){\circle*{2}}
\put(23,65){\circle*{2}}
\put(23,65){\line(2,-3){8}}
\put(11,68){\line(-1,-3){5}}
\put(140,91){\line(-3,-5){7.67}}
\put(132.33,78){\line(3,-2){13.67}}
\put(146,69){\line(-4,-5){7.33}}
\put(138.67,60){\line(5,-3){12.33}}
\put(133,78){\circle*{2}}
\put(133,78){\line(-5,-2){13}}
\put(146,69){\circle*{2}}
\put(146,69){\line(3,1){13}}
\put(139,60){\circle*{2}}
\put(139,60){\line(-3,-2){10}}
\put(115,36){\line(-3,-5){7.33}}
\put(107.67,24){\line(-3,-1){12.67}}
\put(108,24){\circle*{2}}
\put(108,24){\line(1,-3){4}}
\put(112,12){\line(-5,-6){9}}
\put(112,12){\circle*{2}}
\put(112,12){\line(5,1){14}}
\put(126,14.67){\line(5,3){9}}
\put(126,15){\circle*{2}}
\put(126,15){\line(1,-5){2.67}}
\put(45,36){\line(0,-1){13}}
\put(45,23){\line(-6,-5){10}}
\put(45,23){\circle*{2}}
\put(45,23){\line(6,-5){9}}
\put(54,15.67){\line(1,-5){3}}
\put(54,15){\circle*{2}}
\put(54,15){\line(2,1){11}}
\put(35,15){\circle*{2}}
\put(35,15){\line(-2,1){10}}
\put(35,15){\line(-1,-5){2.67}}
\put(70,14){\vector(1,0){20}}
\put(27,50){\vector(2,-3){11.33}}
\put(40,87){\vector(2,1){19}}
\put(104,97){\vector(2,-1){19}}
\put(126,33){\vector(3,4){12.67}}
\put(84,125){\vector(-2,3){0}}
\put(98,115){\vector(3,-1){0}}
\put(82,116){\vector(2,1){0}}
\put(69,113){\vector(4,-1){0}}
\put(78,104){\vector(1,-2){0}}
\put(88,96){\vector(3,-2){0}}
\put(74,97){\vector(-2,-1){0}}
\put(23,86){\vector(-2,3){0}}
\put(35,75){\vector(3,-1){0}}
\put(26,72){\vector(1,3){0}}
\put(29,56){\vector(2,-3){0}}
\put(20,66){\vector(4,-1){0}}
\put(8,59){\vector(-1,-3){0}}
\put(7,70){\vector(2,-1){0}}
\put(128,76){\vector(3,1){0}}
\put(137,86){\vector(2,3){0}}
\put(140,73){\vector(3,-2){0}}
\put(155,72){\vector(3,1){0}}
\put(142,64){\vector(-3,-4){0}}
\put(147,55){\vector(3,-2){0}}
\put(133,56){\vector(-3,-2){0}}
\put(33,5){\vector(-1,-4){0}}
\put(31,17){\vector(2,-1){0}}
\put(39,18){\vector(4,3){0}}
\put(45,31){\vector(0,1){0}}
\put(51,18){\vector(4,-3){0}}
\put(62,19){\vector(2,1){0}}
\put(56,6){\vector(1,-4){0}}
\put(102,22){\vector(3,1){0}}
\put(112,31){\vector(2,3){0}}
\put(110,18){\vector(1,-3){0}}
\put(107,6){\vector(-3,-4){0}}
\put(122,14){\vector(4,1){0}}
\put(133,19){\vector(2,1){0}}
\put(128,5){\vector(1,-4){0}}
\end{picture}
\end{equation}
\caption{The oriented pentagon equation\labl{5giraffe}}
\end{figure}
\be\labl{TwDu=DuTw}
Tw_a Du_a = Du_a Tw_a
\end{equation}
\begin{equation}\labl{braidwith1}
\begin{array}{rcl}
\triupioolpt CAB & @>Br_{AB}^{\pm1}>> & \triupioorpt CBA \\
del_A\searrow\nquad  && \nquad \swarrow del_A \\
& \stoyakd CB  &
\end{array}
\end{equation}
\begin{equation}\label{b2ttt}
Br^2_{BC} = Tw_A\ Tw_B^{-1} \ Tw_C^{-1} : \triupioo ABC \aow \triupioo ABC
\end{equation}
\begin{equation}\label{t=bb1}
\begin{array}{ccccccc}
\triupioo CAB  & \buildrel Br_{AB}\over\arow & \triupioo CBA &
\buildrel du_B\over\arow & \triupioi C{B\pti}A & \buildrel du_C\over\arow
& \triupooi{C\pti}{B\pti}A\\
Tw_A\bigg\uparrow \qquad &&&&&& \quad \bigg\downarrow Br_{A,C\pti}\nquad \\
\triupioo CAB &\buildrel rot\over\lfarow &\triupoio ABC
&\buildrel du_B\over\lfarow & \triupoii A{B\pti}C
& \buildrel du_C\over\lfarow & \triupooi A{B\pti}{C\pti}
\end{array}
\end{equation}
\begin{equation}\label{t=bb2}
\begin{array}{ccccccc}
\triupioo CAB & \buildrel Br_{AB}\over\arow & \triupioo CBA
& \buildrel du_A\over\arow & \triupiio CB{A\pti} & \buildrel du_C\over\arow
& \triupoio{C\pti}B{A\pti} \\
Tw_B \bigg\uparrow \qquad &&&&&& \quad \bigg\downarrow Br_{C\pti,B} \nquad \\
\triupioo CAB&\stackrel{rot^{-1}}\lfarow &\triupooi BCA
&\buildrel du_A\over\lfarow & \triupoii BC{A\pti}
&\buildrel du_C\over\lfarow & \triupoio B{C\pti}{A\pti}
\end{array}
\end{equation}
\begin{equation}\label{hexagon_or}
\begin{array}{ccccc}
\tarahoriooor XABC{} & \stackrel{Br_{BC}^{\pm1}}{\arrow} &
\tarahoriooor XACB{} & \stackrel{fus}{\arrow} & \taraverioood XACB{} \\
fus \bigg\downarrow \qquad &&&& \qquad \bigg\downarrow Br_{AC}^{\pm1} \\
\taraverioood XABCY & \stackrel{Br_{YC}^{\pm1}}{\arrow} &
\tarahoriooor XCABY & \stackrel{fus}{\arrow} & \taraverioood XCAB{}
\end{array}
\end{equation}
\begin{equation}\label{ST3=S2_or}
(S_M Tw_M)^3 =S^2_M : \tennisu {}M \aow \tennisu {}M ,
\end{equation}
\be\label{S2=Br-1Tw-1_or}
\begin{array}{ccc}
\tennisu XM  & \buildrel S_M^2\over\arrow & \tennisu XM \\
 _XBr_{MM}^{-1}\bigg\downarrow\qqquad & & \qquad \bigg\uparrow rev \\
\tennisd XM  & \buildrel Tw_M^{-1}\over\arrow & \tennisd XM
\end{array}
\end{equation}
and \eqref{mainSdiag_or} (see Figure~\ref{orientA_2}).
\begin{figure}[htbp]
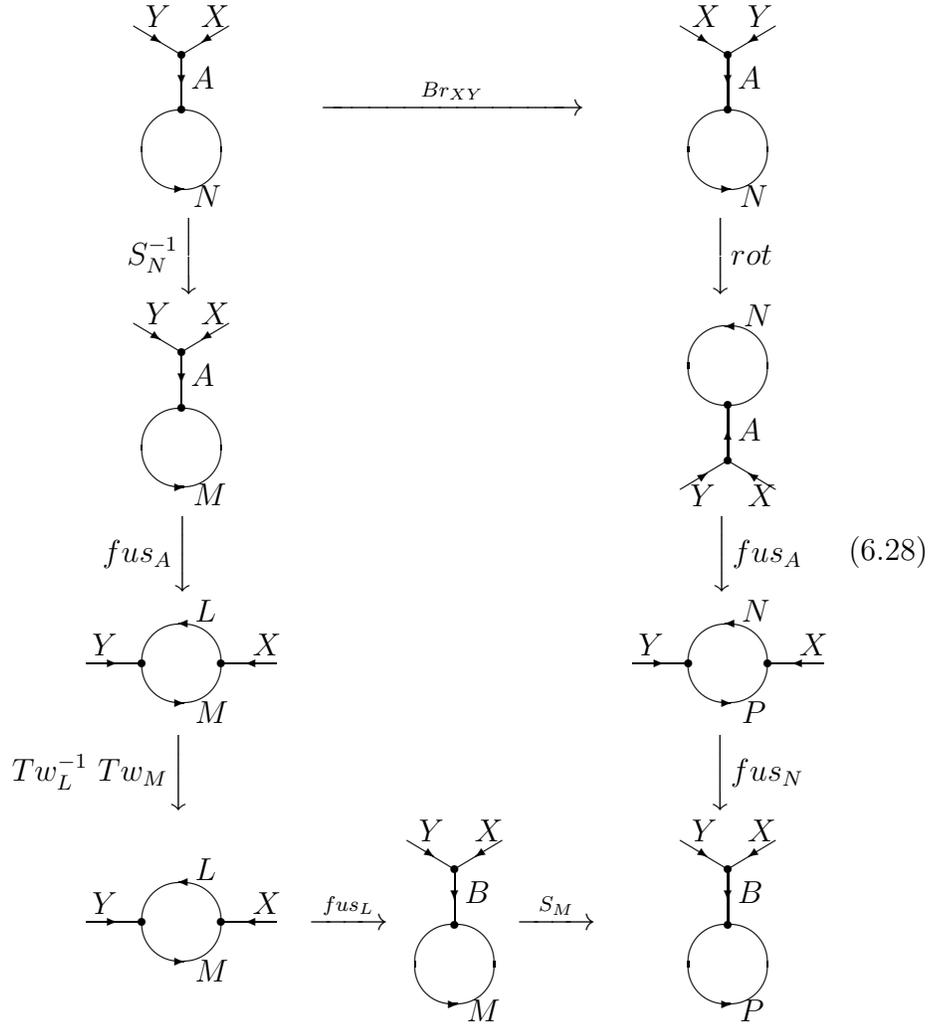

\be\label{mainSdiag_or}
\begin{array}{ccc}
\celodownor YXAN & \stackrel{Br_{XY}}{\arrrrow} & \celodownor XYAN \\
S_N^{-1}\bigg\downarrow\quad && \quad\bigg\downarrow rot \\
\celodownor YXAM &&
{\unitlength=0.75pt
\makebox[42 pt][l]{
\raisebox{-39 pt}[45 pt][39 pt]{
\put(24,76){\oval(40,40)[]}
\put(22,96){\vector(-1,0){0}}
\put(24,28){\line(0,1){28}}
\put(24,44){\vector(0,1){0}}
\put(24,28){\line(-5,-3){24}}
\put(14,22){\vector(3,2){0}}
\put(24,28){\line(5,-3){24}}
\put(34,22){\vector(-3,2){0}}
\put(24,28){\circle*{4}}
\put(24,56){\circle*{4}}
\put(5,5){$Y$}
\put(34,5){$X$}
\put(29,38){$A$}
\put(32,96){$N$}
}}}
\\
fus_A\bigg\downarrow\qquad && \qquad\bigg\downarrow fus_A \\
\sunor YXLM && \sunor YXNP \\
\nqquad Tw^{-1}_L\ Tw_M\bigg\downarrow\qqquad &&\qquad\bigg\downarrow fus_N\\
\sunor YXLM & \stackrel{fus_L}{\arow} \celodownor YXBM
\stackrel{S_M}{\arow} & \celodownor YXBP
\end{array}
\end{equation}
\caption{A relation for $A_2$ oriented net\label{orientA_2}}
\end{figure}
\end{defn}

\begin{thm} \label{desoONtoN}
The functor $deso:ON\to N$ is an equivalence.
\end{thm}

\sEction{Extensions of the category of nets}\label{Extensions} 
Let us look for central extensions of the category $RG$, or equivalently
$N$, or $ON$. Introduce a new type of morphisms---central charge.

\begin{defn}
Let $C_\Gamma:\Gamma\to\Gamma$ for each connected oriented net $\Gamma$ be a
morphism commuting with other generators
$du$, $rev$, $ins$, $del$, $fus$, $Tw$, $Br$, $S$. We assume that
\be\label{empty}
C_\O =\id_\O :\O \to\O .
\end{equation}
If $\Gamma_1$, $\Gamma_2$, $\Gamma$ are connected and
$g:\Gamma_1 \sqcup \Gamma_2 \to\Gamma$ is a glueing we assume a commutative
diagram
\be\label{centralC}
\begin{CD}
\Gamma_1\sqcup\Gamma_2 @>C^k_{\Gamma_1}\sqcup C^l_{\Gamma_2}>>
\Gamma_1\sqcup\Gamma_2 \\
@VgVV @VVgV \\
\Gamma @>C_\Gamma^{k+l}>> \Gamma
\end{CD}
\end{equation}
\end{defn}

The diagram \eqref{centralC} together with \eqref{empty} implies a similar
diagram with $\Gamma_1\sqcup \dots \sqcup\Gamma_n$.

We look for extensions of $ON$ by the generators $C_\Gamma$ with relations
between $du$, $rev$, $fus$, $Tw$, $Br$, $S$ changed by insertion of some
powers of $C_\Gamma$. We assume that the relations where $ins$, $del$ enter
are not changed. Then the relations involving $du$, $rev$, $fus$, $Tw$, $Br$
also do not change. Indeed, assume that $C^k$ is inserted into diagrams
\eqref{pentagon_or}--\eqref{hexagon_or} and such, and glue these diagrams
with as many $\pointedge{}$ as there are external legs. The generators
$fus$, $Tw$, $Br$ will turn into compositions of $ins$ and $del$, and
commutativity of such diagram would imply $k=0$. Also the diagram
\eqref{mainSdiag_or} after such operation become $S=SC^k$, hence, it also
does not change.

Therefore, only relations \eqref{ST3=S2_or} and \eqref{S2=Br-1Tw-1_or} could
change to
\be\label{(ST)3cen}
(S_M\,Tw_m)^3 = S_M^2 C^k ,
\end{equation}
\be\label{SM2}
S_M^2 = Br^{-1}_{MM}\,Tw^{-1}_M\,rev\,C^m .
\end{equation}
Renormalizing $S$ to $SC^{-k}$ we can restore \eqref{(ST)3cen}. For the sake
of symmetry we shall restore \eqref{SM2} changing \eqref{(ST)3cen} though it
might require half-integer powers of $C$.

\begin{defn}
Let $EN$ be a symmetric monoidal category with left cancellations which
objects are oriented nets and morphisms are generated over $\ON$ by the
isomorphisms $ins$, $del$, $fus$, $Tw$, $Br$, $S$, $C$ with all relations of
the category $ON$ except \eqref{ST3=S2_or} which is substituted by
\begin{equation}\label{ST3=CS2}
(S_M\,Tw_M)^3 = C\,S^2_M : \tennisu {}M \aow \tennisu {}M
\end{equation}
and relations \eqref{empty}, \eqref{centralC}.
\end{defn}

In topology to such extension corresponds the category of
framed \cite{Wit:Jones} surfaces.  Here we define a central extension
$E\Surf\to\Surf$ as a pull-back of the central extension $EN\to ON$ along
the functor $\Surf\to RG \to N \to ON$ obtained in Theorems~\ref{StoRGthm},
\ref{fatNtoRG}, \ref{desoONtoN}. Thus the latter
extends to a functor $E\Surf\to EN$.

\sEction{Ribbon categories}\label{intro} 
{\sl Ribbon} (also {\sl tortile} \cite{Shu}) category is the following
thing: a braided monoidal category $\CC$ \cite{JoyStr:tor} with the tensor
product $\tens$, the associativity $a:X\tens(Y\tens Z)\to (X\tens Y)\tens Z$,
the braiding (commutativity) $c:X\tens Y\to Y\tens X$ and a unity object $I$,
such that $\CC$ is rigid (for any object $X\in\CC$ there are dual objects
$\pti X$ and $X\pti$ with evaluations $\ev:\pti X\tens X\to I$,
$\ev:X\tens X\pti\to I$ and coevaluations $\coev:I\to X\tens\pti X$,
$\coev:I\to X\pti\tens X$) and possess a ribbon twist $\nu$. A ribbon twist
\cite{JoyStr:tor,Res:rib,Shu} $\nu=\nu_X:X\to X$ is a self-adjoint
($\nu_{X\pti}=\nu_X^t$) functorial automorphism such that
$c^2=\nu_X^{-1}\tens\nu_Y^{-1}\circ\nu_{X\tens Y}$.
In a ribbon category there are functorial isomorphisms \cite{Lyu:tan}
\[ 
\unitlength=0.8mm
\linethickness{0.4pt}
\begin{picture}(146.33,35)
\put(23,18){\oval(10,10)[r]}
\put(23,20){\oval(6,6)[lt]}
\put(11,35){\makebox(0,0)[cc]{$X$}}
\put(11,1){\makebox(0,0)[cc]{$X\pti\pti$}}
\put(1,18){\makebox(0,0)[cc]{$u_1^2\ =$}}
\put(61,18){\oval(10,10)[r]}
\put(61,20){\oval(6,6)[lt]}
\put(49,35){\makebox(0,0)[cc]{$X$}}
\put(49,1){\makebox(0,0)[cc]{$X\pti\pti$}}
\put(39,18){\makebox(0,0)[cc]{, $u_{-1}^2\ =$}}
\put(92,20){\oval(6,6)[rt]}
\put(106,35){\makebox(0,0)[cc]{$X$}}
\put(106,1){\makebox(0,0)[cc]{$\pti\pti X$}}
\put(76,18){\makebox(0,0)[cc]{, $u_1^{-2}=$}}
\put(132,20){\oval(6,6)[rt]}
\put(146,35){\makebox(0,0)[cc]{$X$}}
\put(146,1){\makebox(0,0)[cc]{$\pti\pti X$}}
\put(116,18){\makebox(0,0)[cc]{, $u_{-1}^{-2}=$}}
\put(20,20){\line(-2,-3){9.33}}
\put(11,30){\line(2,-3){6.67}}
\put(58,16){\line(-2,3){9.33}}
\put(49,6){\line(3,5){6}}
\put(106,6){\line(-4,5){8}}
\put(135,20){\line(4,-5){11.33}}
\put(146,30){\line(-4,-5){8}}
\put(23,16.50){\oval(6,7)[lb]}
\put(61.50,16){\oval(7,6)[lb]}
\put(92,18){\oval(12,10)[l]}
\put(132,18){\oval(12,10)[l]}
\put(91.50,16){\oval(7,6)[rb]}
\put(95,16){\line(4,5){11.20}}
\put(131.50,16){\oval(7,6)[rb]}
\end{picture}
\]
\[u_0^2=u_1^2\circ\nu^{-1}=u_{-1}^2\circ\nu:X\to X\pti\pti,\qquad
u_0^{-2}=u_1^{-2}\circ\nu^{-1}=u_{-1}^{-2}\circ\nu:X\to\pti\pti X.\]
Changing the category $\CC$ by an equivalent one, we can (and we will) assume
that ${}\pti X =X\pti$, $X\pti\pti = {}\pti\pti X = X$ and
$u_0^2 = u_0^{-2} = \id_X$ (see \cite{Lyu:tan}).

If in addition $\CC$ is additive, it is $k$-linear with $k=\End I$. We assume
in the following that $k$ is a field, in which each element has a square
root. Often $\CC$ will be noetherian abelian category with finite dimensional
$k$-vector spaces $\Hom_{\CC}(A,B)$. In such case there exists a coend
$F=\int X\tens X\pti$ as an object of a cocompletion $\hat{\CC}$
\cite{Lyu:tan} of $\CC$. It is a Hopf algebra
(see \cite{Lyu:mod,LyuMaj,Maj:bra}).
There is a Hopf pairing $\omega:F\tens F\to I$ \cite{Lyu:mod},
\[ 
\unitlength=1mm
\linethickness{0.4pt}
\begin{picture}(61,19)
\put(40,5){\line(0,1){4}}
\put(40,9){\line(4,5){7.33}}
\put(61,9){\line(0,1){9}}
\put(40,13){\line(-4,5){4}}
\put(28,19){\makebox(0,0)[cc]{$F$}}
\put(54,19){\makebox(0,0)[cc]{$F$}}
\put(5,10){\makebox(0,0)[cc]{$\omega\ =$}}
\put(32,9){\oval(24,18)[b]}
\put(45,10){\oval(32,20)[rb]}
\put(20,8){\line(0,1){10}}
\end{picture}
\]
such that
\[\Ker\omega = \Ker^{\text{left}}\omega =
\Ker^{\text{right}}\omega\in\hat{\CC}.\]

The quotient $\f=F/\Ker\omega\in\hat{\CC}$ is also a Hopf algebra and the
first modular axiom is \cite{Lyu:mod}

(M1) $\f$ is an object of $\CC$ (and not only of a cocompletion $\hat{\CC}$)

\noindent (more scrupulously, it means that there exists an exact sequence
$0\to\Ker\omega\to F\to \f \to 0$ in $\hat{\CC}$, where $\f$ is an
object from $\CC\subset\hat{\CC}$).

Being the coend $\int X\tens X\pti$, the object $F\in \hat{\CC}$ has an
automorphism $\und{\nu\tens1} \overset{\text{def}}= \int\nu\tens1$ (notations
are from \cite{Lyu:mod}). The second modular axiom says \cite{Lyu:mod}

(M2) $\und{\nu\tens1}(\Ker\omega)\subset\Ker\omega$

\noindent (more scrupulously, there exist morphisms
$T':\Ker\omega\to\Ker\omega\in\hat{\CC}$,
$T:\f\to\f\in\CC$ such that the diagram
\[
\begin{array}{ccccrcrcc}
0 & \to & \Ker\omega & \to & F & \to & \f & \to & 0\\
&& T'\bigg\downarrow && \und{\nu\tens1} \bigg\downarrow &&
T\bigg\downarrow &&\\
0 & \to & \Ker\omega & \to & F & \to & \f & \to & 0
\end{array}
\]
commutes).

\begin{defn}
A noetherian abelian ribbon category $\CC$ with finite dimensional $k$-vector
spaces of morphisms $\Hom_{\CC}(A,B)$ is called {\sl modular}, if axioms
(M1), (M2) are satisfied.
\end{defn}

It was shown in \cite{Lyu:mod} that in the case of a modular category there
exists a morphism $\mu:I\to\f$, which is the integral of a dual Hopf algebra
$\pti\f\simeq\f$, and
\[ 
\unitlength=0.8mm
\linethickness{0.4pt}
\begin{picture}(143,39)
\put(1,20){\makebox(0,0)[cc]{$\nu^{-1}$}}
\put(9,16){\framebox(4,8)[cc]{}}
\put(27,24){\oval(32,16)[lt]}
\put(30,5){\line(0,-1){4}}
\put(30,11){\line(0,1){28}}
\put(23,32){\line(0,1){5}}
\put(18,36){\makebox(0,0)[cc]{$\mu$}}
\put(35,37){\makebox(0,0)[cc]{$X$}}
\put(42,20){\makebox(0,0)[cc]{$=$}}
\put(49,20){\makebox(0,0)[cc]{$\lambda^{-1}$}}
\put(56,16){\framebox(4,8)[cc]{}}
\put(58,24){\line(0,1){15}}
\put(58,16){\line(0,-1){15}}
\put(63,37){\makebox(0,0)[cc]{$X$}}
\put(65,20){\makebox(0,0)[cc]{$\nu$ ,}}
\put(79,20){\makebox(0,0)[cc]{$\nu$}}
\put(87,16){\framebox(4,8)[cc]{}}
\put(105,24){\oval(32,16)[lt]}
\put(108,5){\line(0,-1){4}}
\put(108,11){\line(0,1){28}}
\put(101,32){\line(0,1){5}}
\put(96,36){\makebox(0,0)[cc]{$\mu$}}
\put(113,37){\makebox(0,0)[cc]{$X$}}
\put(120,20){\makebox(0,0)[cc]{$=$}}
\put(127,20){\makebox(0,0)[cc]{$\lambda$}}
\put(134,16){\framebox(4,8)[cc]{}}
\put(136,24){\line(0,1){15}}
\put(136,16){\line(0,-1){15}}
\put(141,37){\makebox(0,0)[cc]{$X$}}
\put(143,20){\makebox(0,0)[cc]{$\nu^{-1}$}}
\put(23,16){\oval(24,18)[b]}
\put(101,16){\oval(24,18)[b]}
\put(113,26){\line(0,-1){11}}
\put(35,26){\line(0,-1){11}}
\end{picture}
\]
for some invertible constant $\lambda\in k^{\times}$. The pair
$(\mu,\lambda)$ is unique up to a sign. Morphisms $S,S^{-1}:\f\to\f$
\[ \fourier \qqquad,\qqquad \invfourier \]
are inverse to each other. Morphisms $S$ and $T$ (defined via (M2)) yield a
projective representation of a mapping class group of a torus with one hole:
\[ (ST)^3=\lambda S^2, \qquad  S^2=\gamma^{-1},\]
\[ T\gamma=\gamma T, \qquad  \gamma^2=\nu. \]
Here $\gamma:\f\to \f$ is the antipode of the Hopf algebra $\f$,
\[ 
\unitlength=0.70mm
\linethickness{0.4pt}
\begin{picture}(38,40)
\put(11,2){\line(0,1){11}}
\put(11,13){\line(6,5){17}}
\put(28,27){\line(0,1){12}}
\put(32.50,14.50){\oval(11,9)[r]}
\put(32,19){\line(-1,-4){4.33}}
\put(28,13){\line(-6,5){7}}
\put(18,21){\line(-6,5){7}}
\put(11,26.67){\line(0,1){12.33}}
\put(20,40){\makebox(0,0)[cc]{$\f$}}
\put(20,1){\makebox(0,0)[cc]{$\f$}}
\put(1,20){\makebox(0,0)[cc]{$\gamma\ =$}}
\end{picture}
\]

\sEction{From a ribbon category to a modular functor}
\label{to a functor}
Assume that we are given a small ribbon category $(\CC,\tens,a,c,\nu,I)$ for
which the duality map $\cdot\pti:\Ob\CC\to \Ob\CC$ is involutive and
$u_0^2:A\to A\pti\pti=A$ equals $1_A$ for all objects $A\in\CC$. Further,
we assume that $\CC$ is a noetherian abelian category and $\End I=k$ is a
field, thus, $\CC$ is $k$-linear. We assume also that $k$-vector spaces
$\Hom(A,B)$ are finite dimensional for any $A,B\in \CC$. Finally, we assume
that algebra $\f$, quotient of $F$, belongs to $\CC$ and
$\Ker(F\to\f)=\Ker\omega$ is $\underline{\nu\tens1}$-invariant.

\subsEction{Preliminaries about exact functors} 
\begin{lem}\label{lemcoend}
Let $F:\CC\to k\Vect$, $G:\CC^{op}\to k\Vect$ be functors. Then
\[\int^X F(X)\tens \Hom(X,B)\to F(B),\qquad v\tens f\mapsto F(f).v\]
\[\int^X \Hom(B,X)\tens G(X)\to G(B),\qquad f\tens v\mapsto G(f).v\]
are isomorphisms of vector spaces.
\end{lem}

Let $F:(\CC^{k+n})^{op}\times\CC^{n+l}\to k\vect$ be a $k$-linear left exact
(preserving kernels) functor. By the ``parameter theorem for coends'' of
Mac Lane \cite{Mac:cat} a coend of the bifunctor
\[ F(A,-;-;B):(\CC^n)^{op}\times\CC^n\to k \text{-vect}\]
with fixed $A\in\CC^k$, $B\in\CC^l$ is identified as a functor in $A,B$ with
a coend of the bifunctor
\[F':(\CC^n)^{op}\times\CC^n\to
\text{Functors}((\CC^{op})^k\times\CC^l  \to k \text{-vect}).\]
Here functors are $k$-linear functors. But, generally, the coend of the
bifunctor
\[F'':(\CC^n)^{op}\times\CC^n\to
\text{Left Exact Functors}((\CC^{op})^k\times\CC^l  \to k \text{-vect}).\]
is a different thing. In any case we have a morphism
\[ \int^X F'(X,X)\to \int^X F''(X,X) .\]
To stress the difference, we introduce different notations for the usual
coend
\[ \int^X F(A,X;X,B) \equiv \Bigl(\int^X F'(X,X) \Bigr)(A,B) \]
and for the second type of coend, the left exact functor
\[\oint^X F(A,X;X,B) \buildrel {\text{def}}\over =
  \bigl( \int^X F''(X,X)\bigr)(A,B).\]

The Fubini theorem for left exact functors (repeated coend theorem) is
proved exactly in the same way as for arbitrary functors (see Mac Lane's
book \cite{Mac:cat}). Nevertheless, we present a proof here
for the sake of completeness.

\begin{thm}\label{LEcoend}
Let $\CC_1,\CC_2,\CC_3,\CA$ be abelian $k$-linear
categories and let $F:\CC_2^{op}\times\CC_1^{op}\times\CC_1\times\CC_2
\times\CC_3\to \CA$ be left exact $k$-linear functor. Assume that there
exists coend with parameters
\[j_X(U;V,A):F(U,X;X,V,A)\to G(U;V,A)=\int^{X\in\CC_1} F(U,X;X,V,A).\]
Assume that $G$, viewed as a functor $G:\CC_2^{op}\times\CC_2
\to \text{ Functors } (\CC_3\to\CA)$, has a left exact coend
$H\in \text{ LEF }(\CC_3\to\CA)$ (set of Left Exact Functors)
\[k_U(A):G(U;U,A)\to H(A)=\int^U G(U;U,A) \]
(that means that $k$ is dinatural, $H$ is left exact and the pair $(k,H)$ is
universal between all dinatural transformations of $G$ into left exact
functors). Then
\[i_{X,U}(A):F(U,X;X,U,A) \buildrel j_X(U;U,A)\over\arrow
   G(U;U,A)\buildrel k_U(A)\over\arow H(A) \]
is a coend $\oint^{X,U} F(U,X;X,U,A)$ of $F$, viewed as a bifunctor
\[(\CC_1\times\CC_2)^{op}\times(\CC_1\times\CC_2)\to
  \text{ LEF } (\CC_3\to \CA).\]
\end{thm}

We write the statement of the theorem as an isomorphism
\[\oint^{X,U} F(U,X;X,U,A)\simeq \oint^U \! \int^X F(U,X;X,U,A). \]
In particular case $\CC_2=0$ we have

\begin{cor}
Let $G(A)=\int^{X\in\CC_1} F(X;X,A)$ for left exact $k$-linear functor
$F:\CC_1^{op}\times\CC_1\times\CC_3\to \CA$. Then
\[\oint^X F(X;X,A)=\oint^0 G(A) \]
if there exists a left exact functor $\oint^0 G:\CC_3\to\CA$ with
isomorphisms $Nat(G,E)\cong Nat(\oint^0 G,E)$ for any left exact
$E:\CC_3\to\CA$.
\end{cor}

\subsEction{Modular functor}\label{Modular} 
Our goal is to construct a functor $Z:EN\to k$-vect, which will
satisfy to the following conditions:

(i) $Z$ is a symmetric monoidal functor.

(ii) To $D_2$ corresponds
\[Z(D_2;B;C)=\Hom(B,C),\]
to glueing of such corresponds the composition
\[\Hom(A,B)\tens\Hom(B,C)\to\Hom(A,C).\]

(iii) To $D_3$ correspond
\[Z\left(\tridowniio CBA \right) =\Hom(A\tens B,C),\]
\[Z\left(\triupioo ABC \right)=\Hom(A,B\tens C).\]
To glueing of $D_3$ with $D_2$ corresponds an action of $\Hom(-,-)$ on
$\Hom(-\tens -,-)$ or $\Hom(-,-\tens -)$ via composition.

(iv) To $D_1$ correspond
\[Z(D_1;X;\ )=\Hom(X,I),\]
\[Z(D_1;\ ;X)=\Hom(I,X),\]
where $I$ is unity object of $\CC$. To glueing of $D_1$ with $D_2$
correspond compositions
\[\Hom(X,Y)\tens\Hom(Y,I)\to\Hom(X,I),\]
\[\Hom(I,X)\tens\Hom(X,Y)\to\Hom(I,Y).\]

(v) To isomorphisms
\[du_B: \tridowniio CBA \arow \tridownioo C{B\pti}A ,\]
\[du_A: \tridowniio CBA \arow \tridownoio CB{A\pti} \]
correspond adjunctions
\[d_l:\Hom(A\tens B,C)\to \Hom(A,C\tens B\pti),\]
\[d_r:\Hom(A\tens B,C)\to \Hom(B,A\pti\tens C).\]

(vi) For any oriented net $\Gamma$ glueings with $D_2$ makes $Z(\Gamma)$ into
a functor
\[Z(\Gamma):(\CC^{op})^k\times\CC^l\to k{\rm -vect}\]
We assume that $Z(\Gamma)$ is left exact (additive and preserving kernels).

(vii) If $f:\Gamma\to\tilde\Gamma$ is a glueing and $g(\Gamma)=g(\tilde
 \Gamma)$, then the morphism of functors
\[\int^{X\in\CC^k} Z(\Gamma,\dots,X;X,\dots)\to Z(\tilde\Gamma;\dots;
   \dots)\]
is an isomorphism.

(viii) To the morphism
\[ \tarahoriooor XABC{} \stackrel{fus}{\arrow} \taraverioood XABCV \]
corresponds
\begin{multline*}
\Hom(X,A\tens(B\tens C))\simeq \int^{U\in\CC} \Hom(X,A\tens U)\tens
  \Hom(U,B\tens C) \simeq \\
\simeq Z\left(\tarahoriooor XABC{} \right) \buildrel Z(fus)\over\arrow
  Z\left( \taraverioood XABCV \right) \simeq \\
\simeq \int^{V\in\CC} \Hom(X,V\tens C)\tens\Hom(V,A\tens B)\simeq
  \Hom(X,(A\tens B)\tens C),
\end{multline*}
which coincides with $\Hom(X,a_{A,B,C})$.

(ix) To the morphism
\[ \triupioo XAB \stackrel{Br_{AB}}{\arow} \triupioo XBA \]
corresponds
\[\Hom(X,c_{A,B}):\Hom(X,A\tens B)\to\Hom(X,B\tens A).\]

(x) To the morphism
\[\lezhakright XY \buildrel Tw_Y\over\arow \lezhakright XY \]
corresponds
\[\Hom(X,\nu_Y):\Hom(X,Y)\to\Hom(X,Y).\]

(xi) To a tadpole $A_1$, obtained via glueing from $D_3$, corresponds
\begin{multline*}
Z\left(
{\unitlength=0.75pt
\makebox[39 pt][l]{
\raisebox{3 pt}[19 pt][13 pt]{
\put(22,0){\line(-1,0){23}}
\put(13,0){\vector(1,0){0}}
\put(22,0){\line(3,-5){12}}
\put(28,-10){\vector(2,-3){0}}
\put(22,0){\line(3,5){12}}
\put(28,10){\vector(-2,-3){0}}
\put(22,0){\circle*{4}}
\put(2,4){$X$}
\put(35,-17){$M$}
\put(35,10){$M$}
}}}
\ \right)=\Hom(X\tens M,M)\simeq \Hom(X,M\tens M\pti) \to \\
\to \Hom(X,\f)=Z\left(\tennisu XM \right).
\end{multline*}

 $Z$ can be viewed as a monoidal functor
\[\{ \text{extended nets} \}\to\hat{\CC}_{*,*}.\]

Now we shall construct such a functor step by step, proving the following

\begin{thm}
There exists one and only one up to equivalence functor
$Z:EN\to k\vect$, satisfying assumptions (i)-(xi) above. It has also
property:

(xii) Let $f:\Gamma\to\tilde\Gamma$ be a glueing and let the boundary of each
connected component of $\tilde\Gamma$ not be empty. The functor
\[Z(\Gamma;\dots,X;X,\dots):
(\CC^{op})^{k+n}\times\CC^{n+l}  \to k{\rm -vect}\]
can be represented as a bifunctor
\[Z'(\Gamma)(X,X):(\CC^n)^{op}\times(\CC^n)\to \hat{\CC}_{k,l},\]
where $\hat{\CC}_{k,l}$ is the category of left exact functors
$(\CC^{op})^k\times\CC^l \to k{\rm -Vect}$.
The coend of this bifunctor is mapped to $Z(\tilde\Gamma)\in\hat{\CC}_{k,l}$
\[\int^{X\in\CC^k} Z'(\Gamma)(X,X)\to Z(\tilde\Gamma).\]
We claim that this is an epimorphism in $\hat{\CC}_{k,l}$.
\end{thm}

Proof is the matter of the remaining part of this chapter.

\subsEction{A functor on the category of oriented nets}\labl{functorCZ}
Let $\ON_>$ be a symmetric monoidal category of glueings of oriented nets,
having at least one end at each connected component. Let us construct a
functor $\CZ:\ON_> \to k\vect$.
Fix the value of the functor $\CZ$ on elementary objects---$D_1,D_2,D_3$
as in (ii)--(iv). So it is fixed on disjoint unions $\bigsqcup_i X_i$
of such objects: $\CZ(\bigsqcup X_i)=\tens_i \CZ(X_i)$. (In fact, it is fixed
only up to a permutation of tensor multiplicands. To fix it completely, we
could define nets with chosen total ordering of the set
$V_3\sqcup V_2\sqcup V_1$, and add to $\ON$ new morphisms, which
change only that ordering. Obtained category is equivalent to the old one.
We shall not remind about such tricks in the following.)

The following condition determines the value of $\CZ$ on $\ON_>$:

(xiii) Let $f:\Gamma\to\tilde\Gamma$ be a glueing, and let the boundary of
each connected component of $\tilde\Gamma$ be not empty, then there exists a
coend $\oint \CZ(\Gamma)$ and the morphism
\[\oint^{X\in\CC^k} \CZ(\Gamma;\dots,X;X,\dots)\to
\CZ(\tilde\Gamma;\dots;\dots)\]
is an isomorphism of left exact functors.

Let $\Gamma \in\ON_>$ be a connected net of genus $g$ with $k$ incoming and
$l$ outgoing legs, $k+l>0$. We construct a space $\CZ(\Gamma)$ in the
following way. Let $\Gamma_1$ be the net obtained from $\Gamma$ by cutting
all internal edges. There is a canonical glueing $f:\Gamma_1\to\Gamma$.
We show the existence of the coend
\[\oint^{X\in\CC^n} \CZ(\Gamma_1;A,X;X,B) \in \hat{\CC}_{k,l} \]
and define $\CZ(\Gamma)$ to be that functor.

Cut $\Gamma$ at such edges $w_1,\dots,w_g$ that the obtained net
$\Gamma_2$ is a tree. Introduce ``internal variables''
$W_1,\dots,W_g\in\CC$ corresponding to $w_i$, $1\le i \le g$, and let
$Y_1,\dots,Y_{n-g}\in\CC$ be other ``internal variables'' corresponding
to other edges. Cutting along them breaks $\Gamma_2$ into pieces,
forming $\Gamma_1$. We have the whole collection of variables
$\{X_1,\dots,X_n\}=\{W_1,\dots,W_g,Y_1,\dots,Y_{n-g}\}$. We apply
\thmref{LEcoend}. The coend\linebreak[4]
 $\int^{Y\in\CC^{n-g}}\CZ(\Gamma_1;A,W,Y;Y,V,B)$ exists and gives the
functor $\CZ(\Gamma_2;A,W;V,B)$. It is isomorphic to a functor
\[ \Hom(A_1\tens\dots\tens A_k,B_1\tens\dots\tens B_l\tens
(V_1\tens W_1\pti)\tens\dots\tens (V_g\tens W_g\pti))\]
with some parenthesis for tensor product. This is proved by induction on
the number of vertices of the tree. Inductive step use Lemma~\ref{lemcoend}
similarly to calculation
\begin{multline*}
\int^X\Hom(A,X\tens B)\tens\Hom((C\tens X)\tens D),E)
\buildrel d_B\over\simeq \\
\simeq \int^X\Hom(A\tens B\pti,X)\tens\Hom((C\tens X)\tens D),E)
\simeq\Hom((C\tens(A\tens B\pti))\tens D,E).
\end{multline*}

By Theorem~\ref{LEcoend}
\[\oint^{X\in\CC^n} \CZ(\Gamma_1;A,X;X,B)\simeq
   \oint^{W\in\CC^g} \CZ(\Gamma_2;A,W;W;B) \simeq \]
\[\simeq\oint^{W\in\CC^g} \Hom(A_1\tens\dots\tens A_k,B_1\tens\dots\tens
   B_l\tens(W_1\tens W_1\pti)\tens\dots\tens (W_g\tens W_g\pti)) \]
if the latter exists. We show that, indeed, it exists and equals
\[\Hom(A_1\tens\dots\tens A_k,B_1\tens\dots\tens B_l\tens
   (\int^{W_1\in\CC} W_1\tens W_1\pti) \tens\dots\tens
   (\int^{W_g\in\CC} W_g\tens W_g\pti)) \]
for $k+l>0$.

Using duality isomorphisms, we can assume that $k=1$. Assume that we are
given functorial morphisms
\begin{multline*}
i_{W_1\dots W_g}(A,B):\Hom(A,B_1\tens\dots\tens B_l\tens C\tens
   (W_1\tens W_1\pti)\tens\dots\tens (W_g\tens\pti W_g))\to \\
\to G(A;B_1,\dots,B_l)
\end{multline*}
which define dinatural transformation. Here $C$ is an object of
$\hat\CC$ and $G:\CC^{op}\times\CC^l\to k$-Vect is a left exact functor.
We prove by induction on $g$ that they all factorize through the unique
morphism
\[\Hom(A,B_1\tens\dots\tens B_l\tens C\tens(\int^{W_1} W_1\tens W_1\pti)
  \tens\dots\tens (\int^{W_g} W_g\tens W_g\pti))\to G(A;B). \]

If $g\ge 1$ denote by $D$ the product $B_1\tens\dots\tens B_l\tens C
\tens (W_1\tens W_1\pti)\tens\dots\tens (W_{g-1}\tens W_{g-1}\pti)$ and
denote by $G_1(A)$ the left exact functor $G(A;B_1,\dots,B_l)$ with fixed
$B_1,\dots,B_l$. There exists tautologically $T\in\hat{\CC}$, such that
$G_1(A)=\Hom(A,T)$. Dinaturality implies commutativity of the diagram
 \[
\begin{CD}
\Hom(A,D\tens(V_g\tens W_g\pti)) @>\Hom(A,D\tens f\tens W_g\pti)>>
     \Hom(A,D\tens(W_g\tens W_g\pti)) \\
@V{\Hom(A,D\tens(V_g\tens f^t))}VV @VVi_{\dots,W_g}V \\
\Hom(A,D\tens(V_g\tens V_g\pti)) @>i_{\dots,V_g}>>  \Hom(A,T)
\end{CD}
\]
Hence, the following diagram in $\hat{\CC}$ commutes for any
$f:V_g\to W_g$:
\[
\begin{CD}
D\tens(V_g\tens W_g\pti) @>D\tens f\tens W_g\pti>> D\tens(W_g\tens W_g\pti)\\
@V{D\tens (V_g\tens f^t)}VV   @VVV \\
D\tens (V_g\tens V_g\pti) @>>>  T
\end{CD}
\]
Consequently, there is a morphism in $\hat{\CC}$
\[D\tens\int^W W\tens W\pti\cong \int^W D\tens (W\tens W\pti)\to T.\]
The first isomorphism here follows from the fact that $\tens$ preserves
colimits in $\hat{\CC}$. Thus, for any collection of objects
$B_1,\dots,B_l,W_1,\dots,W_{g-1}$ we obtained morphisms
\begin{multline*}
j_{B_1,\dots,B_l,W_1,\dots,W_{g-1}}(A):\Hom(A,B_1\tens\dots\tens B_l
  \tens C\tens (W_1\tens W_1\pti)\tens\dots\tens \\
\tens (W_{g-1}\tens W_{g-1}\pti)\tens (\int^{W_g} W_g\tens W_g\pti))
 \to G(A;B_1,\dots,B_l)
\end{multline*}
functorial in $A$. Morphisms $i_W(A;B)$ are factorized through $j_{B,W}(A)$
and the latter are characterized by that property.

One can prove that $j_W(A;B)\equiv j_{B,W}(A)$ is functorial in $A$
and $B_i$ and that $j_W$ is dinatural in $W$.

Constant tensor multiple $F=\int^{W_p} W_p\tens W_p\pti$ can be adjoint
to $C$. We finish computation of $\oint \CZ(\Gamma_1)$ by induction.

Thus, we defined a functor $\CZ:\ON \to k$-Vect.
\[\CZ(\Gamma)\buildrel {\text{def}}\over = \oint \CZ(\Gamma_1) \simeq
  \Hom(A_1\tens\dots\tens A_k,B_1\tens\dots\tens B_l \tens F
  \tens\dots\tens F).\]
This functor satisfies (i)--(iv), (vi), (vii).

Now we prove the property (xiii) for $\CZ$. We assume that $\tilde\Gamma$ is
connected with non-empty boundary. Any glueing $f:\Gamma\to\tilde\Gamma$
can be factorized into $\Gamma\buildrel f_1\over\to\Gamma'
\buildrel f_2\over\to\tilde\Gamma$, where genus of $\Gamma'$ equals genus of
$\Gamma$ and $\Gamma'$ is  connected. Let variables $X_i$ correspond to
circles, glued by $f_1$, and let $Y_j$ correspond to ones, glued by $f_2$.
Considered morphism $\CZ(f)$ factorizes as
\[\oint \CZ(\Gamma)\simeq\oint^Y \! \int^X \CZ(\Gamma)
\simeq\oint^Y \CZ(\Gamma')\to \CZ(\Gamma) \]
by Theorem~\ref{LEcoend} and (vii). This is an isomorphism, because
\[\oint^Y\Hom(A_1\tens\dots\tens A_k,B_1\tens\dots\tens B_l
   \tens(Y_1\tens Y_1\pti)\tens\dots\tens (Y_p\tens Y_p\pti)
   \tens F\tens\dots\tens F)\to \]
\[\to\Hom(A_1\tens\dots\tens A_k,B_1\tens\dots\tens B_l\tens F
   \tens\dots\tens F\tens\dots\tens F)\]
is an isomorphism.

\subsEction{Relations for insertions and reversals} 
We represent morphisms $X=ins,del:\Gamma\to\Gamma'$ from \secref{oriented}
by the following procedure. Let $\Gamma=\Gamma_1\cup\sigma$,
$\Gamma'=\Gamma_1\cup\sigma'$, where $X=ins$, $del:\sigma\to\sigma'$ is a
standard morphism, and $X\vert_{\Gamma_1}=\id$. Then we define $\CZ(X)$ as
\[\CZ(\Gamma)\simeq \oint \CZ(\Gamma_1)\tens \CZ(\sigma)
  \buildrel \oint \id\tens \CZ(X)\over\arrow
  \oint \CZ(\Gamma_1)\tens \CZ(\sigma')\simeq \CZ(\Gamma'),\]
an isomorphism of coends, induced by isomorphism of underlying bifunctors.

Now we construct a functor $\CZ$ on $\ON_>$ extended by $ins$ and $del$
satisfying the axioms. To deletion or insertion of a 2-vertex correspond
isomorphisms of Lemma~\ref{lemcoend}. To deletion or insertion of a 1-vertex
correspond isomorphisms which are glueings of identity with
\[\CZ\left(\torchokdr XY \aow \lezhakright XY \right)=\Hom(X\tens I,Y)
   \buildrel \Hom(r_X^{-1},Y)\over\arrrow \Hom(X,Y)\]
\[\CZ\left(
{\unitlength=0.75pt
\makebox[36 pt][l]{
\raisebox{-4 pt}[15 pt][7.5 pt]{
\put(0,0){\line(1,0){40}}
\put(8,0){\vector(-1,0){0}}
\put(28,0){\vector(-1,0){0}}
\put(20,0){\line(0,1){20}}
\put(20,8){\vector(0,-1){0}}
\put(20,20){\circle*{4}}
\put(20,0){\circle*{4}}
\put(3,4){$Y$}
\put(28,4){$X$}
}}}
\aow \lezhakleft YX \right)=\Hom(I\tens X,Y)
   \buildrel \Hom(l_X^{-1},X)\over\arrrow \Hom(X,Y)\]
\[\CZ\left(\torchokur XY \aow \lezhakright XY \right)=\Hom(X,Y\tens I)
   \buildrel \Hom(X,r_Y)\over\arrrow \Hom(X,Y)\]
\[\CZ\left(\torchokul YX \aow \lezhakleft YX \right)=\Hom(X,I\tens Y)
   \buildrel \Hom(X,l_Y)\over\arrrow \Hom(X,Y)\]

The orientation reversing morphism $du$ is realized on external legs as a
glueing of identity with duality adjunctions
\[\CZ(du_A):\CZ\left(\triupioo XAB \right)=\Hom(X,A\tens B)
  \buildrel d_r^{-1}\over\aow \Hom(A\pti\tens X,B)=
   \CZ\left(\triupioi X{A\pti}B \right),\]
\[\CZ(du_B):\CZ\left(\triupioo XAB \right)=\Hom(X,A\tens B)
  \buildrel d_l^{-1}\over\aow \Hom(X\tens B\pti,A)=
   \CZ\left(\triupiio XA{B\pti} \right),\]
\[\CZ(du_X):\CZ\left(\tridowniio AYX \right)=\Hom(X\tens Y,A)
  \buildrel d_r\over\aow \Hom(Y,X\pti\tens A)=
   \CZ\left(\tridownoio AY{X\pti} \right),\]
\[\CZ(du_Y):\CZ\left(\tridowniio AYX \right)=\Hom(X\tens Y,A)
  \buildrel d_l\over\aow \Hom(X,A\tens Y\pti)=
   \CZ\left(\tridownioo A{Y\pti}X \right).\]
The orientation reversing morphism $rev$ for internal arrow is obtained
from the diagram of isomorphisms:
\begin{gather*}
\int^X \CZ\left(\nquad\trirightr X{}{} \right)\tens
\CZ\left(\trileftr X{}{} \nquad \right) \buildrel du_X\tens du_X\over\arrow
  \int ^{X\in\CC} \CZ\left(\nquad \trirightl{X\pti}{}{} \right)\tens
  \CZ\left(\trileftl{X\pti}{}{} \nquad \right) \\
\begin{array}{ccc}
\wr\Big\vert && \wr\Big\vert \\
\CZ\left(\tarahorr {}{}{}{}{} \right) & \buildrel rev\over\aow
  \CZ\left(\tarahorl {}{}{}{}{} \right) \overset f\simeq &
  \int^{Y\in\CC} \CZ\left(\nquad \trirightl Y{}{} \right)\tens
  \CZ\left(\trileftl Y{}{} \nquad \right)
\end{array}
\end{gather*}
where the isomorphism $f$ is that from Lemma~\ref{lemcoend}. This shows that
the relation~\eqref{gluerev} is satisfied. Similarly for \eqref{1ptdudurev},
\eqref{2ptdudurev}.

The existence of $\CZ(rev): \CZ(\Gamma) \to \CZ(\Gamma')$ for arbitrary net
$\Gamma$ is guaranteed by

\begin{prop}\label{proptipti}
Let $B:\CC^{op}\times\CC\to k\vect$ be a bifunctor. Then
$B_1(X,Y)=B(Y\pti,X\pti)$ is also a bifunctor. Let
\[B(X,X) \buildrel i_X\over\arow \int^X B(X,X), \qquad
B_1(Y,Y)  \buildrel j_Y\over\arow \int^Y B_1(Y,Y)\]
be their coends. Then there exists the unique isomorphism $\alpha$ of coends,
which make the diagram
\[
\begin{CD}
B(X\pti,X\pti) @>i_{X\pti}>> \int^X B(X,X) \\
@|  @VV\alpha V \\
B_1(X,X)  @>j_X>>  \int^Y B_1(Y,Y)
\end{CD}
\]
commute for any $X\in\CC$.
\end{prop}

\begin{prop}
The identity~\eqref{dddddd} is satisfied, that is, the diagram
\begin{equation*}\labl{6dHom}
\begin{array}{ccccc}
&& \Hom(C,A\tens B) && \\
& d_r \nearrow & & \searrow d_l^{-1} & \\
\Hom(A\pti\tens C,B) &&&& \Hom(C\tens B\pti,A) \\
d_l^{-1}\big\uparrow &&&& \big\downarrow d_r \\
\Hom(A\pti,B\tens C\pti) &&&& \Hom(B\pti,C\pti\tens A) \\
& d_r \nwarrow && \swarrow d_l^{-1} & \\
&& \Hom(B\pti\tens A\pti,C\pti)
\end{array}
\end{equation*}
is commutative.
\end{prop}

The obtained functor satisfies (i)-(vii) and (xiii).

\subsEction{Nets without ends} 
We define also $\CZ(\Gamma)$ for a connected net without ends via insertion
of a 1-vertex
\[\CZ(\Gamma)\buildrel \CZ(Ins)\over\arrow \CZ(\Gamma_\bullet) \simeq
\CZ(\Gamma_I) .\]
Previous subsection shows that different choices give isomorphic answers.
The property (xiii) is still true.

\subsEction{Relations for fusing} 
Now we construct a functor $\CZ$ on the category $\ON$ extended by $ins$,
$del$, $fus$. We put
\begin{multline*}
\CZ(fus):\CZ\left(\tarahoriooor MABC{} \right)\simeq \int^{X\in \CC}
  \Hom(M,A\tens X)\tens\Hom(X,B\tens C) \simeq \\
\simeq\Hom(M,A\tens(B\tens C))
\buildrel \Hom(M,a_{A,B,C})\over\arrrow \Hom(M,(A\tens B)\tens C) \simeq \\
\simeq\int^{Y\in\CC} \Hom(M,Y\tens C)\tens\Hom(Y,A\tens B)
  \simeq \CZ\left(\taraverioood MABC{} \right).
\end{multline*}
The fusing pentagon~\eqref{pentagon_or} follows from the associativity
pentagon in $\CC$. We have to prove commutativity of
diagrams~\eqref{fus1?}, \eqref{fus2?}.

\begin{prop}
The relations~\eqref{fus1?}, \eqref{fus2?} are satisfied, that is, the
diagrams
\begin{equation*}\labl{realfus1?}
\begin{CD}
\Hom(A,B\tens(C\tens D))
\text{\makebox[0mm][l]{\put(13,0){$\stackrel{\Hom(A,a_{B,C,D})}\arrow$}}}
@. \Hom(A,(B\tens C)\tens D) \\
@V\wr VV  @VV\wr V \\
\int^X\Hom(A,B\tens X)\tens\Hom(X,C\tens D) @.
   \int^X\Hom(A,X\tens D)\tens\Hom(X,B\tens C) \\
@V\int 1\tens d_r^{-1}VV  @VV\int 1\tens d_l^{-1}V \\
\int^X\Hom(A,B\tens X)\tens \Hom(C\pti\tens X,D) @.
   \int^X\Hom(A,X\tens D)\tens\Hom(X\tens C\pti,B) \\
@V\int d_l^{-1}\tens d_lVV  @VV\int d_r^{-1}\tens d_rV \\
\int^X\Hom(A\tens X\pti,B)\tens\Hom(C\pti,D\tens X\pti) @.
   \int^X\Hom(X\pti\tens A,D)\tens\Hom(C\pti,X\pti\tens B) \\
@V\wr VV  @VV\wr V \\
\int^Y\Hom(A\tens Y,B)\tens\Hom(C\pti,D\tens Y) @.
   \int^Y\Hom(Y\tens A,D)\Hom(C\pti,Y\tens B) \\
@V\int d_r\tens1VV  @VV\int d_l\tens1V \\
\int^Y\Hom(Y,A\pti\tens B)\tens\Hom(C\pti,D\tens Y) @.
   \int^Y\Hom(Y,D\tens A\pti)\tens\Hom(C\pti,Y\tens B) \\
@V\wr VV  @VV\wr V \\
\Hom(C\pti,D\tens(A\pti\tens B))
\text{\makebox[0mm][l]{\put(6,0)
{$\stackrel{\Hom(C\pti,a_{D,A\pti,B})}\arrow$}}}
@. \Hom(C\pti,(D\tens A\pti)\tens B)
\end{CD}
\end{equation*}
\begin{equation*}\labl{realfus2?}
\begin{CD}
\Hom(D,A\tens(B\tens C))
\text{\makebox[0mm][l]{\put(11,0){$\stackrel{\Hom(D,a_{A,B,C,})}\arrow$}}}
@. \Hom(D,(A\tens B)\tens C) \\
@A\wr AA  @VV\wr V \\
\int^X\Hom(D,A\tens X)\tens\Hom(X,B\tens C) @.
   \int^X\Hom(D,X\tens C)\tens\Hom(X,A\tens B) \\
@A\int d_r\tens1AA  @VV\int 1\tens d_r^{-1}V \\
\int^X\Hom(A\pti\tens D,X)\tens \Hom(X,B\tens C) @.
   \int^X\Hom(D,X\tens C)\tens\Hom(A\pti\tens X,B) \\
@A\int d_l^{-1}\tens1AA  @VV\int d_r^{-1}\tens d_lV \\
\int^X\Hom(A\pti,X\tens D\pti)\tens\Hom(X,B\tens C) @.
 \ \int^X\Hom(X\pti\tens D,C)\tens\Hom(A\pti,B\tens X\pti) \\
@A\wr AA  @VV\wr V \\
\Hom(A\pti,(B\tens C)\tens D\pti) @.
   \int^Y\Hom(Y\tens D,C)\tens \Hom(A\pti,B\tens Y) \\
@A\Hom({A\pti},a_{B,C,D\pti})AA  @VV\int d_l\tens1V \\
\Hom(A\pti,B\tens (C\tens D\pti))
\text{\makebox[0mm][l]{\put(5,0){$\stackrel{\sim}\lfarow$}}}
@. \int^Y\Hom(Y,C\tens D\pti)\tens\Hom(A\pti,B\tens Y)
\end{CD}
\end{equation*}
are commutative.
\end{prop}

\subsEction{Relations for braiding and twists} 
We extend the functor $\CZ$ to the category generated over $\ON$ by $ins$,
$del$, $fus$, $Tw$ and $Br$. We prove the relations involving $Tw$ and $Br$.
All conditions with $Tw$ are obvious or follow from equations
\[ 
\unitlength=0.70mm
\linethickness{0.4pt}
\begin{picture}(135,24)
\put(5,8){\framebox(4,8)[cc]{}}
\put(1,12){\makebox(0,0)[cc]{$\nu$}}
\put(7,8){\line(0,-1){8}}
\put(14,16){\oval(14,16)[t]}
\put(21,16){\line(0,-1){16}}
\put(30,12){\makebox(0,0)[cc]{=}}
\put(53,8){\line(0,-1){8}}
\put(51,8){\framebox(4,8)[cc]{}}
\put(59,12){\makebox(0,0)[cc]{$\nu$\ ,}}
\put(46,16){\oval(14,16)[t]}
\put(39,16){\line(0,-1){16}}
\put(80,8){\framebox(4,8)[cc]{}}
\put(76,12){\makebox(0,0)[cc]{$\nu$}}
\put(82,16){\line(0,1){8}}
\put(89,8){\oval(14,16)[b]}
\put(96,8){\line(0,1){16}}
\put(105,12){\makebox(0,0)[cc]{=}}
\put(126,8){\framebox(4,8)[cc]{}}
\put(135,12){\makebox(0,0)[cc]{$\nu$ .}}
\put(128,16){\line(0,1){8}}
\put(121,8){\oval(14,16)[b]}
\put(114,8){\line(0,1){16}}
\end{picture}
\]

The hexagon~\eqref{hexagon_or} for braiding
\[\Hom(C,A\tens B) \buildrel \Hom(C,c_{AB})\over\arrrow \Hom(C,B\tens A)\]
follows from that one for commutativity $c$. Property~\eqref{b2ttt} follows
from the equation
\[\Hom(C,c_{AB}^2)=\Hom(C,\nu_{A\tens B}\circ\nu_A^{-1}\tens\nu_B^{-1})=
\Hom(\nu_C,\nu_A^{-1}\tens\nu_B^{-1}).\]

\begin{prop}
The relations~\eqref{t=bb1}, \eqref{t=bb2} are satisfied. That is, the
diagrams
\[
\begin{array}{rcl}
\Hom(C,A\tens B) & \buildrel \Hom(C,c)\over\arrow \Hom(C,B\tens A)
   \buildrel d_r^{-1}\over\aow & \Hom(B\pti\tens C,A)  \\
&& \qqquad \Big \downarrow d_l \\
\Hom(C,\nu\tens B)
\raisebox{0pt}[0pt][0pt]{\put(3,-20){\vector(0,1){40} }} \qqquad &&
\Hom(B\pti,A\tens C\pti) \\
&& \qqquad \Big\downarrow \Hom(B\pti,c) \\
\Hom(C,A\tens B) & \buildrel d_l\over\lfarrow \Hom(C\tens B,A)
   \buildrel d_r^{-1}\over\lfaow  & \Hom(B\pti,C\pti\tens A)
\end{array}
\]
\[
\begin{array}{rcl}
\Hom(C,A\tens B) & \buildrel \Hom(C,c)\over\arrow \Hom(C,B\tens A)
   \buildrel d_l^{-1}\over\aow & \Hom(C\tens A\pti,B)  \\
&& \qqquad \Big \downarrow d_r \\
\Hom(C,A\tens\nu)
\raisebox{0pt}[0pt][0pt]{\put(3,-20){\vector(0,1){40} }} \qqquad &&
\Hom(A\pti,C\pti\tens B) \\
&& \qqquad \Big \downarrow \Hom(A\pti,c) \\
\Hom(C,A\tens B) & \buildrel d_r\over\lfarrow \Hom(A\pti\tens C,B)
   \buildrel d_l^{-1}\over\lfaow  & \Hom(A\pti,B\tens C\pti)
\end{array}
\]
commute.
\end{prop}

\subsEction{A quotient functor} 
Let $g:\Gamma'\to\Gamma$ be a glueing and $gf_1= gf_2:\Gamma'\to\Gamma''$.
Then $\CZ(g)\CZ(f_1)= \CZ(g) \CZ(f_2)$. The morphism
$\CZ(g):\CZ(\Gamma';A,X;X,B) \to \CZ(\Gamma;A;B)$ can be regarded as a
morphisms of left exact functors belonging to $\hat\CC_{k,l}$
\[ \CZ(g): \oplus_{X\in\CC^n} \CZ(\Gamma';A,X;X,B) \aow \CZ(\Gamma;A;B) .\]
It is an epimorphism if $k+l>0$. Hence, by definition $\CZ(f_1) = \CZ(f_2)$.

In case $k=l=0$ we add one more leg $A$ to our nets, obtaining new morphisms
$\bar g\bar f_1= \bar g\bar f_2$. By the above considerations
$\CZ(\bar f_1) = \CZ(\bar f_2): \CZ(\bar\Gamma;A;) \to \CZ(\bar\Gamma'';A;)$.
Setting $A=I$ and applying isomorphisms we change the end $A$ to a 1-vertex.
Deleting it we again deduce $\CZ(f_1) = \CZ(f_2)$.

When $g$ is any other generator it is invertible, hence, in all cases we
proved that $gf_1= gf_2$ implies $\CZ(f_1) = \CZ(f_2)$. This means that the
functor $\CZ$ is, in fact, defined on a category having the left cancellation
property.

Now we construct a functor $Z$ as an epimorphic image of $\CZ$ on the
category with left cancellations $\ON$ extended by
$ins$, $del$, $fus$, $Tw$, $Br$ to $k\vect$.

\begin{prop}
There exists a unique up to equivalence functor $Z$ on the category $\ON$
extended by $ins$, $del$, $fus$, $Tw$, $Br$ to $k\vect$ satisfying
conditions (i)--(xi) of Section~\ref{Modular}. It has also the property
(xii). There is an epimorphism $\CZ\to Z$.
\end{prop}

\subsEction{Construction of switches} 
Now we extend the functor $Z$ to $EN$. We shall show that
there are exactly two such extensions, which differ by a sign of $Z(S)$.

Morphisms $S^{\pm 1}$ in $EN$ must be represented by functorial in
$X$ isomorphisms
\[\Hom(X,\f)\simeq Z\Biggl(\tennisu X{} \!\! \Biggr) \buildrel Z(S^{\pm 1})
\over\arrow Z\Biggl(\tennisu X{} \!\! \Biggr) \simeq\Hom(X,\f). \]
Hence, they are induced by automorphisms $S^{\pm 1}:\f\to \f$ in $\CC$.

\begin{prop}
The axiom \eqref{mainSdiag_or} is satisfied, or equivalently the diagram
at Figure~\ref{realrela} made of morphisms of left exact functors in $X,Y$
(quotients are taken in the category of left exact functors) with
$K=\Ker(\int^M M\tens M\pti \to \f)$
\begin{figure}[htbp]
\[
\begin{CD}
\Hom(Y\tens X,\f)
\text{\makebox[0mm][l]{\put(20,0){$\stackrel{\Hom(c_{XY},\f)}\arrrow$}}}
@. \Hom(X\tens Y,\f)  \\
@V\Hom(Y\tens X,S^{-1})VV  @VV\wr V \\
\Hom(Y\tens X,\f) @. \Hom(X\tens Y,\int^N N\tens N\pti)/\Hom(X\tens Y,K) \\
@V\wr VV  @VV\wr V \\
\frac{\nquad{\displaystyle\Hom(Y\tens X,\int^M M\tens M\pti)}\nquad}
{\displaystyle\Hom(Y\tens X,K)} @.
     \oint^N\Hom((X\tens Y)\tens N,N)/\Hom(X\tens Y,K) \\
@V\wr VV  @VV\Hom(a,N)V \\
\frac{\nquad\oint^M{\displaystyle\Hom((Y\tens X)\tens M,M)}\nquad}
{\displaystyle\Hom(Y\tens X,K)} @.
     \oint^N\Hom(X\tens (Y\tens N),N)/\Hom(X\tens Y,K) \\
@V\Hom(a,M)VV  @VV\wr V \\
\frac{\nquad\oint^M{\displaystyle\Hom(Y\tens (X\tens M),M)}\nquad}
{\displaystyle\Hom(Y\tens X,K)} @.
\frac{\nquad\oint^{N,P}{\displaystyle\Hom(X\tens P,N)
\tens\Hom(Y\tens N,P)}\nquad} {\displaystyle\Hom(X\tens Y,K)}  \\
@V\Hom(Y\tens\nu^{-1}_{X\tens M},\nu)VV @VV\wr V \\
\frac{\nquad\oint^M{\displaystyle\Hom(Y\tens(X\tens M),M)}\nquad}
{\displaystyle\Hom(Y\tens X,K)} @.
       \oint^P\Hom(Y\tens(X\tens P),P)/\Hom(X\tens Y,K) \\
@V\Hom(a^{-1},M)VV  @VV\Hom(a^{-1},P)V \\
\qquad \frac{\nquad\oint^M{\displaystyle\Hom((Y\tens X)\tens M,M)}\nquad}
{\displaystyle\Hom(Y\tens X,K)}\qquad @.
     \quad \oint^P\Hom((Y\tens X)\tens P,P)/\Hom(Y\tens X,K)\quad \\
@V\wr VV  @VV\wr V \\
\frac{\nquad{\displaystyle\Hom(Y\tens X,\int^M M\tens M\pti)}\nquad}
{\displaystyle\Hom(Y\tens X,K)} @.
    \Hom(Y\tens X,\int^P P\tens P\pti)/\Hom(Y\tens X,K) \\
@V\wr VV  @VV\wr V \\
\Hom(Y\tens X,\f)
\text{\makebox[0mm][l]{\put(20,0){$\stackrel{\Hom(Y\tens X,S)}\arrrow$}}}
@. \Hom(Y\tens X,\f)
\end{CD}
\]
\caption{The realization of a relation for a torus with
two holes\label{realrela}}
\end{figure}
is commutative if and only if
\begin{equation}\label{*}
\invfourier
\end{equation}
for some morphism $\mu:I\to \f$.
\end{prop}

The relation~\eqref{S2=Br-1Tw-1_or} implies that
\[
\begin{array}{ccc}
\int^X X\tens X\pti & @>\int c>> \int^X X\pti\tens X @>\int1\tens\nu>> &
\int^X X\pti\tens X \\
i_X\bigg\downarrow\quad && \quad\bigg\downarrow i_{X\pti} \\
\f & \stackrel{S^{-2}}\arrrrow & \f
\end{array}
\]
by \propref{proptipti}. That is,
\be\label{S2=gamma}
S^{-2}=\gamma: \f\to\f
\end{equation}
(see \secref{intro}). So we find
\be\label{!}
\quad
\unitlength=0.7mm
\begin{picture}(112,46)
\put(88,25){\oval(20,18)[b]}
\put(102,21.50){\oval(20,19)[t]}
\put(112,22){\line(0,-1){18}}
\put(92,4){\line(0,1){8}}
\put(98,35){\line(0,1){9}}
\put(78,25){\line(0,1){19}}
\put(88,45){\makebox(0,0)[cc]{$\f$}}
\put(102,3){\makebox(0,0)[cc]{$\f$}}
\put(102,31){\line(0,1){8}}
\put(107,37){\makebox(0,0)[cc]{$\mu$}}
\put(68,24){\makebox(0,0)[cc]{$=$}}
\put(44,27){\oval(20,18)[b]}
\put(28.50,22){\oval(21,20)[t]}
\put(39,14){\line(0,-1){10}}
\put(18,22){\line(0,-1){18}}
\put(28,32){\line(0,1){8}}
\put(44,46){\makebox(0,0)[cc]{$\f$}}
\put(29,3){\makebox(0,0)[cc]{$\f$}}
\put(23,38){\makebox(0,0)[cc]{$\mu$}}
\put(8,24){\makebox(0,0)[rc]{$S= \gamma^{-1} S^{-1} =$}}
\put(57,28){\line(-3,2){23}}
\put(57,32){\oval(10,8)[r]}
\put(55,43){\line(-5,-3){9}}
\put(43,36){\line(-3,-2){6}}
\put(57.50,32){\oval(7,8)[lt]}
\end{picture}
\end{equation}
Theorem 6.13 from \cite{Lyu:mod} states
that if morphisms (\ref{*}) and (\ref{!}) are inverse to each other, then
$\mu$ is an integral of the Hopf algebra $\f$. It is unique up to a constant.
The normalizing constant for
the integral $\mu$ is fixed up to a sign by \eqref{S2=gamma}. The relation
\[(ST)^3=\lambda S^2 \]
for $T=\int 1\tens\nu:\f\to \f$ and some constant $\lambda$ is proven in
\cite{Lyu:mod}. Hence, putting the central charge $C$ equal this constant
$\lambda$ on $Z(A_1)$ we get the relation~\eqref{ST3=CS2}. For arbitrary
net $\Gamma$ of genus $g$ we set $Z(C_\Gamma)=\lambda^g$.

Thus, we obtained a functor $EN \to k$-vect satisfying all conditions of
\secref{Modular}. It is unique up to a choice of sign of the normalizing
constant.

\ifx\undefined\bysame
\newcommand{\bysame}{\leavevmode\hbox to3em{\hrulefill}\,}
\fi

\bibliographystyle{amsplain}

\begin{thebibliography}{\fontsize{11}{13pt}\selectfont[22]}
\fontsize{11}{13pt}\selectfont

\bibitem[1]{Bir:mcg}
J.~S. Birman, {\em Mapping class groups and their relationship to braid
groups}, Commun. Pure Appl. Math. {\bf 22} (1969), 213--238.

\vskip 6pt

\bibitem[2]{Dehn}
M.~Dehn, {\em Die Gruppe der Abbildungsklassen},
Acta Math. {\bf 69} (1938), 135--206.

\vskip 6pt

\bibitem[3]{Gro:esq}
A.~Grothendieck, {\em Esquisse d'un programme}, 1984, unpublished.

\vskip 6pt

\bibitem[4]{JoyStr:tor}
A.~Joyal and R.~Street, {\em Tortile Yang-Baxter operators in tensor
categories}, J. Pure Appl. Alg. {\bf 71} (1991) 43--51.

\vskip 6pt

\bibitem[5]{Ko:inv}
T.~Kohno, {\em Topological invariants for 3-manifolds using representations
of mapping class groups I}, Topology {\bf 31} (1992), n.~2, 203--230.

\vskip 6pt

\bibitem[6]{Ko:3man}
\bysame, {\em Three-manifold invariants derived from conformal field theory
and projective representations of modular groups},
Int. J. of Modern Phys. B {\bf 6} (1992), n.~11--12, 1795--1805.

\vskip 6pt

\bibitem[7]{Lic:3}
W.~B.~R. Lickorish, {\em A representation of orientable combinatorial
3-manifolds}, Annals Math. {\bf 76} (1962), n.~3, 531--540.

\vskip 6pt

\bibitem[8]{Lic:gen}
\bysame, {\em A finite set of generators for the homeotopy group
of a 2-manifold}, Proc. Camb. Phil. Soc. {\bf 60} (1964), 769--778.

\vskip 6pt

\bibitem[9]{Lyu:tan}
V.~V. Lyubashenko, {\em Tangles and Hopf algebras in braided categories},
to appear in J. Pure Appl. Alg.

\vskip 6pt

\bibitem[10]{Lyu:mod}
\bysame, {\em Modular transformations for tensor categories},
to appear in J. Pure Appl. Alg.

\vskip 6pt

\bibitem[11]{Lyu:rib=mod}
\bysame, {\em Ribbon categories as modular categories}, preprint.

\vskip 6pt

\bibitem[12]{LyuMaj}
V.~V. Lyubashenko and S.~Majid, {\em Braided groups and quantum Fourier
transform}, to appear in J. Algebra.

\vskip 6pt

\bibitem[13]{Mac:cat}
S.~Mac Lane, {\em Categories for the working mathematician},
Springer-Verlag, 1971.

\vskip 6pt

\bibitem[14]{Mag}
W.~Magnus, {\em \"Uber Automorphismen von Fundamentalgruppen berandeter
Fl\"achen}, Math. Ann. {\bf 109} (1934), 617--646.

\vskip 6pt

\bibitem[15]{Maj:bra}
S.~Majid, {\em Braided groups}, Preprint, DAMTP/90-42, 1990.

\vskip 6pt

\bibitem[16]{MooSei}
G.~Moore and N.~Seiberg, {\em Classical and Quantum Conformal Field Theory},
Commun. Math. Phys., {\bf 123} (1989), 177--254.

\vskip 6pt

\bibitem[17]{Res:rib}
N.~Reshetikhin, {\em Quasitriangular Hopf algebras, solutions to the
Yang-Baxter equations and link invariants},
Algebra and Analysis {\bf 1} (1989), n.~2, 169--194.

\vskip 6pt

\bibitem[18]{ResTur:3}
N.~Reshetikhin and V.~G. Turaev, {\em Invariants of 3-manifolds via link
polynomials and quantum groups}, Invent. math. {\bf 103} (1991), 547--597.

\vskip 6pt

\bibitem[19]{Scott}
G.~P. Scott, {\em Braid groups and the group of homeomorphisms of a surface},
Proc. Camb. Phil. Soc. {\bf 68} (1970), 605--617.

\vskip 6pt

\bibitem[20]{Shu}
M.~C. Shum, {\em Tortile tensor categories}, Macquarie preprint, 1989.

\vskip 6pt

\bibitem[21]{Tur:q3}
V.~Turaev, {\em Quantum invariants of 3-manifolds},
Publ. de l'IRMA 509/P-295, October 1992.

\vskip 6pt

\bibitem[22]{Waj}
B.~Wajnryb, {\em A simple presentation for the mapping class group of
an orientable surface}, Israel J. Math. {\bf 45} (1983), no.~2--3, 157--174.

\vskip 6pt

\bibitem[23]{Wal}
K.~Walker, {\em On Witten's 3-manifold invariants}, Preprint, February 1991.

\vskip 6pt

\bibitem[24]{Wit:Jones}
E.~Witten, {\em Quantum field theory and the Jones polynomial},
Commun. Math. Phys. {\bf 121} (1989), 351--399.

\end{thebibliography}

\end{document}